\documentclass[aps,%
superscriptaddress,
 amsmath,amssymb,
floatfix,
twocolumn
]{revtex4-2}

\usepackage[utf8]{inputenc}
\usepackage[T1]{fontenc}
\usepackage{etoolbox}

\usepackage[dvipdfm]{graphicx} 
\usepackage{amsmath}
\usepackage{physics}
\usepackage{xcolor,colortbl}
\usepackage{bm}
\usepackage{mathtools}
\usepackage[ruled,lined]{algorithm2e}
\usepackage{siunitx}
\usepackage{chemformula}
\usepackage{titlesec}
\usepackage[acronym]{glossaries}
\usepackage{enumitem}
\usepackage{grffile}
\usepackage{booktabs} 
\usepackage{listings}
\usepackage{xcolor}
\usepackage{ulem}

\usepackage{cancel}
\usepackage{multirow}
\usepackage{makecell}
\usepackage{pifont}
\newcommand{\gcheck}{\textcolor{green!50!black}{\checkmark}}

\definecolor{codegreen}{rgb}{0,0.6,0}
\definecolor{codegray}{rgb}{0.5,0.5,0.5}
\definecolor{codepurple}{rgb}{0.58,0,0.82}
\definecolor{backcolour}{rgb}{0.95,0.95,0.92}

\lstdefinestyle{mystyle}{
    backgroundcolor=\color{backcolour},   
    commentstyle=\color{codegreen},
    keywordstyle=\color{magenta},
    numberstyle=\tiny\color{codegray},
    stringstyle=\color{codepurple},
    basicstyle=\ttfamily\footnotesize,
    breakatwhitespace=false,         
    breaklines=true,                 
    captionpos=b,                    
    keepspaces=true,                 
    numbers=left,                    
    numbersep=5pt,                  
    showspaces=false,                
    showstringspaces=false,
    showtabs=false,                  
    tabsize=2
}

\usepackage{booktabs}
\usepackage[table]{xcolor}

\definecolor{cA}{RGB}{198,239,206}  
\definecolor{cB}{RGB}{170,235,190}  
\definecolor{cC}{RGB}{235,250,240}  
\definecolor{cD}{RGB}{255,247,173}  
\definecolor{cE}{RGB}{255,235,190}  
\definecolor{cF}{RGB}{255,220,190}  
\definecolor{cG}{RGB}{255,200,200}  
\definecolor{cH}{RGB}{255,170,170}  
\definecolor{cFail}{RGB}{224,224,224} 

\newcommand{\MFO}{MACE-POLAR-1-OMOL }

\newcommand{\MOMOL}{MACE-OMOL }

\newcommand{\MFP}{MACE-POLAR-1 }
\newcommand{\MFPM}{MACE-POLAR-1-M }
\newcommand{\MFPL}{MACE-POLAR-1-L }

\newcommand{\UMAS}{UMA-S-1P1 }
\newcommand{\UMAM}{UMA-M-1P1 }
\newcommand{\UMASOMOL}{UMA-S-1-OMOL }
\newcommand{\UMAMOMOL}{UMA-M-1-OMOL }
\newcommand{\ORBMOL}{ORBMOL }
\newcommand{\wb}{$\omega$B97M-V}

\providecommand{\xmark}{\ding{55}} 
\newcommand{\rcross}{\textcolor{red!70!black}{\xmark}}
\newcommand{\R}{\mathbb{R}}

\begin{document}
\title{MACE-POLAR-1: A Polarisable Electrostatic Foundation Model for Molecular Chemistry}
\author{Ilyes Batatia}
\affiliation{Engineering Laboratory, University of Cambridge, Trumpington St, Cambridge, UK}
\author{William J. Baldwin}
\affiliation{Engineering Laboratory, University of Cambridge, Trumpington St, Cambridge, UK}
\author{Domantas Kuryla}
\affiliation{Yusuf Hamied Department of Chemistry, University of Cambridge, Lensfield Road, Cambridge, UK}
\affiliation{Engineering Laboratory, University of Cambridge, Trumpington St, Cambridge, UK}
\author{Joseph Hart}
\affiliation{Engineering Laboratory, University of Cambridge, Trumpington St, Cambridge, UK}
\affiliation{Cavendish Laboratory, University of Cambridge, J. J. Thomson Avenue, Cambridge, UK}
\author{Elliott Kasoar}
\affiliation{Scientific Computing Department, Science and Technology Facilities Council, Daresbury Laboratory, Keckwick Lane, Daresbury WA4 4AD, UK}
\affiliation{Engineering Laboratory, University of Cambridge, Trumpington St, Cambridge, UK}
\author{Alin M. Elena}
\affiliation{Scientific Computing Department, Science and Technology Facilities Council, Daresbury Laboratory, Keckwick Lane, Daresbury WA4 4AD, UK}
\author{Harry Moore}
\affiliation{Engineering Laboratory, University of Cambridge, Trumpington St, Cambridge, UK}
\affiliation{Ångström AI, San Francisco, California, USA}
\author{Miko{\l}aj J. Gawkowski}
\affiliation{Department of Physics and Astronomy, University College London, 7-19 Gordon St, London WC1H 0AH, United Kingdom}
\author{Benjamin X. Shi}
\affiliation{Initiative for Computational Catalysis, Flatiron Institute, 160 5th Avenue, New York, NY 10010}
\author{Venkat Kapil}
\affiliation{Department of Physics and Astronomy, University College London, 7-19 Gordon St, London WC1H 0AH, United Kingdom}
\author{Panagiotis Kourtis}
\affiliation{School of Natural and Environmental Sciences, Newcastle University, Newcastle upon Tyne, UK}
\author{Ioan-Bogdan Magd\u{a}u}
\affiliation{School of Natural and Environmental Sciences, Newcastle University, Newcastle upon Tyne, UK}
\author{Gabor Csanyi}
\affiliation{Engineering Laboratory, University of Cambridge, Trumpington St, Cambridge, UK}
\affiliation{Max Planck Institute for Polymer Research, Ackermannweg 10, Mainz, Germany}
\date{\today}

\begin{abstract}
Accurate modelling of electrostatic interactions and charge transfer is fundamental to computational chemistry, yet most machine learning interatomic potentials (MLIPs) rely on local atomic descriptors that cannot capture long-range electrostatic effects. We present a new electrostatic foundation model for molecular chemistry that extends the MACE architecture with explicit treatment of long-range interactions and electrostatic induction. Our approach combines local many-body geometric features with a non-self-consistent field formalism that updates learnable charge and spin densities through polarisable iterations to model induction, followed by global charge equilibration via learnable Fukui functions to control total charge and total spin. This design enables an accurate and physical description of systems with varying charge and spin states while maintaining computational efficiency and ease of training. Trained on the OMol25 dataset of 100 million hybrid DFT calculations, our models achieve chemical accuracy across diverse benchmarks, with accuracy competitive with hybrid DFT on thermochemistry, reaction barriers, conformational energies, and transition metal complexes. Notably, we demonstrate that the inclusion of long-range electrostatics leads to a large improvement in the description of non-covalent interactions and supramolecular complexes over non-electrostatic models, including sub-kcal/mol prediction of molecular crystal formation energy in the X23-DMC dataset and a fourfold improvement over short-ranged models on protein-ligand interactions. Our model also demonstrates an improved description of ions and redox reactions of transition metals in solution. The model's ability to handle variable charge and spin states, respond to external fields, provide interpretable spin-resolved charge densities, and maintain accuracy from small molecules to protein-ligand complexes positions it as a versatile tool for computational molecular chemistry and drug discovery.
\end{abstract}

\maketitle

\section{Introduction}
Electrostatic interactions govern the structure, dynamics, and function of molecular systems throughout chemistry and biology. From the intricate folding of proteins stabilised by salt bridges, to the specific recognition of substrates by enzymes through complementary charge distributions, electrostatics dictate molecular behaviour at many scales. In drug discovery, the electrostatic complementarity between a ligand and its protein target is a primary determinant of binding affinity and specificity. In materials science, charge transfer and polarisation phenomena at interfaces control the performance of catalysts, batteries, and electronic devices. Despite this fundamental importance, the accurate and efficient modelling of electrostatic interactions remains one of the central challenges in computational chemistry.
Machine learning interatomic potentials (MLIPs) have emerged as a transformative tool for computational chemistry, reaching the accuracy of quantum mechanics (QM) at a fraction of the computational cost. The leading MLIP architectures~\cite{batatia_mace_2023, batzner20223NequIP, fu2025learningsmoothexpressiveinteratomic, rhodes2025orbv3atomisticsimulationscale, Mazitov2025}, most notably message-passing neural networks, have achieved success by learning complex and accurate relationships between the local atomic environment of an atom and its contribution to total energy. Although these local descriptors excel at capturing short-range quantum effects such as covalent bonding, Pauli repulsion, and short-range electrostatics, they are, by construction, unable to model the long-range nature of electrostatics or dispersion. In many systems, long-range electrostatic interactions are screened, and local models can accurately reproduce many observables of complex chemical systems. However, the absence of long-range interactions becomes a critical failure point for charged molecules, ionic materials, and large biomolecular complexes, where long-range electrostatics is a dominant physical interaction. Moreover, this limitation prevents these models from responding to external electric fields, correctly modelling charge transfer between distant fragments, or describing polarisation in extended systems. 

To address this, several strategies have been developed to incorporate electrostatics into MLIPs; for comprehensive reviews see Refs.~\cite{Olexandr_lr_review, behler_4gnn_review_2021, Baldwin2026SCF, grasselli2026longrangeelectrostaticsatomisticmachine}. The simplest approaches predict partial charges or atomic multipoles directly from local geometry. Early examples include the 3\textsuperscript{rd} Generation Neural Network (3GNN)~\cite{Morawietz2012ACharges, Artrith2011}, the polarisable multipolar electrostatic potential~\cite{Mills2011} and PhysNet~\cite{physnet2019}. The LODE~\cite{Grisafi2019} computes long-range features using equivariant descriptors as sources and different algebraic decays to model electrostatics and dispersion effects.
Although these models include long-range electrostatic interactions through a Coulomb term, they cannot handle systems with varying total charge. Subsequent architectures introduced mechanisms to enforce a specific total charge, either through global charge embeddings, as in Deep Potential Long Range (DPLR)~\cite{Zhang2022AInteractions, Zhang2024}, the Latent Ewald Summation (LES) method~\cite{Cheng2025}, or the SO3LR model~\cite{Kabylda2025}; through multi-hypothesis total-charge equilibration in FENNIX~\cite{Pl__2023, PL_2025}; or through learned Fukui functions as in AIMNet-NSE~\cite{aimnetnse}. However, predicting charges purely from local geometry has fundamental limitations: such models cannot capture induced polarisation between well-separated subsystems or long-range charge transfer~\cite{behler_4gnn_review_2021, kocer2024machinelearningpotentialsredox}.

More sophisticated approaches employ charge equilibration (QEq) schemes~\cite{qeq1985, qeq1986, Rappe1991ChargeSimulations}, as first demonstrated in the 
CENT architecture~\cite{cent2015, cent2017, cent2019} and subsequently in 4G-HDNN~\cite{4gnn_ko_2020}, kQEq~\cite{kqeq_og2022}, and BAMBOO~\cite{Gong2025}. These models define an energy functional quadratic in charges, with electronegativities and hardnesses predicted from local geometry, and obtain charges by minimising this functional, similar to the self-consistency loop of density functional theory (DFT). Although these models can capture induced polarisation through their self-consistency loop, classical QEq suffers from fundamental deficiencies, including incorrect fractional charge separation upon dissociation~\cite{Jensen2023UnifyingModels, Perdew1982Density-FunctionalEnergy, Vondrak2025PushingLimits} and unphysical cubic scaling of polarisability with system size~\cite{LeeWarren2008OriginMethods, nonlinear_pol_fq}, leading to metal-like over-screening in large insulating systems~\cite{conducting_molecules} that makes them inadequate for biomolecules. Alternative self-consistent machine learning approaches based on electronic structure theory~\cite{Thomas2025}, such as SCFNN~\cite{scfnn} using iteratively updated Wannier function centres, eMLP~\cite{eMLP} treating Wannier centres as pseudo-atoms, and BpopNN~\cite{bpopnn} inspired by orbital-free DFT, can capture phenomena like induced polarisation and long-range charge transfer, but are significantly more cumbersome to train and deploy because of their self-consistent loop and often require constrained DFT data for training.

A central goal of atomistic modelling is to obtain interatomic potentials that can be deployed out of the box across broad chemical space, while retaining near--\textit{ab initio} accuracy for energies and forces. Foundation force fields~\cite{batatia2024foundationmodelatomisticmaterials, Deng2023, wood2025umafamilyuniversalmodels} pursue this by pretraining on large, chemically diverse datasets and transferring across phases and chemistries.  The release of large-scale datasets and foundation models in both materials and molecular chemistry has transformed computational chemistry by democratising the use of MLIPs for a wide range of chemists.

In molecular chemistry, the development of transferable potentials has followed a distinct trajectory shaped by the challenges of variable charge, spin, and long-range interactions. Moreover, the accurate description of molecular chemistry often requires high levels of electronic structure theory (hybrid DFT or even wavefunction methods), which significantly increases the cost of generating large and diverse datasets. Early transferable MLIPs such as ANI-1 and ANI-2x established broad coverage across neutral closed-shell small organic molecules.\cite{smith2017ani1,devereux2020ani2x,smith2020ani1x} MACE-OFF~\cite{kovacs2025maceoff}, trained on the SPICE dataset~\cite{eastman2022spicedatasetdruglikemolecules}, demonstrated that pre-trained short-range MLIPs can reach \textit{ab initio} accuracy across a wide range of chemical systems, from molecular liquids and crystals to drug-like molecules and biopolymers. AIMNet~\cite{aimnet2019} and AIMNet2~\cite{anstine2025aimnet2, Kalita2025} introduced neural spin equilibration to handle charged and open-shell species, enabling accurate treatment of radicals and small-molecule reactivity. SO3LR~\cite{kabylda2025so3lr} combined learned short-range interactions with explicit long-range electrostatics and dispersion computed from local charges, improving generalisation to condensed-phase environments. Domain-focused foundation models targeting biomolecular simulations, such as FeNNix-Bio~\cite{Pl__2023,PL_2025}, have achieved accuracies competitive with classical force fields for protein and nucleic acid dynamics. The release of the OMol25 dataset~\cite{levine2025openmolecules2025omol25} has represented a breakthrough for large-scale pretraining in molecular chemistry, owing to its unprecedented size and chemical coverage, computed throughout at the $\omega$B97M-V range-separated hybrid level of theory. Short-range MLIPs trained on OMol25---including UMA~\cite{wood2025umafamilyuniversalmodels}, MACE-OMol~\cite{levine2025openmolecules2025omol25}, OrbMol~\cite{rhodes2025orbv3atomisticsimulationscale}, and MACE-MH-1~\cite{batatia2025crosslearningelectronicstructure}---have demonstrated unprecedented accuracies on molecular benchmarks, establishing MLIPs as a credible replacement for semi-empirical methods and, increasingly, for DFT itself in molecular simulations.

In this work, we introduce a new family of electrostatic foundation models, \MFP, that provide a physics-based treatment of long-range electrostatic interactions through induction effects while retaining computational efficiency. Our model builds on the proven accuracy of the MACE architecture for short-range interactions and incorporates long-range electrostatics through a novel non-local field update. Sequential polarisable updates to a non-local charge density refine atomic multipoles in response to the electrostatic potential; after each update, the total charge and spin of the system are equilibrated using learnable, environment-dependent Fukui functions, a mechanism inspired by conceptual density functional theory and the AIMNet-NSE model~\cite{aimnetnse}. This design captures the essential physics of polarisation and charge transfer while avoiding the cost and potential instability of self-consistent field iterations.
We demonstrate the capabilities of this approach by training \MFP models on 100 million diverse molecular structures from the OMol25 dataset and conducting an extensive benchmark campaign spanning thermochemistry, reaction barriers, conformational energies, transition metal complexes, protein--ligand interactions, supramolecular complexes, molecular crystals, redox chemistry in solution, and molecular liquids. Our results show that the explicit inclusion of physics-based electrostatics dramatically improves the description of charged systems and non-covalent complexes, achieving state-of-the-art accuracy on challenging benchmarks such as molecular crystal lattice energies and protein--ligand binding. The model's ability to handle variable charge and spin states, respond to external fields, provide interpretable spin-resolved charge densities, and maintain accuracy from small molecules to protein--ligand complexes positions it as a versatile tool for computational molecular chemistry, drug discovery and bio-simulations.

\section{Theory and Methods}

\subsection{Theoretical Foundation}

The accurate description of molecular systems requires capturing both short-range and long-range interactions. While message-passing neural networks excel at learning local quantum mechanical interactions, they cannot, by construction, capture interactions beyond their cutoff radius. Our approach addresses this fundamental limitation through explicit treatment of long-range interactions while maintaining the flexibility and generalisability of neural network potentials for short-range interactions.

The total energy of an atomistic system is decomposed into short-range and long-range contributions:
\begin{equation}
E_{\text{total}} = E_{\text{local}} + E_{\text{non-local}} + E_{\text{electrostatic}}
\end{equation}
where $E_{\text{local}}$ captures most of the contributions to the energy, including covalent bonding, Pauli repulsion, and short-range electrostatic effects, and is predicted by a local MLIP model; $E_{\text{electrostatic}}$ describes smeared long-range Coulombic interactions; and $E_{\text{non-local}}$ accounts for a learned correction that captures residual non-local effects beyond pure electrostatics, including dispersion. In the next section, we explain how we restrict the flexibility of $E_{\text{non-local}}$ by parametrising it as a function of local geometry and a non-local charge density. The locality of the $E_{\text{local}}$ term depends on the choice of hyper-parameters for the local MLIP, and it typically describes interactions between $10$ and $20$~\AA. For most MLIP models, this depends on the number of layers used for message passing and the receptive field at each layer. See Table~\ref{tab:mpnn_ranges} for the receptive field of the models tested in this paper.

\subsection{Model Architecture}

Our model extends the MACE~\cite{batatia_mace_2023} architecture with explicit long-range interactions through a field-dependent induction mechanism. The key addition is a non-self-consistent field formalism that updates atomic multipoles based on the local electric field while maintaining computational efficiency. For a detailed description of the MACE architecture, see~\cite{batatia_mace_2023, Kovcs2023}. For a full exposition of the design space of electrostatics extensions to the MACE architecture, see our companion paper on the design space of self-consistent electrostatics extensions to MLIPs~\cite{Baldwin2026SCF}, which also presents a self-consistent version of the field charge equilibration used here.

We first predict local features using $(T)$ MACE layers, forming node features at each layer $h_{i,klm}^{(t)}$, where $i$ is the atom index, $k$ is the channel index of the MACE features, and $lm$ are the usual spherical indices. These node features encode rich many-body information about the geometry and chemistry of the local environment of atom $i$. We also read out a local contribution to the energy $E_{\text{local}}$. In the rest of the paper, we denote feature vectors in bold, with the $(klm)$ indices implicit, for example $(\textbf{h}_{i}^{(t)})_{klm} := h_{i,klm}^{(t)}$. 
To model both the electrostatic and non-local contributions to the energy, we use a physics-based model of long-range interactions. Below, we outline how we introduce an explicit spin-charge density in the model as a proxy feature for long-range interactions, and use it to compute electrostatic and non-local contributions. The main motivation to learn a physics-based model of long-range interactions is to ensure that it can be fitted to small and medium-sized systems accessible from ab initio data and extrapolated appropriately to larger systems.

\begin{figure*}
    \includegraphics[width=\linewidth]{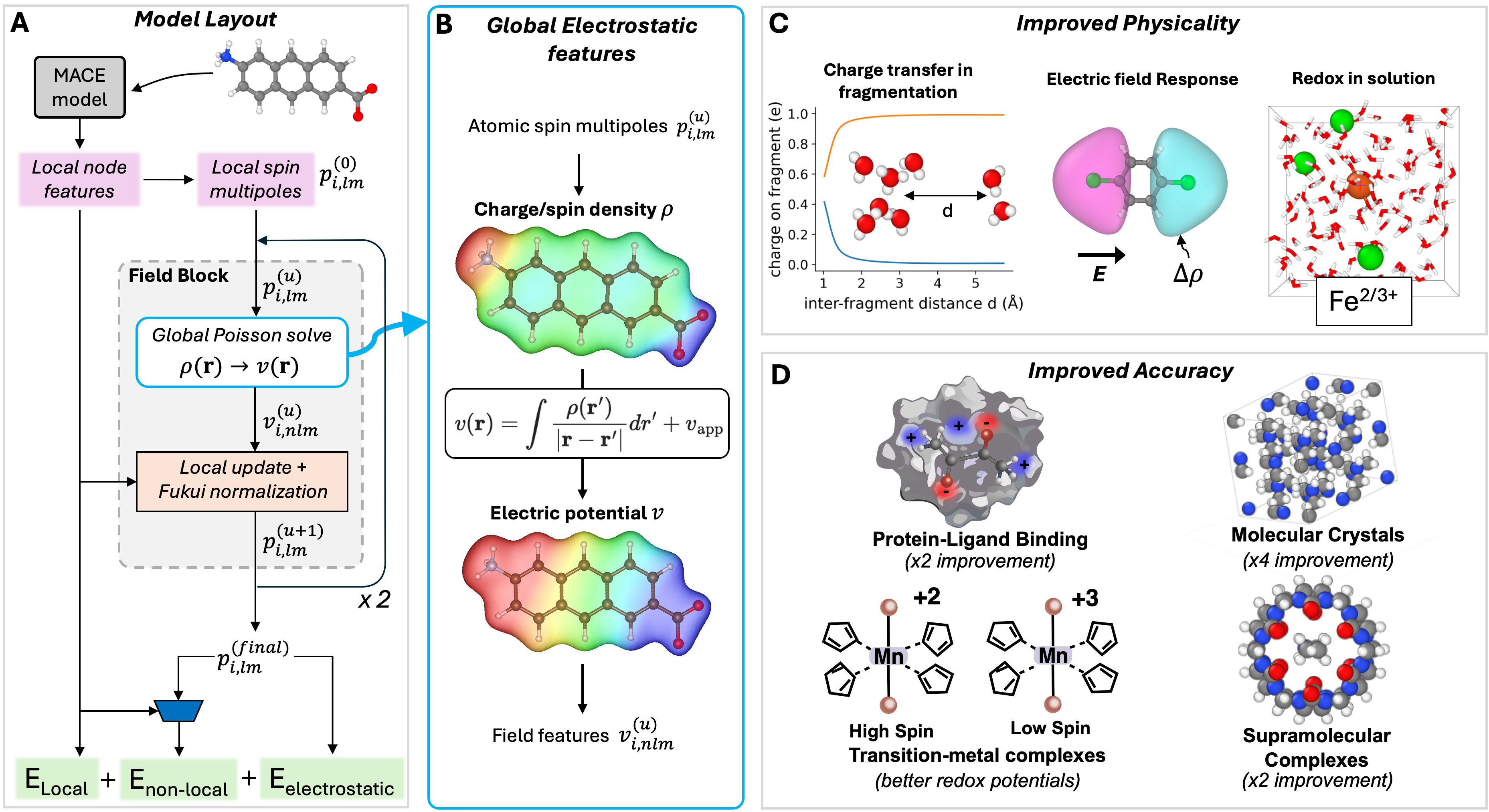}
      \caption{\textbf{Overview of the \MFP architecture and benchmarked applications.}
      (A) Model architecture. Atomic positions and species are passed to a MACE model, which predicts local node features, a local energy contribution, and an initial set of spin-charge multipoles. The local features richly encode semi-local geometry and chemistry. Then, the spin-charge multipoles are iteratively refined through a long-range operation. First, a physically inspired global convolution maps the spin-charge multipoles into atom-centred electrostatic features $v_{i,nlm}$. Then, a local operation followed by a global normalisation predicts a new set of spin-charge multipoles. The process is iterated twice, before the final set of multipoles is used to compute a Coulomb energy and an additional learned non-local energy contribution. 
      (B) Construction of long-range electrostatic features. Information is propagated between distant atoms by constructing a smooth charge density $\rho(\mathbf{r})$ from the spin-charge multipoles and convolving with the Coulomb kernel to give an electric potential $v(\mathbf{r})$. Then, one can project the potential onto atom-centred functions to give equivariant electrostatic features. 
      (C) Physical extrapolation capabilities: charge localisation when fragmenting a cluster, response of the model charge density to an electric field, and correct prediction of the oxidation state of transition metal ions in water.
      (D) Application domains benchmarked: protein-ligand binding, molecular crystals, transition-metal redox potentials, and supramolecular complexes.}
      \label{fig:summary}
\end{figure*}

\subsubsection{Coarse-graining the electron density}

In our long-range extension, we introduce a learnable spin-resolved charge
density as the central non-local feature of the model that will contribute to the total energy. We refer to this quantity as the spin-charge density. Before defining it
formally, we explain its physical content. Throughout this paper, we use atomic units ($e=\hbar=m_e=4\pi\epsilon_0=1$) unless stated otherwise.

In electronic structure, the electron density of a molecular
system, $n(\mathbf{r})$, is a positive function over $\mathbf{r} \in \R^{3}$ that integrates
to the total number of electrons~$N_{\text{el}}$ and gives the expected number of electrons per unit volume at each point in space. The charge density is the difference between the positive nuclear charges and the electron density:
\begin{equation}
\rho(\mathbf{r})
  = \sum_i Z_i\,\delta(\mathbf{r} - \mathbf{R}_i)
  - n(\mathbf{r}),
\label{eq:charge-dens-def}
\end{equation}
where $Z_i$ is the nuclear charge of atom $i$ at position $\mathbf{R}_i$.
The charge density integrates to $\sum_i Z_i - N_{\text{el}} = Q$, the total
charge.
Because the nuclear charge $Z_i$ around each atom is
almost entirely screened by the approximately $Z_i$ electrons that
surround it, the charge density is close to zero near each nucleus in
a neutral molecule: it represents only the residual net
charge arising from bonding, charge transfer, and polarisation. In
this sense, $\rho(\mathbf{r})$ is analogous to a deformation density
in crystallography.
In our model, rather than learning the electron density
$n(\mathbf{r})$ directly, we learn a smooth coarse-graining of the charge density $\rho(\mathbf{r})$.
This has a fundamental consequence for the energy decomposition. The
dominant electrostatic contributions to the total
energy---nuclear--nuclear repulsion and nuclear--electron
attraction---are short-ranged because they are locally screened, and
are absorbed entirely into $E_{\text{local}}$, which the MACE model
learns from data. The explicit $E_{\text{electrostatic}}$ term
accounts only for the much smaller long-range Coulomb interaction of
this smooth residual density. The $E_{\text{non-local}}$ captures residual long-range interactions beyond pure electrostatics, including dispersion, as a function of this smooth charge density and local geometry.

\subsubsection{Multipoles of the Charge and Spin Density}
We now define the spin-charge density formally. For a system
of $N$ atoms at positions $(\mathbf{R}_1, \ldots, \mathbf{R}_N)$
with chemical species $(Z_1, \ldots, Z_N)$, total charge $Q$, and
total spin $S$, the spin-charge density
\begin{align}
    \rho^{\uparrow \downarrow}(\mathbf{r})
    := \{\rho^{\uparrow}(\mathbf{r}),\;
    \rho^{\downarrow}(\mathbf{r})\} \in \mathbb{R}^2
\end{align}
gives the spin-up ($\uparrow$) and spin-down ($\downarrow$)
components of the residual density at each point
$\mathbf{r} \in \mathbb{R}^3$ as defined in Equation~\ref{eq:charge-dens-def}.
Each spin channel of the charge density is therefore defined as,
\begin{align}
    \rho^{\uparrow}(\mathbf{r})
    := \sum_i \frac{Z_i}{2}\,\delta(\mathbf{r} - \mathbf{R}_i)
  - n^{\uparrow}(\mathbf{r}) \\
  \rho^{\downarrow}(\mathbf{r})
    := \sum_i \frac{Z_i}{2}\,\delta(\mathbf{r} - \mathbf{R}_i)
  - n^{\downarrow}(\mathbf{r})
\end{align}
with $n^{\uparrow}(\mathbf{r})$ the electron density of spin-up electrons, which integrates to the number of spin-up electrons $N_{\text{el}}^{\uparrow}$, and $n^{\downarrow}(\mathbf{r})$ the electron density of spin-down electrons, which integrates to the number of spin-down electrons $N_{\text{el}}^{\downarrow}$. On each atom, half the nuclear charge, $Z_i/2$, is used to screen each spin channel of the electron density.
Throughout the paper, we leave the dependence
on the atomic configuration, total charge, and total spin implicit and we write $\mathbf{\rho^{\uparrow \downarrow}}(\mathbf{r}, \mathbf{R}_1, Z_{1}, \ldots, \mathbf{R}_N, Z_{N}, Q, S):=\mathbf{\rho^{\uparrow \downarrow}}(\mathbf{r})$. This density may be non-local as its
value at a point $\mathbf{r}$ can depend on the positions and species of
atoms far from $\mathbf{r}$.

Because the nuclear point charges and core electron density have 
been coarse-grained away through the screening described above, 
the residual charge density $\rho(\mathbf{r})$ is a smooth, slowly 
varying function that lacks the rapid oscillations and cusps of the 
full electronic density. Only this low-frequency component of the 
charge density gives rise to long-range electrostatic interactions; 
the high-frequency structure associated with core electrons and 
nuclear cusps produces interactions that decay rapidly with distance 
and is therefore captured by the local MACE model within 
$E_{\text{local}}$. A low-order multipole expansion with broad 
Gaussian smearing is consequently sufficient to represent $\rho(\mathbf{r})$ for the purpose of computing 
long-range electrostatics.

We expand the spin-charge density using atomic multipoles and Gaussian type orbitals (GTOs):
\begin{align}
\label{eq:rho-up}
\mathbf{\rho^{\uparrow}}(\mathbf{r}) &= \sum_{i,lm} p_{i,lm}^{\uparrow} \phi_{n=1,lm}(\mathbf{r} - \mathbf{r}_i) \\
\label{eq:rho-down}
\mathbf{\rho^{\downarrow}}(\mathbf{r}) &= \sum_{i,lm} p_{i,lm}^{\downarrow} \phi_{n=1,lm}(\mathbf{r} - \mathbf{r}_i) \\
\phi_{nlm}(\mathbf{r} - \mathbf{r}_i) &=  C_{nl} |\mathbf{r} - \mathbf{r}_i|^{l} \exp\left(-\frac{|\mathbf{r} - \mathbf{r}_i|^2}{2\sigma_{n}^2}\right) Y_{lm}\!\left(\widehat{\mathbf{r} - \mathbf{r}_i}\right) \label{eq:gaussian-basis}
\end{align}
where $p_{i,lm}^{\uparrow}$ and $p_{i,lm}^{\downarrow}$ are the atomic multipole coefficients of spin-up and spin-down channels, respectively, $C_{nl}$ is a normalisation constant, $\sigma_n$ is the Gaussian smearing width and $Y_{lm}$ are spherical harmonics. As mentioned above, we made implicit the dependency of the multipole coefficients on the atomic positions, atomic species, and total charge and spin. The charge density $\rho$ and the spin density $s$ can be computed from the spin-charge density as shown in Eq.~\ref{eq:spin-charge-dens}, by taking the sum or the difference of the two channels of the spin-charge density,
\begin{align}
\label{eq:spin-charge-dens}
\mathbf{\rho}(\mathbf{r}) &= \mathbf{\rho^{\uparrow}}(\mathbf{r}) +  \mathbf{\rho^{\downarrow}}(\mathbf{r}), \quad s(\mathbf{r}) = \mathbf{\rho^{\uparrow}}(\mathbf{r}) - \mathbf{\rho^{\downarrow}}(\mathbf{r}).
\end{align}
Throughout, we use the superscript $^{\uparrow \downarrow}$ to represent the spin channels on each quantity. For example, the two channels of spin-charge density will be written as $\rho^{\uparrow \downarrow} = \{\rho^{\uparrow}, \rho^{\downarrow}\}$ to simplify notation. 
The $p_{i, 00}^{\uparrow \downarrow}$ features correspond to up and down atomic charges, the $\{p_{i, 1m}^{\uparrow \downarrow}\}_{m \in [-1, 0, 1]}$ to (up and down) atomic dipoles, and for $l>1$ one obtains the higher-order multipoles, resolving the spin-charge density at higher and higher resolutions.
The spin and charge densities need to obey total normalisation constraints,
\begin{align}
\int s(\mathbf{r})d\mathbf{r} = S, \quad \int \rho(\mathbf{r})d\mathbf{r} = Q,
\end{align}
where $S$ corresponds to the total number of unpaired electrons and $Q$ is the total charge. We adopt the convention $S = N^{\downarrow}_{\text{el}} - N^{\uparrow}_{\text{el}} \geq 0$, which in the absence of spin-orbit coupling is equivalent to the opposite convention by the spin-reversal 
symmetry of the non-relativistic Hamiltonian and has no effect on any observable.
Restricting the non-local contribution to the total energy to be a function of a non-local spin-charge density and local geometry descriptors prevents the model from learning overly flexible non-local energy terms that would not extrapolate well. The use of two scalar fields for the spin-resolved density corresponds to a coarse-graining of collinear unrestricted spin-DFT (spin-polarised Kohn–Sham DFT). In particular, the model can correctly predict a non-zero spin density even for closed-shell systems, which makes the model smoother and enables a better description of reactivity. The inclusion of spin-orbit coupling would require introducing a magnetisation vector field, which we leave for future work.

\subsubsection{Local guess to the spin-charge density}

Using the MACE node features $h_{i,klm}^{(t)}$ in each layer $t$, we first predict multipoles on each atom:
\begin{equation}
\tilde{p}_{i,lm}^{(0),\uparrow \downarrow} =\sum_{t}^{T} \sum_{k} W_{lk}^{(t)}\, h_{i,klm}^{(t)}
\end{equation}
where T is the number of layers of the local MACE model, and $W$ is a learnable weight matrix. 
These multipoles correspond to a local baseline, capturing the geometrical and chemical dependence of the spin-charge density that can be well described within the receptive field of the local model.

Then, we equilibrate the monopoles $\tilde{p}_{i, 00}^{(0), \uparrow \downarrow}$ to integrate to the correct total charge and total spin using the Fukui mechanism,
\begin{align}
f_i^{(0), \uparrow \downarrow} &= \text{MLP}(h_{i,k00}^{(T)}) \\
p_{i, 00}^{(0), \uparrow} &= \tilde{p}_{i, 00}^{(0),\uparrow} + \frac{f_i^{(0),\uparrow}}{\sum_j f_j^{(0), \uparrow}} \left( \frac{Q + S}{2} - \sum_j \tilde{p}_{j, 00}^{(0),\uparrow} \right) \label{eq:fukui-eq-up}\\
p_{i, 00}^{(0), \downarrow} &= \tilde{p}_{i, 00}^{(0), \downarrow} + \frac{f_i^{(0),\downarrow}}{\sum_j f_j^{(0), \downarrow}} \left( \frac{Q - S}{2} - \sum_j \tilde{p}_{j, 00}^{(0), \downarrow} \right)
\label{eq:fukui-eq-down}
\end{align}
where $f_i^{(0)}$ are locally predicted Fukui features, $Q$ is the target total charge and $S$ the total spin. The values of $\frac{Q + S}{2}$ and $\frac{Q - S}{2}$ are the target spin-resolved normalisation constraints for the $\uparrow$ and $\downarrow$ residual-charge channels, respectively. This equilibration is equivalent to AIMNet's ``neural charge equilibration''~\cite{aimnetnse}. We use the term "Fukui" here to emphasise the connection between the $f_i$ features and the usual Fukui functions in conceptual DFT. In the supplementary information section~\ref{sec:fukui-derivation}, we derive an explicit connection of the Fukui features to conceptual DFT and show that the Fukui features are equal to the partial derivative of the monopole coefficients with respect to the chemical potential $f_{i}^{\uparrow \downarrow} := \frac{\partial p_{i}^{\uparrow \downarrow}}{\partial \mu}$. The equations~\ref{eq:fukui-eq-up} and \ref{eq:fukui-eq-down} can therefore be understood as a first-order Taylor expansion of the monopoles,
\begin{align}
    \Delta p_{i,00}^{\uparrow \downarrow} \approx \frac{\partial  p_{i,00}^{\uparrow \downarrow}}{\partial Q^{\uparrow \downarrow}} \Delta Q^{\uparrow \downarrow} = \frac{\frac{\partial p_{i}^{\uparrow \downarrow}}{\partial \mu}}{\sum_j \frac{\partial p_{j}^{\uparrow \downarrow}}{\partial \mu}} \Delta Q^{\uparrow \downarrow}
\end{align}
where the quantities $\Delta Q^{\uparrow} = \frac{Q + S}{2} - \sum_j \tilde{p}_{j, 00}^{(0),\uparrow}$ and $\Delta Q^{\downarrow} = \frac{Q - S}{2}- \sum_j \tilde{p}_{j, 00}^{(0),\downarrow}$ are the spin-resolved channel deficits. As total charge and spin normalisation only affect the monopoles, the rest of the multipoles (e.g. dipoles) are equal to the unequilibrated multipoles: $\{p_{i,lm}^{(0)}:=\tilde{p}^{(0)}_{i,lm}\}_{l>0}$.

\subsubsection{Long-range Polarisable Field Updates and Fukui Equilibration}

In order to capture polarisation effects and long-range charge transfer, the model performs a series of long-ranged updates to the spin-charge density, which are inspired by a self-consistent field loop~\cite{Baldwin2026SCF}. Each global update consists of three steps, which we call (1) long-ranged electrostatic feature construction, (2) local multipole update, and (3) Fukui equilibration. We now describe each of these steps.

In the following, $u=1,...,U$ denotes the update iteration number (similar to layer number $t$ in the local part of the model), $\|$ denotes the concatenation of vectors, the different $W$ represent learnable weight matrices and MLP stands for multilayer perceptron.

1. \textbf{Long-Ranged Electrostatic Feature Construction} 

For each layer $u$, we first compute long-ranged electrostatic features using the spin-charge density $\rho^{(u), \uparrow \downarrow}(\mathbf{r'})$.  This is done by computing the spin-resolved electrostatic potential $v^{(u),\uparrow\downarrow}(\mathbf{r})$ generated by the spin-charge density convolved with a $1/|\mathbf{r}|$ Coulomb kernel:
\begin{align}
v^{(u),\uparrow\downarrow}(\mathbf{r}) &= \int \frac{\rho^{(u), \uparrow \downarrow}(\mathbf{r'})}{|\mathbf{r} - \mathbf{r'}|} d\mathbf{r'} + \frac{1}{2}v_{\text{app}}(\mathbf{r})
\label{eq:potentials}
\end{align}
where the integral is over the whole space $\R^3$ (or a three-dimensional flat torus for periodic systems). The potential $v_{\text{app}}$ corresponds to the potential generated by an externally applied electric field and is zero in the absence of an external field.
Then, atom-centred electrostatic features are computed by projecting the electrostatic potential onto atom-centred Gaussian basis functions:
\begin{align}
v_{i,nlm}^{(u), \uparrow \downarrow} &= \frac{1}{\mathcal{N}_{nl}}\int \phi_{nlm}(\mathbf{r} - \mathbf{r}_i) v^{(u), \uparrow \downarrow}(\mathbf{r})d\mathbf{r} 
\label{eq:field-features}
\end{align}
where  $\phi_{nlm}(\mathbf{r} - \mathbf{r}_i)$ are the same atom-centred Gaussian
basis functions as in Equation~5.
One can choose to use more basis functions $\phi_{nlm}$ for the potential than what was used for the spin-charge density in Equations~\eqref{eq:rho-up}-\eqref{eq:gaussian-basis}, to get a richer description of the potential. The smearing width and maximum angular momentum can also be chosen differently to those used in the spin-charge density expansion. In a practical implementation, we do not compute $\rho$ or $v$ on a grid in order to compute the integrals in Equations~\ref{eq:potentials} and \ref{eq:field-features}, but use analytical Gaussian-orbital integrals or reciprocal-space transforms. Details of the projected-potential evaluation in open and periodic boundary conditions are given in SI Sections~\ref{sec:si_realspace_electrostatics} and~\ref{sec:si_periodic_electrostatics}. 
The electrostatic features are inherently non-local due to the integral over the whole space. These features are closely related to the LODE features~\cite{Grisafi2019}. The main differences are that we constrain the sources of these electrostatic features to be the spin-charge density multipoles, and we consider only the Coulomb potential.

2. \textbf{Multipoles Update}. Following this, the (non-local) electrostatic features are combined with local node features to predict an updated set of spin-charge multipoles $p^{(u+1),  \uparrow \downarrow}_{i,lm}$. This begins by jointly embedding the electrostatic features with the MACE local features using an element-agnostic biased linear embedding:
\begin{align}
V_{i,klm}^{(u)} &= \sum_{n,\uparrow \downarrow} W_{knl \uparrow \downarrow}^{(u)} v_{i,nlm}^{(u), \uparrow \downarrow} +  \sum_{\uparrow \downarrow} W_{kl\uparrow \downarrow} p_{i,lm }^{(u), \uparrow \downarrow} \\ &+ \sum_{\tilde{k}} W_{k\tilde{k}}^{(u)} h_{i,\tilde{k}lm}^{(T)} + b_{k,lm}^{(u)} \delta_{lm,00}
\end{align}
The object $V_{i,klm}^{(u)}$ jointly describes the non-local electric field, spin-charge density and the local geometry.
We use these features to update the spin-charge density through a series of nonlinear transformations:
\begin{align}
\label{eq:update-dot-product}
&  d_{i,lk}^{(u)}
  =
  \sum_{k'} W^{(u),\mathrm{dot}}_{lkk'}
  \sum_{m=-l}^{l}
  h_{i,klm}^{(T)}\,V_{i,k'lm}^{(u)}\\
\label{eq:update-tp}
& a_{i,lk}^{(u)} := \mathrm{MLP}^{(u)}\!\left([\mathbf d_i^{(u)} \| \mathbf e(z_i)]\right)_{lk} \\
&g_{i,klm}^{(u)}= \sum_{k'} W^{(u),\mathrm{tp}}_{lkk'}\,
a_{i,lk'}^{(u)}\, h_{i,k'lm}^{(T)} \\
  \label{eq:readout-rho-update}
&\Delta p_{i,lm}^{(u), \uparrow \downarrow} = \text{ReadoutMLP}(\mathbf{g}_i^{(u)})_{lm} \\
\label{eq:readout-fukui}
&f^{(u), \uparrow \downarrow}_{i} = \text{ReadoutMLP}(\mathbf{g}_i^{(u)}) \\
\label{eq:update-rho}
&\tilde{p}^{(u),  \uparrow \downarrow}_{i,lm} = p^{(u),  \uparrow \downarrow}_{i,lm} + \Delta p_{i,lm}^{(u),  \uparrow \downarrow}
\end{align}
with $(\mathbf{g}_i^{(u)})_{klm} := g_{i,klm}^{(u)}$.
The tilde on $\tilde{p}^{(u),  \uparrow \downarrow}_{i,lm}$ indicates that these are ``un-equilibrated'' and the monopoles do not yet sum to the correct total charge and spin. The update MLP in Equation~\ref{eq:update-tp} consists of 3 layers with 64 channels and SiLU activation. Both readouts in Equations~\ref{eq:readout-fukui} and \ref{eq:readout-rho-update} use a two-layer gated nonlinearity with hidden irreps \texttt{64x0e + 32x1o}, employing SiLU for the scalar gate and sigmoid for the equivariant gate. 
Importantly, the above operations incorporate non-local information through the electrostatic features, but the computation itself is local since the new multipoles $\tilde{p}^{(u),  \uparrow \downarrow}_{i,lm}$ on atom $i$ are updated based only on the geometry and the electrostatic features around atom $i$.

3. \textbf{Fukui Equilibration}: 
After each prediction of the spin-charge density (including the initial local prediction), we renormalise the total charge and total spin using learnable Fukui functions.
\begin{equation}
p_{i, 00}^{(u+1), \uparrow} = \tilde{p}_{i, 00}^{(u), \uparrow} + \frac{f_i^{(u), \uparrow}}{\sum_j f_j^{(u), \uparrow}} \left( \frac{Q + S}{2} - \sum_j \tilde{p}_{j, 00}^{(u),  \uparrow} \right)
\label{eq:fukui-eq-up-nl}
\end{equation}
\begin{equation}
p_{i, 00}^{(u+1), \downarrow} = \tilde{p}_{i, 00}^{(u),\downarrow} + \frac{f_i^{(u),\downarrow}}{\sum_j f_j^{(u), \downarrow}} \left( \frac{Q - S}{2} - \sum_j \tilde{p}_{j, 00}^{(u),\downarrow} \right)
\label{eq:fukui-eq-down-nl}
\end{equation}
where $f_i^{(u),\uparrow\downarrow}$ are the predicted spin-resolved Fukui functions in Equation~\ref{eq:readout-fukui}. This equilibration is equivalent to Equations~\ref{eq:fukui-eq-up} and~\ref{eq:fukui-eq-down}, except that the Fukui functions are not restricted to purely local features but can incorporate non-local information through the field update. This operation is a global equilibration due to the normalisation in the denominators of Equations \eqref{eq:fukui-eq-up-nl} and \eqref{eq:fukui-eq-down-nl}.
The new spin-charge multipoles $p_{i, 00}^{(u+1)}$ now re-enter at step 1. As with the previous equilibration, the rest of the multipoles (e.g. dipoles) are unchanged, i.e. $\{p_{i,lm}^{(u+1)}:=\tilde{p}^{(u)}_{i,lm}\}_{l>0}$.

\subsubsection{Non-local energy}
$E_{\text{non-local}}$ is a field and charge-dependent correction term that is added to the total energy to account for additional non-local energetic contributions beyond the Coulomb interaction. We compute it using an MLP on the sum of the dot product of local geometry features, spin-charge and electrostatic features,
\begin{align}
&p_{i,klm}^{\text{emb}} = \sum_{\uparrow \downarrow} W^{\text{emb}}_{k} p_{i,lm}^{(u), \uparrow \downarrow}\,\,v^{\text{emb}}_{i,klm} = \sum_{n,\uparrow \downarrow} W^{\text{emb}}_{nk \uparrow \downarrow} v_{i,nlm}^{\uparrow \downarrow} \\
&d^{p}_{i,lk} = \sum_{k'} W^{p,\mathrm{dot}}_{lkk'}
  \sum_{m=-l}^{l}
  h_{i,klm}\,p^{\mathrm{emb}}_{i,k'lm}\\
&d^{v}_{i,lk}
  =
  \sum_{k'} W^{v,\mathrm{dot}}_{lkk'}
  \sum_{m=-l}^{l}
  h_{i,klm}\,v^{\mathrm{emb}}_{i,k'lm}\\
 \label{eq:non-local-energy}
&E_{\text{non-local}} = \sum_i \mathrm{MLP}\!\left([\mathbf d_i^{p}\|\mathbf d_i^{v}]\right)
\end{align}
where $\boldsymbol{p}_i^{\text{emb}}$ and $\mathbf{v}_i^{\text{emb}}$ are linearly scaled features of the charges and electrostatic features and $\|$ is the concatenation of two vectors. We use a 3-layer MLP with SiLU activations and 128 hidden units in Equation~\ref{eq:non-local-energy}.

\subsubsection{Electrostatic Energy from Smeared Multipoles}

The electrostatic energy $E_{\text{electrostatic}}$ is computed from the final multipole charge density coefficients using a Gaussian smearing scheme that ensures smooth blending with the local energy at small distances. In the rest of this section, we only consider the final charge density after $U$ iterations of the long-range update, $\mathbf{\rho}(\mathbf{r}) = \mathbf{\rho}^{(U)}(\mathbf{r})$. The Gaussian basis functions defined in Eq.~\ref{eq:gaussian-basis} enable an exact treatment of electrostatics through their analytically tractable Fourier transforms. As discussed, the charge density of the model is expanded in terms of atomic multipoles and Gaussian basis functions:
\begin{align}
\mathbf{\rho}(\mathbf{r}) &= \sum_{i,lm} p_{i,lm} \phi_{nlm}(\mathbf{r} - \mathbf{r}_i)
\end{align}
where $p_{i,lm} = p_{i,lm}^\uparrow + p_{i,lm}^\downarrow$ and $\phi_{nlm}$ is defined in \eqref{eq:gaussian-basis}. The range of $n$ is just one, since we only use one radial function for the density expansion. The electrostatic energy is defined as \begin{align}
    E_\text{electrostatic} &= E_\text{Hartree} + E_\text{app} \\
    &=\frac{1}{2}\iint \frac{\rho(\mathbf{r})\rho(\mathbf{r}')}{|\mathbf{r}-\mathbf{r}'|} d\mathbf{r} d\mathbf{r}' + \int \rho(\mathbf{r}) v_\text{app}(\mathbf{r}) d\mathbf{r}
\end{align}
where $v_\text{app}$ is an applied potential such as that from an applied field. To evaluate this energy in practice, we use separate implementations for real-space or periodic computations.

\textbf{a. Non-Periodic:}
For isolated molecular systems (open boundary conditions), we employ a direct real-space summation. Substituting the expression for $\rho$ in terms of $p_{i,lm}$, the Hartree term becomes:
\begin{align}
    E_\text{Hartree} &= \frac{1}{2}\sum_{ilm,jl'm'} p_{i,lm}p_{j,l'm'} \mathcal{T}_{ilm,jl'm'} \label{eq:def_E_Hartree} \\
    \mathcal{T}_{ilm,jl'm'} &:=\iint \frac{\phi_{lm}(\mathbf{r}-\mathbf{r}_i)\phi_{l'm'}(\mathbf{r}'-\mathbf{r}_j)}{|\mathbf{r}-\mathbf{r}'|} d\mathbf{r} d\mathbf{r}'
    \label{eq:def_T_coeffs}
\end{align}
For example, the coefficient for the monopole part of this sum is simply:
\begin{equation}
\mathcal{T}_{i00,j00} = \frac{\text{erf}(r_{ij}/\sqrt{2}\sigma)}{r_{ij}}
\label{eq:realspace-monopole}
\end{equation}
which is the familiar Gaussian Coulomb damping. For the other coefficients, we compute the interaction coefficient $\mathcal{T}_{ilm,jl'm'}$ by approximating one $l=1$ Gaussian basis function in \eqref{eq:def_T_coeffs} as the difference of two $l=0$ functions, which are slightly displaced relative to each other. This is equivalent to constructing a dipole from two opposite-sign charges. While it is possible to evaluate $l=1$ integrals analytically, we find this implementation to be very efficient and transparent. In the supplementary information, it is shown that for Gaussian $l=1$ multipoles, this approximation becomes exact for all separations $\mathbf{r}_{ij}$---including when two Gaussians overlap---as the relative offset between the two $l=0$ functions approaches zero. In our implementation, we use an offset of 0.02 \AA \space for representing Gaussian dipoles as the sum of two monopoles, allowing us to use \eqref{eq:realspace-monopole} to compute the charge-dipole and dipole-dipole interactions. Full details of the algorithm are presented in the supplementary information.

Note that the sum in the expression above includes $i=j$, meaning that we are including a self-energy term $E_{\text{self}} = \frac{1}{2} \sum_{i,lm} p_{i,lm} \mathcal{T}_{ilm,ilm} p_{i,lm}$. One cannot use the formula \eqref{eq:realspace-monopole} to compute this because $r_{ij}=0$, but these coefficients can be computed directly from \eqref{eq:def_T_coeffs} using the properties of Gaussian integrals. We found that including this term is beneficial since it generally leads to a smoother predicted total electrostatic energy when atoms are close together, as has been discussed previously \cite{eMLP}.

The applied field term can be evaluated similarly:
\begin{align}
    E_\text{app} &= \sum_{ilm} p_{i,lm} B_{i,lm} \\
    B_{i,lm} &= \int \phi_{lm}(\mathbf{r}-\mathbf{r}_i) v_\text{app}(\mathbf{r}) d\mathbf{r}
\end{align}
All the coefficients $B_{i,lm}$ can be computed analytically as long as the applied potential is a homogeneous field.

\textbf{b. Periodic:} For periodic systems, we do not consider applied fields, and compute the Hartree energy in reciprocal space. Provided the integral of $\rho(\mathbf{r})$ over the supercell is zero, we can express the Hartree energy as:
\begin{equation}
E_{\text{electrostatic}} = E_{\text{Hartree}} = \frac{\Omega}{2(2\pi)^6} \sum_{\substack{\mathbf{k} \in \Lambda^\star\\\mathbf{k}\neq\mathbf{0}}} \frac{4\pi}{k^2} |\tilde{\rho}(\mathbf{k})|^2
\label{eq:ewald-energy}
\end{equation}
where $\Omega$ is the volume of the supercell, $\tilde\rho(\mathbf{k})$ is the Fourier series of the charge density, $\Lambda^\star$ is the reciprocal lattice, and the term $\mathbf{k}=\mathbf{0}$ is omitted. This is tractable because the Fourier series of the charge density, $\tilde{\rho}(\mathbf{k})$, can be computed analytically using the properties of Gaussian functions:
\begin{align}
\tilde{\rho}(\mathbf{k}) &= \frac{(2\pi)^3}{\Omega} \sum_{i,lm} p_{i,lm} \tilde{\phi}_{lm}(\mathbf{k}) e^{-i\mathbf{k}\cdot\mathbf{r}_i} \\
\tilde{\phi}_{lm}(\mathbf{k}) &= C_l (-i)^l Y_{lm}(\hat{\mathbf{k}}) I_{l,\sigma}(k)
\label{eq:fourier-basis}
\end{align}
where $Y_{lm}$ are the spherical harmonics, and $I_{l,\sigma}(k)$ is the radial Fourier transform of the Gaussian basis, evaluated via cubic spline interpolation for computational efficiency. Further details are presented in the supplementary information.

If the total charge of a periodic system is not zero, one can still assign an energy using equation \eqref{eq:ewald-energy}, but in order to do meaningful calculations of, for instance, charged defect formation energies, correction terms should be applied similarly to DFT \cite{MakovPayneChargedCellCorrections}.

The gradient of $E_{\text{electrostatic}}$ with respect to atomic positions and multipole coefficients provides the electrostatic forces and field-response terms that enter the training loss and molecular dynamics propagation.

\subsubsection{Summary of the model architecture}

The model is summarised visually in Figure~\ref{fig:summary}A, explaining each step of the \MFP model architecture:

\begin{enumerate}
      \item Start by inputting the atomic positions and species into the local MACE model to produce node features that embed the semi-local geometry and chemistry around each atom.
      \item Read out a local energy contribution from the node features, capturing the most local and semi-local interactions (e.g., bonding, Pauli repulsion, and short-range electrostatics).
      \item Predict an initial local spin-charge density, expanded in an atom-centred spherical Gaussian basis.
      \item From the spin-charge density, compute the electrostatic potential per spin channel and project it into atom-centred electrostatic features.
      \item Use the electrostatic features together with the local geometry and current spin charge density to predict an additive update to the density; also predict two per-atom scalars, the Fukui features.
      \item Normalise the Fukui features so they sum to one over all atoms.
      \item Use the normalised Fukui features to equilibrate the charges via the Fukui equilibration step.
      \item Repeat steps (5–7) several times (two iterations in the current models).
      \item Use the final density to compute both the Coulomb energy and a non-local energy contribution.
      \item Sum the local, Coulomb, and non-local energy terms to obtain the total energy and differentiate w.r.t positions to get the forces.
  \end{enumerate}

\subsection{Training Strategy and Model Hyper-parameters}

\MFO models were trained on the energies and forces of 100 million OMol25 structures using distributed training across 64 NVIDIA H200 GPUs. We trained multiple model variants and benchmark the medium and large variants (\MFPM, \MFPL), which were obtained by varying the number of local MACE interaction layers. The full hyper-parameter settings and loss weights are listed in the SI (Table~\ref{tab:mace_hparams}). For all models, we use the non-linear interaction blocks introduced in~\cite{batatia2025crosslearningelectronicstructure}. We use a weighted sum of L1 loss on the energy and L2 loss on the forces:
\begin{equation}
\mathcal{L} = w_E |E - E_{\text{ref}}| + w_F \|\mathbf{F} - \mathbf{F}_{\text{ref}}\|_2
\end{equation}

We do not include any training on the partial charges that were available in the dataset (L\"owdin and NBO charges) as we found that it deteriorates the performance of the models. At the time of training, the total dipoles were not available for the full dataset and will be considered when the data become available.

\begin{table}[h!]
\centering
\caption{Receptive fields and explicit long-range physics in benchmarked models.}
\label{tab:mpnn_ranges}
\resizebox{\columnwidth}{!}{%
\begin{tabular}{@{}lcc@{}}
\toprule
\textbf{Model} & \textbf{\makecell[ct]{Receptive\\Field}} & \textbf{\makecell[ct]{Long-Range\\Coulomb}} \\
\midrule
\MOMOL        & 18\,\AA & \rcross \\
\MFPM        & 12\,\AA & \gcheck \\
\MFPL          & 18\,\AA & \gcheck \\
\UMASOMOL  & 30\,\AA & \rcross \\
\UMAMOMOL  & 66\,\AA & \rcross \\
\ORBMOL & 30\,\AA & \rcross \\
SO3LR & 13.5\,\AA & \gcheck \\
AIMNet2-NSE & 15\,\AA & \gcheck \\
\bottomrule
\end{tabular}%
}
\hspace{5pt}
\raggedright\footnotesize\emph{Notes:} The effective information range grows with the number of message-passing layers, but without explicit long-range terms the learned interactions remain limited by the graph topology.
\end{table}

\section{Results and Discussion}
\subsection{Benchmarking and Models}

To assess robustness and accuracy, we assembled a comprehensive benchmark suite spanning high-quality reference data in thermochemistry, reaction barriers, non-covalent interactions, conformers, protein-ligand interactions, molecular
crystals, and transition metal complexes. Beyond static single-point benchmarks, we evaluate molecular dynamics and geometry-optimisation tasks, including organic liquid densities, liquid-water radial distribution functions, solvated-ion
spin dynamics, redox potentials in water, and solvated ion pairs.
We also benchmark a series of baseline models: 
\begin{itemize}
    \item \textbf{MACE-OMOL}~\cite{levine2025openmolecules2025omol25}: A local MACE model with the same local architecture and hyper-parameters as \MFPL, trained on the full 100M OMOL dataset, but without the electrostatics part and including a global embedding for the total charge and total spin that feed into the initial node features. This model represents a close ablation that enables us to isolate the improvements from the electrostatic component. 
    \item \textbf{UMA-S/M-1P1}~\cite{wood2025umafamilyuniversalmodels}: eSEN~\cite{fu2025learningsmoothexpressiveinteratomic} model trained on 100M inorganic crystals of OMAT, 230M surfaces/small molecules of OC20/OC22, 100M molecular configurations of OMOL, 25M molecular-crystal configurations of OMC and 29M metal organic frameworks of ODAC. The dataset types, the total charge, and the total spin are embedded in the model as a global input. We benchmark the OMOL variant throughout this paper, with both the S-1.1 and the larger M-1.1 variants referred to as \UMAS and \UMAM.
    \item \textbf{OrbMol}: Orb-v3~\cite{rhodes2025orbv3atomisticsimulationscale} model trained on the 100M OMOL dataset, with a global embedding of total charge and total spin.
    \item \textbf{g-xTB}~\cite{gxtb2025}: A semi-empirical extended tight-binding method parameterised for 103 elements using reference energies and forces at the $\omega$B97M-V/aTZ level. The paper does not report a single total training-set
    size; based on the stated 8,000–25,000 training data points per element (up to $\sim$40,000 for key elements), we estimate approximately 1.1–2.7 million configurations overall.
\end{itemize}

\subsection{Thermochemistry and Reactions}

\begin{figure*}[t!]
\centering
\includegraphics[width=\linewidth]{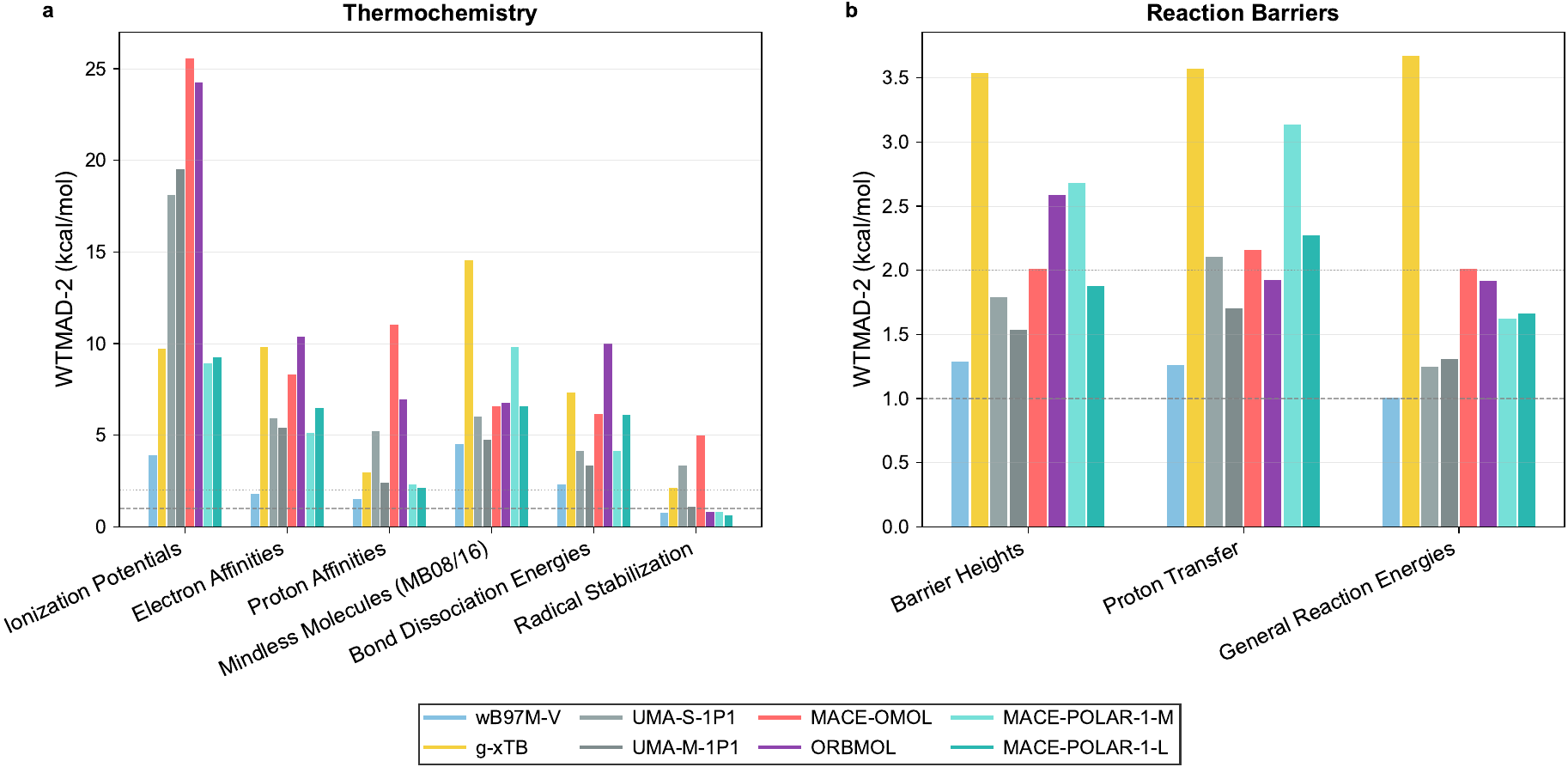}
\caption{\textbf{Thermochemistry and reaction barrier benchmarks on GSCDB138 subsets.}
Bar heights show the weighted total mean absolute deviation (WTMAD-2) in kcal/mol for each model, where lower values indicate better accuracy. WTMAD-2 rescales errors by the characteristic energy scale of each subset to enable fair comparison across datasets of different magnitudes (definition in SI~\ref{sec:wtmad2}). For these summary bars, extreme outliers are filtered by excluding points with $|\Delta E|>100$~kcal/mol (details in SI~\ref{sec:exclusions}).
(a) Thermochemistry subsets grouped by property class: ionisation potentials, electron affinities, proton affinities, and bond dissociation energies.
(b) Reaction-barrier subsets: barrier heights, proton-transfer reactions, and general reaction energies.
Models compared: $\omega$B97M-V (reference hybrid DFT), g-xTB (semi-empirical), \UMAS/\UMAM, \MOMOL, \ORBMOL (local MLIPs), and the electrostatic \MFPM/\MFPL.}
\label{fig:thermochem}
\end{figure*}

\begin{figure*}[t]
\centering
\includegraphics[width=\linewidth]{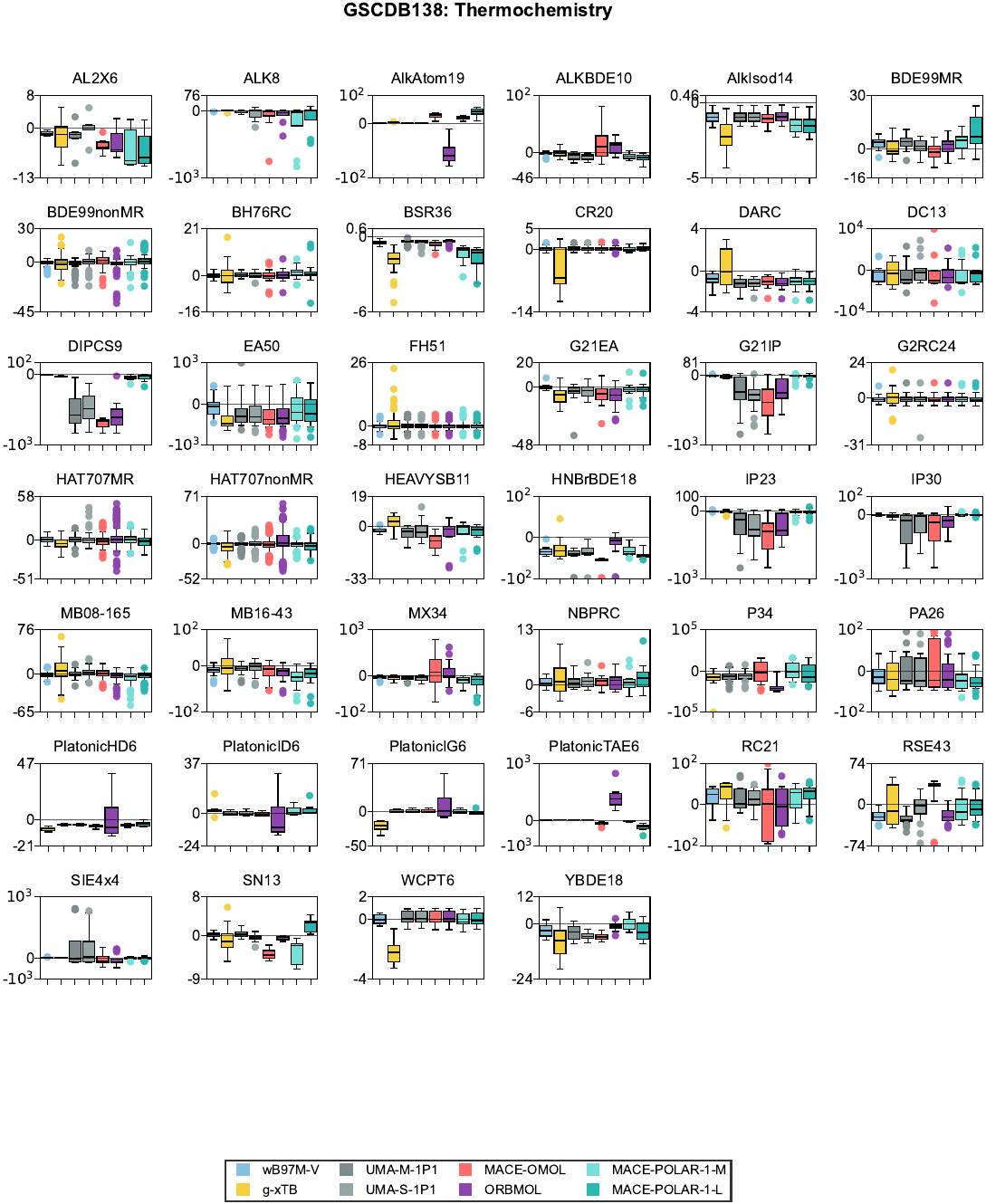}
\caption{Distribution of signed thermochemical energy errors (kcal mol$^{-1}$) for GSCDB138. Total atomisation energy (TAE) datasets (e.g., PlatonicTAE6 and W4‑17 TAE subsets) are excluded by design.}
\label{fig:gscdb138_tc}
\end{figure*}

Thermochemistry and reaction barriers are core tests of bonding, electron redistribution, and transition-state energetics. We evaluate the models on the curated GSCDB138 database~\cite{Liang2025}, an update of GMTKN55~\cite{Goerigk2017} and MGCDB84~\cite{Mardirossian2017} with improved CCSD(T) references and better filtering for spin contamination. The tested systems are typically small ($\lesssim$20 atoms), where long-range electrostatic effects are expected to be small. Reference data are CCSD(T) extrapolated to CBS or F12-quality. We report performance using the weighted mean absolute deviation 2 (WTMAD-2), which rescales each subset by its energetic scale to yield a statistically representative aggregate (as in the GMTKN convention; see SI~\ref{sec:wtmad2}); benchmark lists and citations are provided in SI~\ref{sec:tc_benchmarks}. Full per-set results appear in Fig.~\ref{fig:gscdb138_tc} (SI). A small number of configurations are excluded per model due to evaluation failures or outlier filtering; criteria, counts, and full identifier lists are provided in SI~\ref{sec:exclusions}.

\begin{table*}[t]
\centering
\small
\setlength{\tabcolsep}{3pt}
\renewcommand{\arraystretch}{1.1}
\caption{WTMAD-2 (kcal/mol) for thermochemistry and reaction-barrier subsets with only H, C, N, O, F, P, S, and Cl elements to support SO3LR and AIMNet2-NSE. Datasets for each subset are listed in SI Sec.~\ref{sec:tc_benchmarks}.}
\label{tab:wtmad_tc_bh_subgroups}
\resizebox{\textwidth}{!}{%
\begin{tabular}{l c c c | c c c c c c | c c c}
\toprule
Subset & $N_{\mathrm{sets}}$ & wB97M-V & g-xTB & \UMAS & \UMAM & \MOMOL & \ORBMOL & \MFPM & \MFPL & SO3LR & AIMNet2-NSE  & FENNIX-BIO-2 \\
\midrule
TC: Bond Energies + HAT & 7 & 1.57 & 5.30 & 3.16 & 2.18 & 4.68 & 5.63 & 2.75 & 3.80 & 26.46 & 9.27 & 14.18 \\
TC: Reaction Energies & 6 & 0.94 & 3.61 & 1.06 & 1.14 & 1.77 & 1.31 & 1.51 & 1.47 & 18.66 & 3.37 & 6.76 \\
TC: Thermochemistry & 9 & 1.22 & 5.31 & 2.22 & 2.47 & 3.27 & 2.94 & 2.39 & 2.34 & 27.22 & 4.87 & 11.15 \\
BH: Barrier + Proton & 10 & 1.27 & 3.56 & 1.79 & 1.56 & 2.06 & 2.62 & 2.69 & 1.91 & 14.27 & 6.75 & 18.32 \\
\bottomrule
\end{tabular}%
}
\end{table*}

\subsubsection{Thermochemistry} Figure~\ref{fig:thermochem}a shows WTMAD-2 results for the models across thermochemistry subsets grouped into major categories. Ionisation potentials and electron affinities probe the description of varying electron numbers, including self-interaction and delocalisation errors and open-shell, spin-polarised energetics. The electrostatic models roughly halve the error of other MLIPs on ionisation potentials by representing total charge and spin through a physical spin-charge density with learnable equilibration, rather than relying solely on flexible global embeddings that generalise poorly when electrons are added or removed. We also see consistent gains on electron affinities compared to the local baseline \MOMOL. We observe large improvements on proton affinities, with the \MFP reaching accuracies close to the reference DFT, \wb. The \MFP models bridge the gap to the semi-empirical g-xTB on these properties. The radical stabilisation sets show consistent gains, reflecting that improved extrapolation yields a better description of radicals. Mindless molecules (MB08/16) are synthetic, out-of-distribution structures made to test robustness far beyond the training set by constructing diverse and chemically plausible molecules through random atomic placement and subsequent geometry optimisation; surprisingly, all MLIPs remain relatively close to \wb{}, underscoring robust extrapolation.

\subsubsection{Reaction Barriers} Figure~\ref{fig:thermochem}b aggregates datasets probing reaction barrier heights of small molecules, proton transfer, and general reaction energies.
Barrier heights test the energetics of transition states, which directly link to reaction kinetics, while reaction energies capture thermodynamic gaps between reactants and products; both require accurate description of local energetics. Proton-transfer barriers are highly relevant in biochemistry and energy materials (e.g., fuel cells and proton-conducting membranes) and involve coupled proton motion and hydrogen-bond rearrangement. Across all barrier categories, the MLIPs perform well relative to hybrid DFT, reaching accuracies on the order of a few kcal/mol; differences are modest because these small-system barriers are dominated by local structural chemistry rather than long-range charge redistribution.

\subsubsection{Comparison to other pre-trained models on a subset of elements} In order to compare the OMol-trained models with other molecular foundation MLIPs like SO3LR~\cite{kabylda2025so3lr}, AIMNet2-NSE~\cite{Kalita2025}, and Fennix-2-bio~\cite{PL_2025} that do not cover the full set of elements in GSCDB138, we curated a subset of the thermochemistry and reactivity benchmarks including those that contain only H, C, N, O, F, P, S, and Cl
elements. These three models were trained on smaller databases than the OMol25 dataset: SO3LR was trained on 4M configurations at the PBE0+MDB level of theory, FENNIX-BIO-2 was trained on 2.2M at the $\omega$B97M-D3(BJ) level, and AIMNet2-NSE was trained on 33M configurations at the $\omega$B97M-D3(BJ) level.
Table~\ref{tab:wtmad_tc_bh_subgroups} summarises the WTMAD-2 values for these thermochemistry and reaction-barrier subsets. We observe that the OMol models significantly outperform the other three models, reflecting the benefit of the scale of the OMol dataset. AIMNet2-NSE, which is trained on roughly one-third of the OMol dataset, is the model that performs best after the OMol models, which shows the crucial importance of a large amount of high-quality data. Moreover, both the SO3LR and FENNIX-BIO-2 models are mainly optimised for speed in order to be used in bio-simulations and therefore use less expressive but faster architectures than the other models, which is reflected in their diminished accuracy on broad chemistries.

\subsection{Non-Covalent Interactions}

\begin{figure*}[ht!]
\centering
\includegraphics[width=\textwidth]{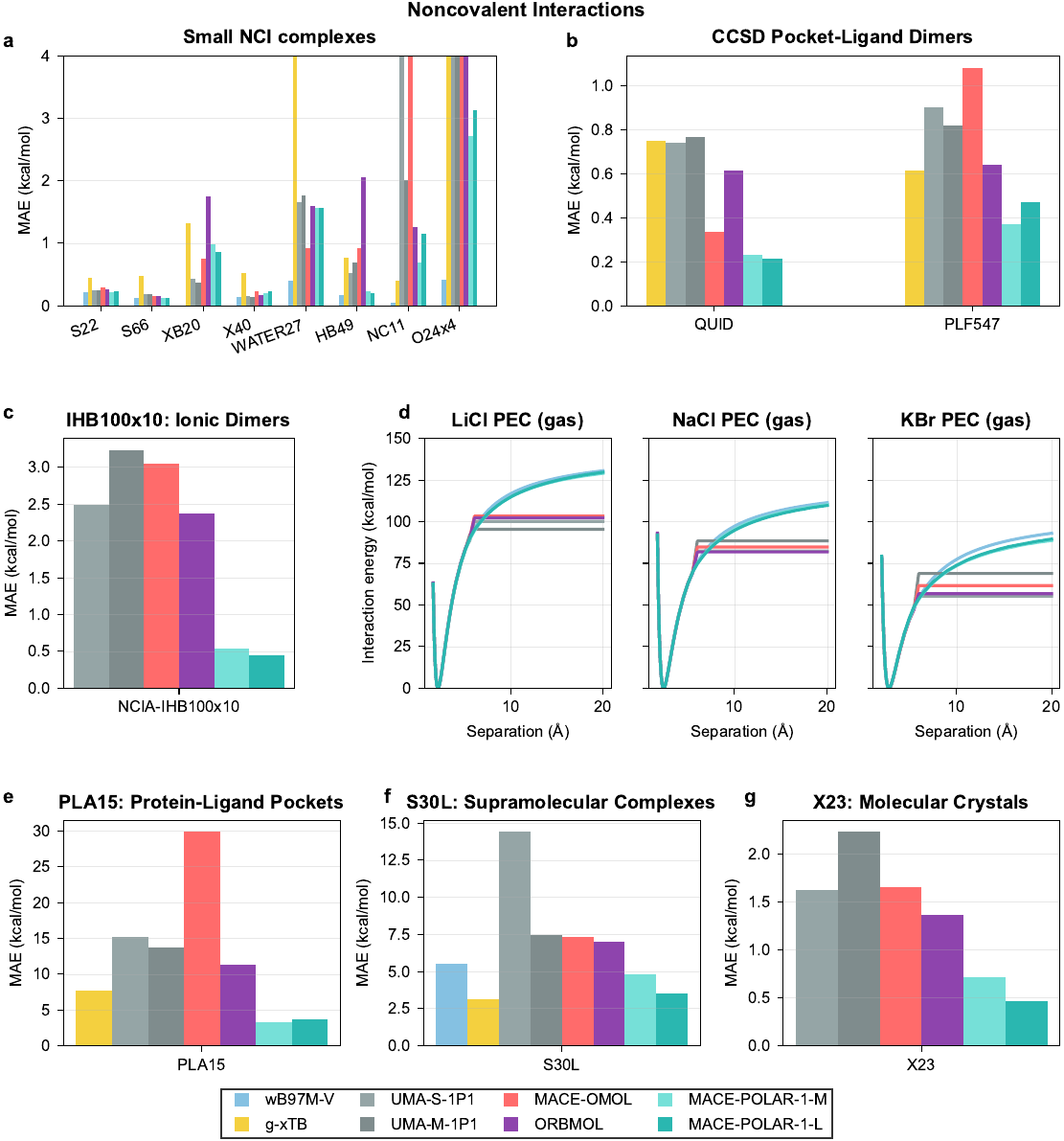}
\caption{\textbf{Comprehensive evaluation of non-covalent interaction accuracy.}
(a) Bar heights show mean absolute errors in kcal/mol for small-molecule non-covalent interaction datasets with CCSD(T)/CBS references: S22 and S66 (hydrogen bonding and dispersion), XB20 (halogen bonding), X40 (mixed interactions), WATER27 (water clusters), HB49 (diverse hydrogen bonds), NC11 (charge-transfer complexes), and O24x4 (potential energy curves).
(b) Mean absolute errors for protein-ligand fragment benchmarks: QUID (quantum-chemistry dimers from pocket-ligand motifs) and PLF547 (protein-ligand fragments with MP2-F12 + DLPNO-CCSD(T) references).
(c) Mean absolute errors on the IHB100x10 ionic hydrogen bond dataset from NCI Atlas, where electrostatic polarisation is critical.
(d) Potential energy curves for gas-phase alkali halide dissociation (LiCl, NaCl, KBr), plotting energy versus interatomic distance to test long-range $1/r$ Coulombic behaviour.
(e) Mean absolute errors for PLA15 complete protein-ligand active sites (259--584 atoms).
(f) Mean absolute errors for S30L supramolecular host-guest complexes (up to 200 atoms, charge states $-1$ to $+4$).
(g) Mean absolute errors for X23-DMC molecular crystal lattice energies with diffusion Monte Carlo references.
In all bar charts, lower values indicate better accuracy.}
\label{fig:nci}
\end{figure*}

Accurate treatment of non-covalent interactions (NCIs) is essential throughout chemistry and biology, from drug binding to protein folding. We evaluated the tested MLIPs on comprehensive benchmark sets of non-covalent interactions in small neutral systems, ionic dimers, and large supramolecular systems. Improving NCIs is a primary goal of adding long-range electrostatic interactions.

Figure~\ref{fig:nci}a aggregates results on CCSD(T)/CBS benchmarks from GSCDB138, testing small non-covalent systems and probing hydrogen bonding, halogen bonding, dispersion, charge transfer, and open-shell and charge-neutral potential-energy curves. In all these benchmarks, the maximum separation distances between clusters fall within the 6~\AA{} cutoff of the local models; hence, the models can, in principle, reproduce the DFT reference. All MLIPs are broadly competitive, but the electrostatic \MFP models consistently rank among the best, improving over the local \MOMOL baseline and often over \UMAS/\UMAM and \ORBMOL. This reflects better long-range electrostatics.

Panels (c) and (d) isolate electrostatics-dominated tests: ionic hydrogen bonds (IHB100x10) and alkali-halide dissociation curves. The \MFP variants reduce errors by a factor of three for ionic hydrogen bonds relative to local baselines and capture the expected long-range 1/r Coulombic behaviour in LiCl/NaCl/KBr dissociation, whereas local models flatten at long range due to their hard cutoff distance.

To test non-covalent interactions in supramolecular complexes, we use the S30L dataset~\cite{Sure2015} containing 30 host-guest complexes with up to 200 atoms and charge states ranging from -1 to +4. Various types of non-covalent interactions are present in the dataset, including hydrogen-halogen bonding, $\pi-\pi$ stacking, and non-polar dispersion. Figure~\ref{fig:nci}f shows a 40\% improvement in MAE from 7.31 kcal/mol for \MOMOL to 3.52--4.78 kcal/mol for \MFPL and \MFPM, respectively, with the largest gains observed for charged host-guest systems and $\pi$-stacked complexes. The $\omega$B97M-V functional performs worse than g-xTB on this test. This is likely due to the poor description of three-body dispersion effects that are important in $\pi$ interactions and not present in the VV-10 dispersion correction.

\subsection{Molecular Crystals}

\begin{figure*}[ht!]
\centering
\includegraphics[width=\textwidth]{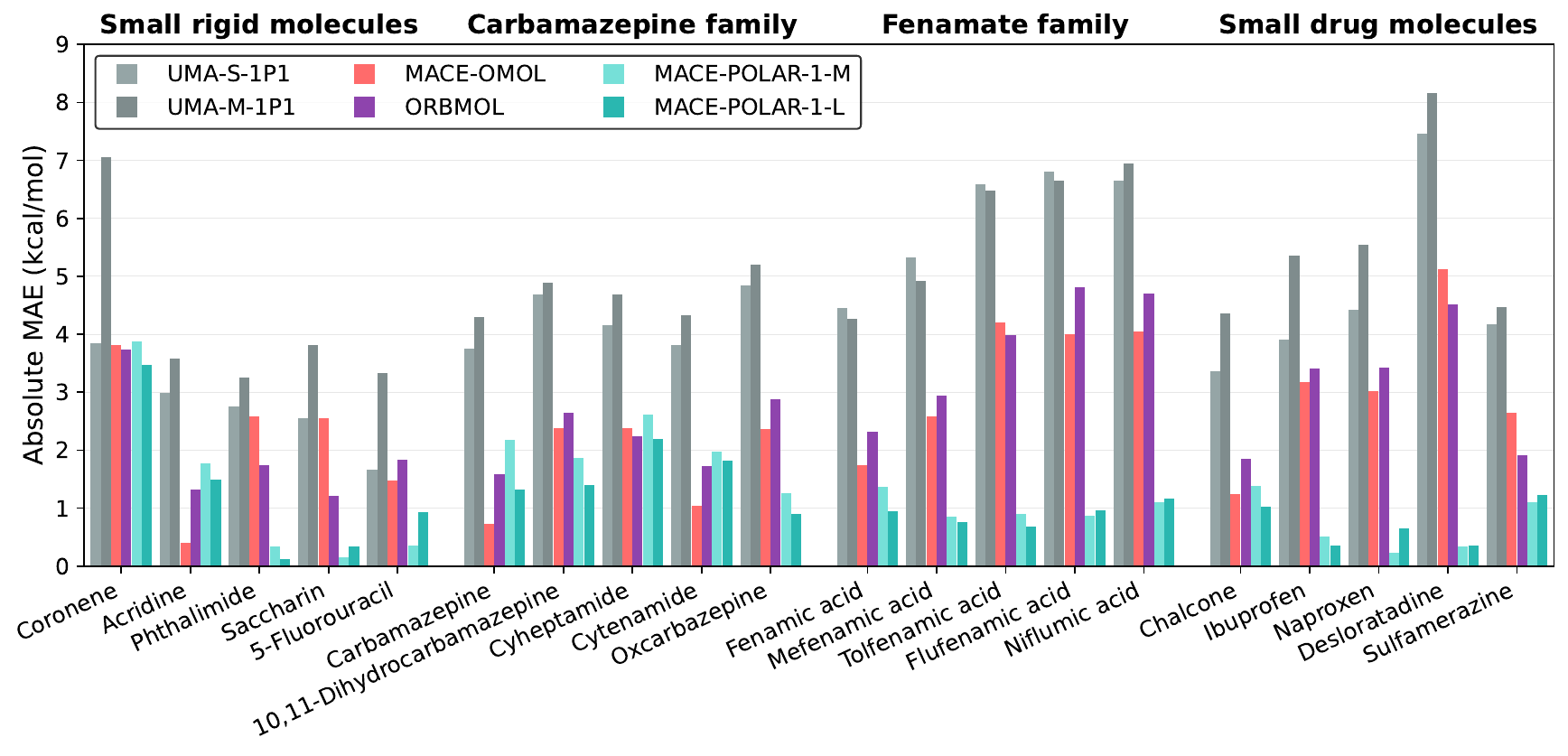}
\caption{\textbf{Absolute lattice energy errors for CPOSS209 molecular crystals.}
Bar heights show mean absolute errors in kcal/mol for predicted lattice formation energies, grouped by molecular family. The dataset comprises 209 experimental and predicted polymorphs from 20 small drug molecules. Reference calculations are performed at the $\omega$B97M-D3(BJ) level with 1-body CCSD(T) corrections. Lower values indicate better accuracy.}
\label{fig:cposs-absolute-mae}
\end{figure*}

We further benchmark the models on molecular-crystal systems.
Molecular crystals are central in pharmaceuticals, semiconductors, and agrochemicals. Crystal-structure prediction (CSP) remains a long-standing challenge in computational chemistry. One important subtask in CSP is the stability ranking of molecular crystals.
Lattice energies provide a first-order proxy for crystal stability, making them an essential property to predict accurately for CSP.
The OMOL dataset contains no periodic structures and therefore represents a significant extrapolation test for the OMOL-only trained models (\MOMOL, \MFP models, and \ORBMOL). The extrapolation from clusters to bulk is a stringent test of the physicality of the learned long-range interactions for the \MFP models, as the electrostatic interaction is formally infinite in the crystal. Both UMA models have been trained on a large number of periodic structures, including 20 million molecular crystals at the PBE level of theory in the OMC dataset~\cite{gharakhanyan2025openmolecularcrystals2025}, and it is therefore less of an extrapolation, even though we are using the OMOL task and not the OMC task.

We benchmark the models on the X23-DMC~\cite{DellaPia2024} benchmark containing 23 different neutral molecular crystals computed with Diffusion Monte Carlo, a high-level theory that has shown very good accuracy on molecular crystals. 
Figure~\ref{fig:nci}g reports MAEs (kcal/mol) on the lattice formation energies of the 23 molecular crystals of the dataset for all models. The \MFP variants are the only models below 1 kcal/mol, with \MFPL delivering the best accuracy with 0.46 kcal/mol, representing a three-fold improvement over the local baseline \MOMOL. This highlights the benefit of explicit electrostatics for extrapolating from gas-phase molecules to bulk molecular crystals.

Furthermore, we benchmark the models on a more challenging CPOSS209 dataset of organic molecular crystals~\cite{cposs209}, containing 209 experimental and predicted polymorphs of 20 small drug molecules and precursors, each with 6--17 polymorphs. The CPOSS209 crystal and gas phases, optimised at the PBE-TS level of theory, are used to determine molecular lattice energies, and we use them to perform additional reference $\omega$B97M-D3(BJ) calculations with 1-body CCSD(T) corrections. Figure~\ref{fig:cposs-absolute-mae} shows the absolute MAE achieved by the models for all molecular crystals in the CPOSS209 dataset. We note the excellent performance of the \MFPM and \MFPL models, which attain 1.06 kcal/mol and 1.22 kcal/mol, respectively, significantly improving on the local models, including MACE-OMOL (2.73 kcal/mol). However, all models perform with similar accuracy when we consider an error metric based on the MAE of relative lattice energies, as shown in Figure~\ref{fig:cposs-relative-mae}. This error metric is more relevant to the CSP stability ranking task, and our findings may indicate substantial error cancellation when computing relative lattice energies, as subtracting the lowest polymorph as a baseline may also remove the influence of subtle long-range interactions that are similar between polymorphs. The absolute lattice energy is therefore a good proxy to demonstrate that the \MFP models correctly predict the relative energetics of polymorphs for the right reasons and not error cancellation.

\subsection{Protein Fragments}

The accurate description of protein-ligand interactions is central to drug discovery, where intermolecular interactions between ligands and their protein targets determine binding affinity and specificity. These interactions are typically weak compared to intramolecular interactions and therefore challenging to capture from data for MLIPs. We evaluate our models using three complementary benchmarks derived from protein-ligand complexes: QUID~\cite{puleva2025quid}, PLF547, and PLA15~\cite{kriz2020benchmarking}. 

We first consider the Quantum Interacting Dimer (QUID) benchmark introduced by Tkatchenko and co-workers~\cite{puleva2025quid}, which targets ligand--pocket motifs with high-level reference data. QUID contains 170 dimers (42 equilibrium and 128 non-equilibrium) with up to 64 atoms spanning H, C, N, O, F, P, S, and Cl, built from nine large pocket-like monomers from the Aquamarine dataset and two small ligand fragments (benzene and imidazole). The reference interaction energies are computed using LNO-CCSD(T). We observe that the \MFP models perform best, significantly improving over the other models. All the dimers are within the local cutoff of the models (6 \AA{}) and therefore within the range of interactions that local models can capture. The improvement therefore reflects better capture of intermolecular interactions. These interactions are usually weak, and the inclusion of physics-based electrostatic interactions enables better learning of such weak signals by creating a strong physical prior.

The PLF547 dataset contains 547 complexes of ligands with protein fragments (amino acid side chains and backbone segments), providing detailed insight into the individual contributions to binding. These fragments were generated by cutting bonds between C$\alpha$ and C$\beta$ for side chains and between C and C$\alpha$ for backbone segments, with appropriate hydrogen capping. The benchmark interaction energies are based on MP2-F12/cc-pVDZ-F12 calculations with DLPNO-CCSD(T) corrections, providing near-CCSD(T)/CBS quality references.
On the PLF547 dataset (Fig.~\ref{fig:nci}b), \MFPL and \MFPM achieve MAEs of 0.37 and 0.47 kcal/mol, respectively, representing a substantial improvement over \MOMOL (1.08 kcal/mol), \UMAS (0.90 kcal/mol), and \UMAM (0.82 kcal/mol). As in the QUID benchmark, these dimers are within the local cutoffs, and therefore this test probes the ability of the models to capture subtle intermolecular interactions within their cutoff.

The PLA15 dataset extends this analysis to complete active site models of proteins, capturing many-body effects including mutual polarisation between protein residues. These 15 protein-ligand complexes contain 259--584 atoms, representing realistic drug binding sites. The benchmark energies combine pairwise MP2-F12 + DLPNO-CCSD(T) contributions of all dimers with DFT-D3 calculations to account for many-body polarisation effects.
On PLA15 (Fig.~\ref{fig:nci}e), \MFPL and \MFPM achieve MAEs of 3.35--3.68 kcal/mol, respectively, dramatically outperforming \MOMOL (29.9 kcal/mol). This order-of-magnitude improvement highlights the critical importance of long-range electrostatics in large biomolecular systems. The protein environment creates complex electrostatic fields that significantly modulate ligand binding, effects that cannot be captured by local descriptors alone. \UMAS and \UMAM show intermediate performance with MAEs of 15.13 kcal/mol and 13.67 kcal/mol, respectively. The large interaction range of \UMAM (66~\AA) enables it to capture the full active site; however, its performance is worse than the shorter-ranged \UMAS. This result highlights that the flexibility of message passing may not be suitable for accurately learning long-range interactions, and demonstrates that the more constrained physics-based approach performs better.

The excellent performance on both fragment-level and complete active site benchmarks demonstrates that \MFP models accurately capture the hierarchy of interactions in protein-ligand binding: from individual hydrogen bonds to the collective electrostatic environment of the binding pocket. This capability is essential for computational drug design, where an accurate ranking of binding affinities directly impacts lead optimisation success rates.

\subsection{Transition Metals}

\begin{figure*}[t!]
\centering
\includegraphics[width=\linewidth]{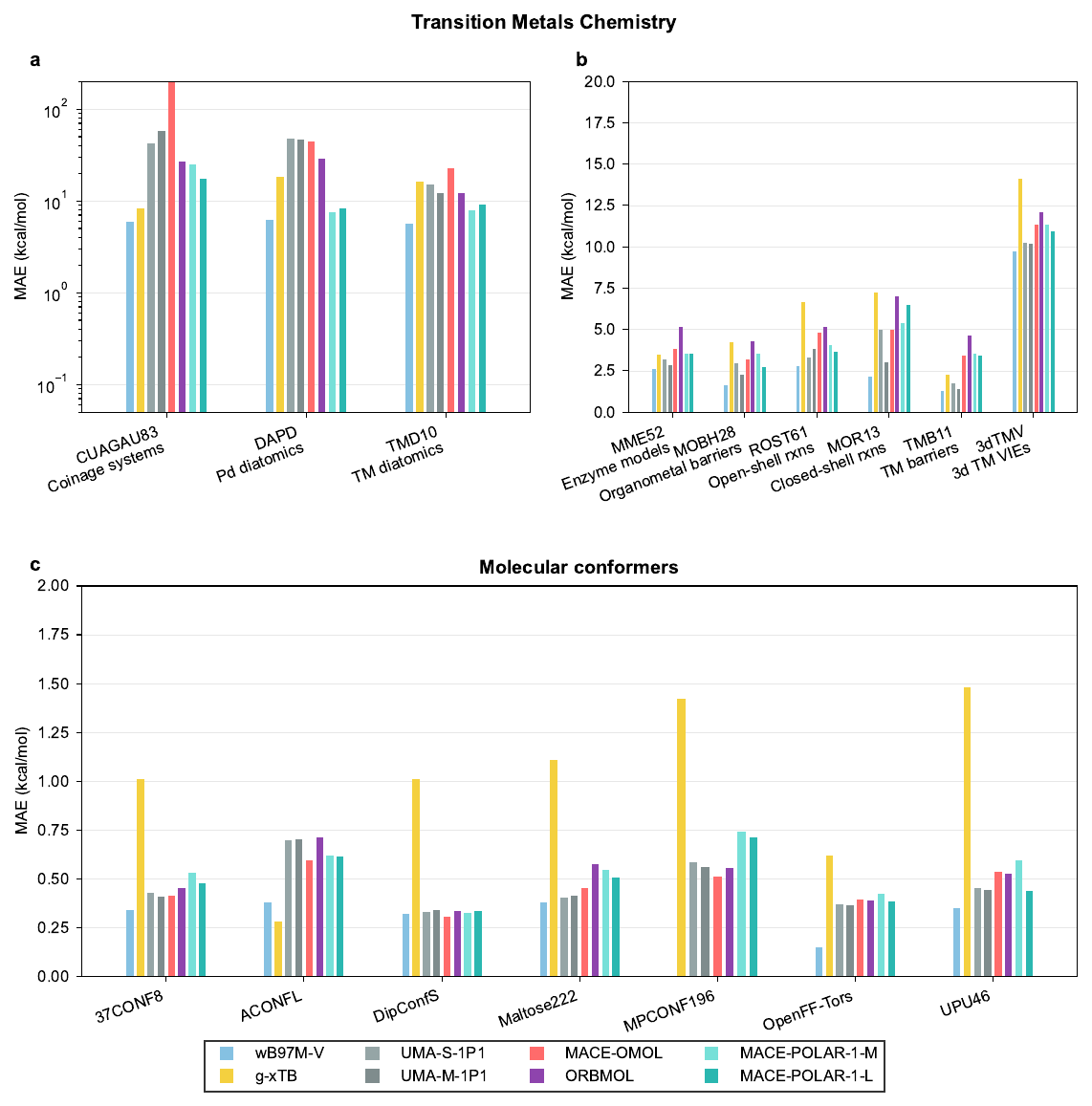}
\caption{\textbf{Accuracy on transition metal complexes and molecular conformers.}
Bar heights show mean absolute errors in kcal/mol for each model; lower values indicate better accuracy.
(a) Transition metal datasets on logarithmic scale due to large error ranges: CUAGAU83 (coinage metal Cu, Ag, Au complexes), DAPD (palladium diatomics), MOBH28 (organometallic barrier heights), and TMD10 (transition metal diatomics).
(b) Transition metal datasets on linear scale: 3dTMV (vertical ionisation energies, ph-AFQMC references), MME52 (metalloenzyme models, DLPNO-CCSD(T) references), ROST61 (open-shell reactions), MOR13 (closed-shell reactions), and TMB11 (barrier heights).
(c) Conformational energy benchmarks: 37CONF8 (small organics), ACONFL (n-alkane conformers), DipConfS (amino acids and dipeptides), Maltose222 (carbohydrates), MPCONF196 (medicinal fragments), OpenFF-Tors (torsional profiles), and UPU46 (RNA backbone fragments).
All reference values are CCSD(T) or equivalent. The GSCDB138 transition metal sets use updated references with spin-contaminated structures removed.}
\label{fig:tm}
\end{figure*}

Transition metal systems present unique challenges due to variable oxidation states, open-shell electronic structures, and complex coordination environments. We benchmarked the models on a series of tests of transition-metal chemistry taken from the GSCDB benchmark. All references are CCSD(T).

Figure~\ref{fig:tm}a (log scale) shows results for ionisation and bonding energies of transition metals, with a wide spread of errors across coinage complexes (CUAGAU83), Pd diatomics (DAPD), and TM diatomics (TMD10). CUAGAU83 probes coinage-metal complexes (Cu, Ag and Au); \MOMOL fails dramatically (MAE $>10^4$ kcal/mol), indicating a hole in the potential, while \MFPM and \MFPL greatly reduce the error to the 10~kcal/mol range, comparable to \ORBMOL and below \UMAS and \UMAM. On the DAPD subset that focuses on palladium diatomics, \MFPM and \MFPL are the best performing MLIPs, close to $\omega$B97M-V and well ahead of the local baselines. TMD10 covers transition-metal diatomics with weighted MAE; \MFPM/\MFPL again lead the MLIPs and track the DFT reference most closely.

Figure~\ref{fig:tm}b contains a series of benchmarks on diverse transition metal complexes from the GSCDB~\cite{Liang2025}.
MME52~\cite{Wappett2023} benchmarks metalloenzyme model reaction energies and barriers, MOBH28 targets organometallic barrier heights, ROST61~\cite{maurer2021assessing} probes open-shell reactions, MOR13 evaluates closed-shell reactions and TMB11 measures transition-metal barrier heights. All MLIPs cluster within a narrow band, with accuracy comparable to the underlying DFT; only \ORBMOL shows significantly higher errors. \MFPM/\MFPL improve on \MOMOL but remain above \UMAS.
The 3dTMV benchmark~\cite{Neugebauer2023} probes vertical ionisation energies (VIEs) at the ph-AFQMC reference level; MLIPs cluster around 10--12 kcal/mol, with \MFPL slightly improving over \MFPM and \MOMOL and g-xTB remaining higher, while $\omega$B97M-V is lowest. This accuracy is notable given the multireference character in this test set, where DFT itself is not highly accurate.

\subsection{Conformers}

Conformational energies are critical for applications ranging from drug design to molecular crystals. We evaluated \MFP models and other MLIPs across diverse conformer benchmarks spanning small organics to RNA fragments (Fig.~\ref{fig:tm}c). The MLIPs achieve remarkable accuracy with MAEs below 0.5 kcal/mol for most systems, approaching thermal-fluctuation limits at room temperature.
We use the following conformer benchmarks. 37CONF8~\cite{sharapa2019robust} targets diverse small organic conformers. ACONFL~\cite{ehlert2022conformational, werner2023accurate} focuses on longer n-alkane conformers and their torsional landscapes. DipConfS~\cite{plett2024toward} covers amino acids and dipeptides with multiple backbone and side-chain rotamers. Maltose222~\cite{marianski2016assessing} benchmarks carbohydrate conformers centred on maltose. MPCONF196~\cite{rezac2018mpconf196,plett2023mpconf196water} collects conformers of medicinal chemistry fragments. OpenFF-Tors~\cite{behara2024openff} evaluates torsional profiles for drug-like fragments across diverse chemistries. UPU46~\cite{kruse2015quantum} probes RNA backbone fragment conformers.

Analysis of error distributions (Fig.~\ref{fig:tm}c) reveals that the MLIP models achieve very similar accuracy, with little variation depending on the test set. Overall, they largely outperform semi-empirical approaches, reaching well within chemical accuracy for most subsets and showing an overall accuracy close to the underlying $\omega$B97M hybrid DFT. These results highlight the excellent coverage of the OMOL dataset for conformers. The inclusion of electrostatic interactions does not materially impact the accuracy on conformers as these primarily probe covalent intramolecular interactions or short-range non-covalent interactions that are well described by local models.

\subsection{Water Properties}

Water represents a stringent test for molecular models due to its complex hydrogen-bonding network and anomalous thermodynamic properties.
The accurate description of water is essential for biological and chemical applications, where aqueous solvation dominates reaction thermodynamics and protein dynamics.
The density profile as a function of temperature arises from a delicate balance between hydrogen bond directionality and molecular packing that challenges both classical force fields and \textit{ab initio} methods. We evaluate the temperature-dependent density of liquid water, a critical test of the models' ability to capture many-body effects and long-range electrostatic interactions in condensed phases.

\begin{figure}
    \centering
\includegraphics[width=\columnwidth]{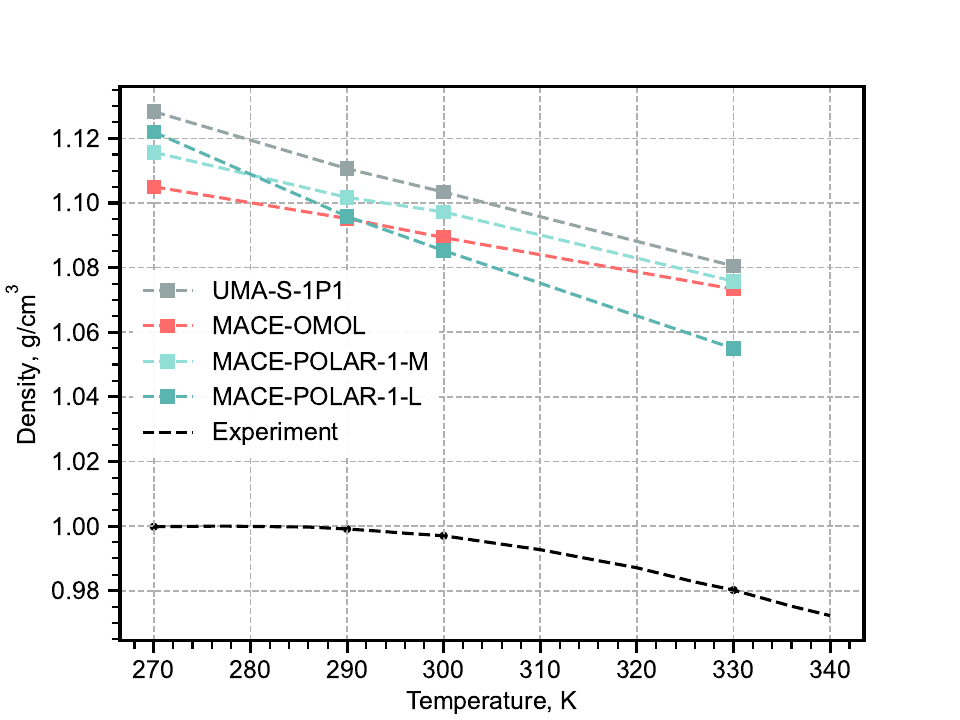}
\caption{\textbf{Liquid water density as a function of temperature.}
Data points show the equilibrium density (g/cm$^3$) from NPT molecular dynamics simulations of 333 H$_2$O molecules at 1~atm pressure. Each point represents the mean density from 500~ps of production dynamics following 500~ps of equilibration. Coloured symbols correspond to different models as indicated in the legend. Experimental data (black circles) show the characteristic density maximum near 277~K arising from competition between thermal expansion and the tetrahedral hydrogen-bond network.}
    \label{fig:water}
\end{figure}

We performed isothermal-isobaric (NPT) molecular dynamics simulations of 333 water molecules at temperatures ranging from 270 to 330 K and 1 atm pressure. The initial structure was taken from the GitHub repository associated with Ref.~\cite{Weber2025MPNICE} and equilibrated with NVT Langevin dynamics implemented in the Atomic Simulation Environment (ASE) \cite{ASE} for 50 ps. Constant pressure molecular dynamics was performed using the
Martyna-Tobias-Klein barostat implemented in ASE, with characteristic timescales of the thermostat and barostat of 50 and 500 fs, respectively. Each temperature point was equilibrated for 500 ps followed by 500 ps production runs for density calculations.
Figure~\ref{fig:water} shows the temperature-dependent density profiles for \MOMOL, \UMAS, \MFPM, and \MFPL compared to experimental data. We did not test \ORBMOL because at the time of writing it did not support stress computation. All models capture the qualitative decrease in density with increasing temperature, and overall agree on the density at room temperature around 1.08--1.10 g/cm$^3$.
The deviation from the experimental value of 1.00 g/cm$^3$ is likely due to the functional of the training data rather than the model architectures themselves. This interpretation is supported by several observations: (1) all models overestimate the density despite having substantially different architectures; (2) VV10-based functionals are reported to overstructure liquid water (e.g., SCAN has a density of 1.05 and SCAN+rVV10 of 1.16)~\cite{wiktor2017scanrvv10}; (3) VV10 lacks explicit three-body dispersion, while three-body interactions are known to be important in liquid-water simulations and can bias equilibrium densities~\cite{pruitt2016threebodywater}; and (4) the models perform well on water-cluster interaction benchmarks (e.g., WATER27 within GSCDB138), suggesting that short-range water energetics are well captured~\cite{manna2017water27,Liang2025}. Further investigations are needed to clarify the relative contributions of these effects.

\begin{figure}
    \centering
\includegraphics[width=\columnwidth]{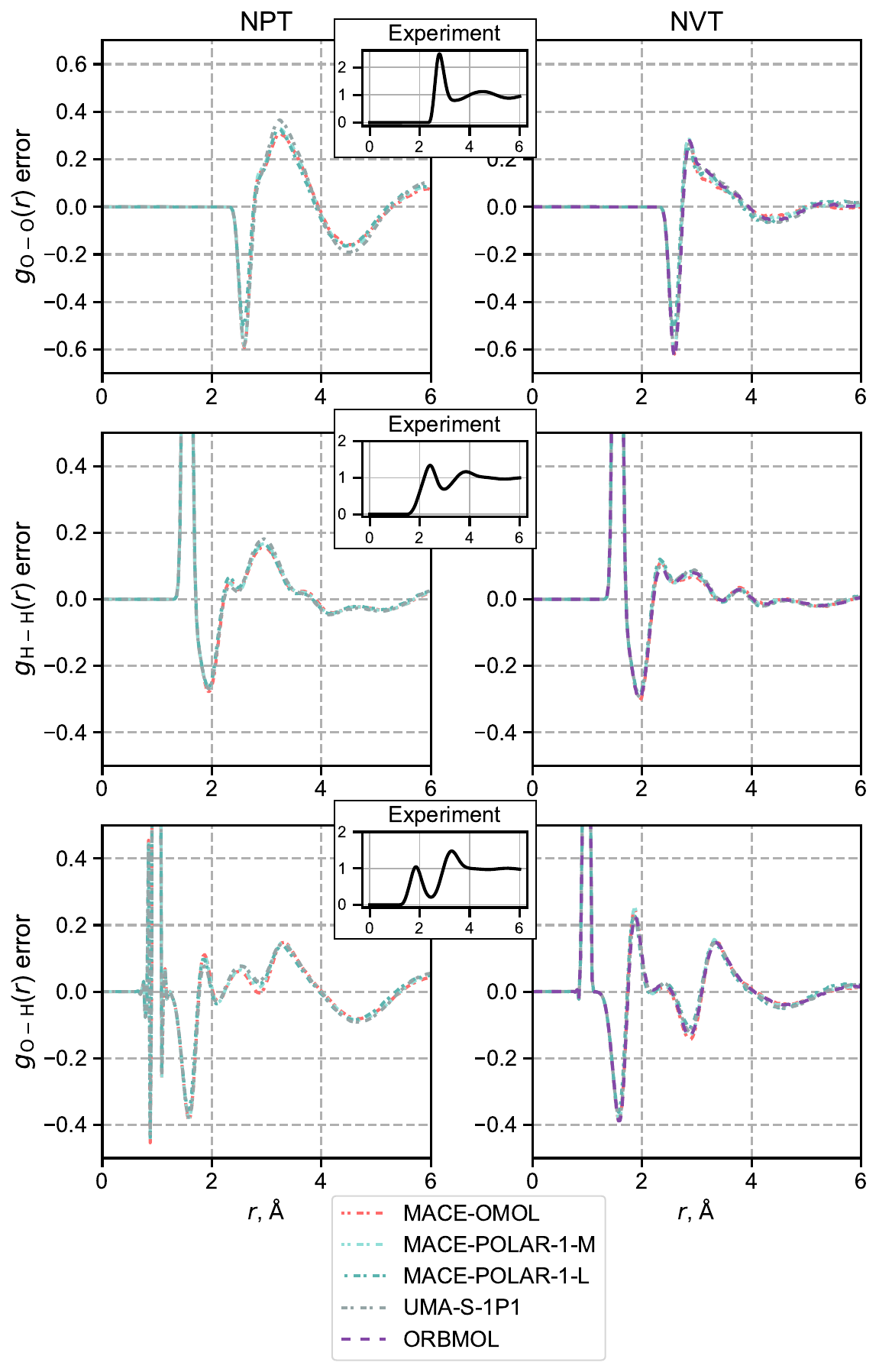}
    \caption{\textbf{Differences between predicted and experimental radial distribution functions for O-O, H-H, and O-H pairs in liquid water at 300~K.}
The radial distribution function $g(r)$ measures the probability of finding an atom pair at distance $r$ relative to an ideal gas. Left panels show differences between predicted and experimental $g(r)$ from a 500 ps NPT simulation; right-panel dashed lines show differences in $g(r)$ from NVT simulations at the experimental density of 0.997~g~cm$^{-3}$. The inset indicates the reference experimental neutron diffraction data. Coloured lines correspond to different models as indicated in the legend.}
    \label{fig:water_rdf}
\end{figure}

We use the same NPT production runs at 300 K to compute radial distribution functions (RDFs) for the oxygen-oxygen, hydrogen-hydrogen, and oxygen-hydrogen pairs and compare them against the experimental RDF available at the same temperature \cite{soper_radial_2013}, as shown in Figure \ref{fig:water_rdf}. While the models agree closely with each other, the simulated RDFs deviate from experiment in peak heights and second-shell structure, indicating that short-range packing and intermediate-range ordering remain imperfect. These deviations are also likely due to the \wb{} functional, as all models agree closely, and the functional lacks three-body dispersion effects known to be important in water structure.
While the models overestimate the water density, simulations at constant volume at the correct density are also informative. We therefore performed NVT simulations at a density of 0.997 g cm$^{-3}$ and a temperature of 300 K, using the final 400 ps of 500 ps runs to compute RDFs. As shown in Figure \ref{fig:water_rdf}, the water structure shifts noticeably at the experimental density, more closely reproducing the experimental structure, but residual deviations from experiment remain.

\subsection{Organic liquid densities}

\begin{figure}[h!]
    \centering
\includegraphics[width=\columnwidth]{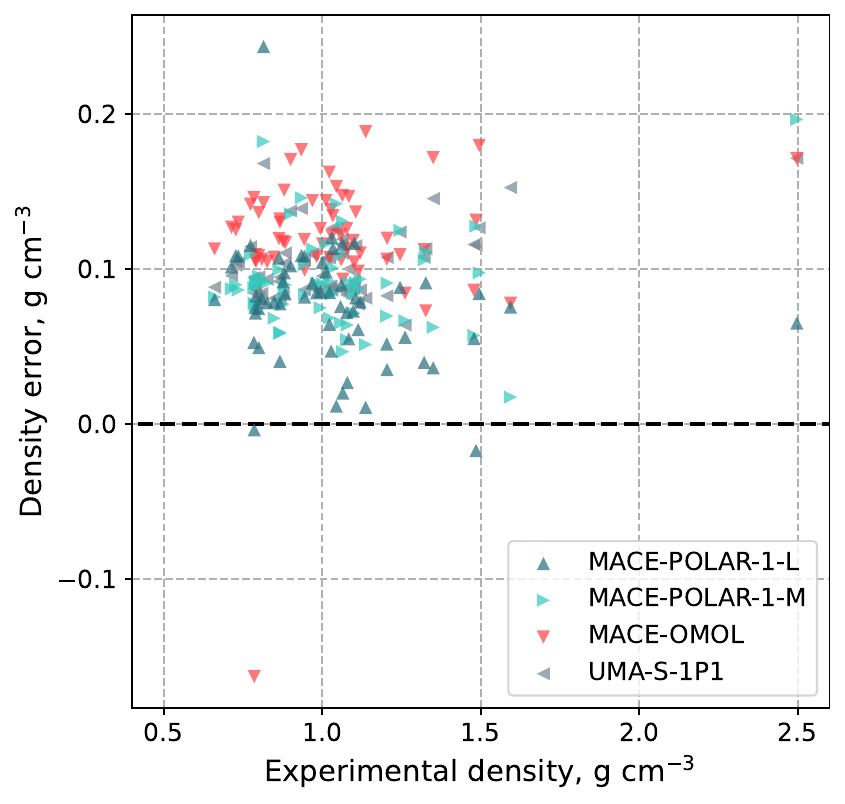}
    \caption{\textbf{Absolute errors on experimental densities for 62 organic liquids.}
Each point represents the absolute error on the equilibrium density (g/cm$^3$) of one molecular liquid computed from NPT molecular dynamics at 1~atm with approximately 1000 atoms for 500~ps of equilibration followed by 500~ps of simulation. The dataset spans alkanes (n-hexane to n-dodecane), alcohols (methanol to octanol), ethers (diethyl ether, THF), aromatics (benzene, toluene, xylenes), ketones (acetone, MEK), and polar solvents (acetonitrile, DMF, DMSO). The diagonal black dashed line indicates perfect agreement; points above the line indicate overestimated density.}
    \label{fig:liquid_densities}
\end{figure}

Equilibrium densities of liquids are determined by a subtle balance between attractive and repulsive intermolecular forces~\cite{magduau2023machine}. We validated four models---\MOMOL, \MFPM, \MFPL, and \UMAS---on the densities of 62 organic liquids covering both aliphatic and aromatic compounds, and common functional groups such as alcohols, ethers, esters, ketones, nitriles, halogens, and others. We did not evaluate \UMAM as it was too slow at the time of writing.
For all liquids, excluding 1-octanol, we used initial structures, experimental temperatures and densities provided in Ref.~\cite{Weber2025MPNICE} and equilibrated each model by running NVT molecular dynamics at the experimental temperatures for 50 ps. For 1-octanol, the system was initialised at low density using Packmol \cite{packmol}. To obtain a starting frame for the NPT simulation, each model was equilibrated at 298 K while shrinking the cubic cell parameter by 0.05 \AA\ in 500 fs intervals, until the system reached 98\% of the experimental density. Isothermal-isobaric ensemble molecular dynamics runs were performed for 1 ns using the Martyna-Tobias-Klein barostat \cite{mtkbarostat} implemented in ASE, with the thermostat and barostat damping parameters of 50 and 500 ps, respectively. For each system and model, only the final 500 ps of the MD trajectory were used to compute the equilibrium density.
Figure~\ref{fig:liquid_densities} shows that all tested models predict liquid densities close to experiment but systematically overestimate them, similar to the water density in the previous section. Mean absolute errors over the entire test set, given in Table \ref{tab:liquids_mae}, show that \MFPL performs the best, with a density MAE of 0.077 g cm$^{-3}$, while \MOMOL is the least accurate on liquid densities, with an MAE of 0.126 g cm$^{-3}$. The calculated densities shown in Figure \ref{fig:liquid_densities} are also tabulated in Table \ref{tab:liquid_densities}.

\begin{table}[h!]
\centering
\caption{Mean absolute errors (MAE) in predicted organic liquid densities.}
\label{tab:liquids_mae}
\resizebox{0.9\columnwidth}{!}{%
\begin{tabular}{@{}lc@{}}
\toprule
\textbf{Model} & \textbf{Density MAE, g cm$^{-3}$} \\
\midrule
\MOMOL &  0.126\\
\MFPM & 0.091 \\
\MFPL & \textbf{0.077} \\
\UMAS & 0.102 \\

\bottomrule
\end{tabular}%
}
\hspace{5pt}
\end{table}

\subsection{Solvated Ions and Changes in Oxidation State}
\begin{figure*}
    \centering
\includegraphics[width=0.9\linewidth]{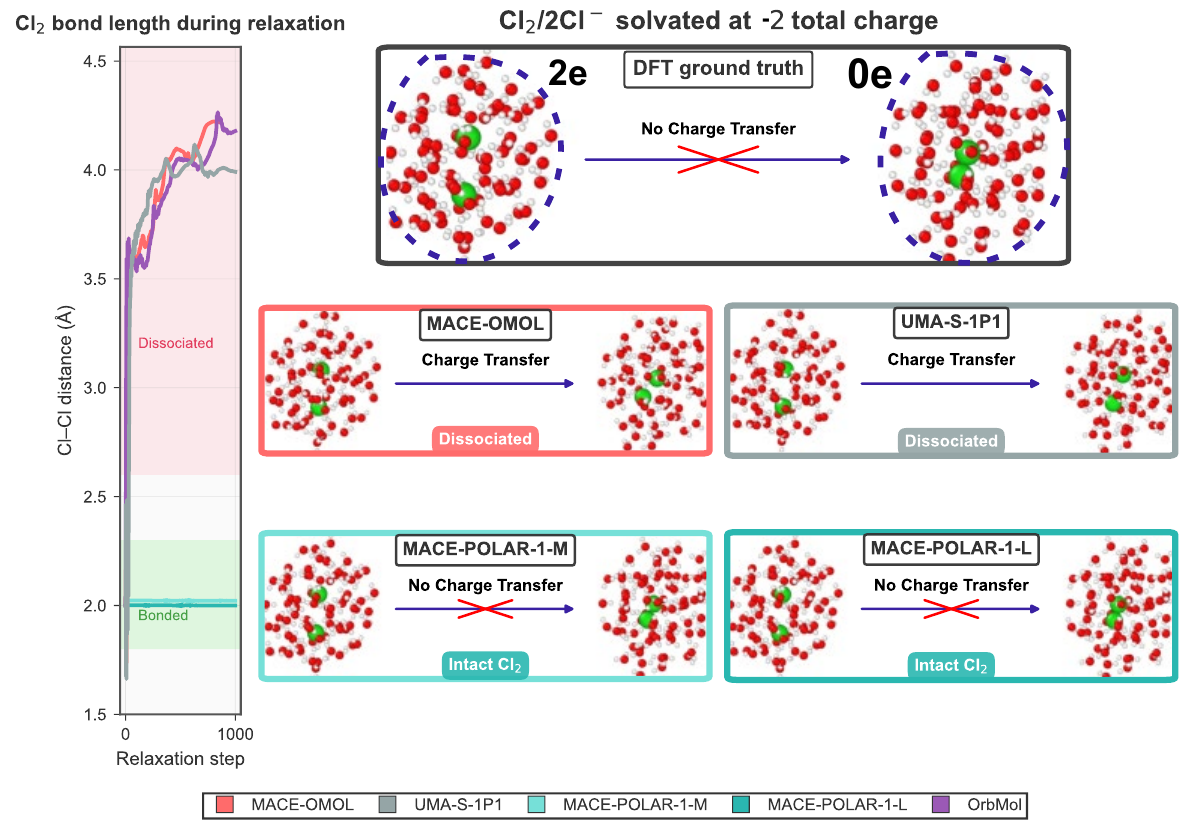}
 \caption{\textbf{Stability of solvated Cl$_2$ during geometry relaxation.}
A neutral Cl$_2$ molecule is solvated in water alongside two distant Cl$^-$ ions (total system charge $-2$). Because the chloride ions lie beyond the message-passing cutoff ($>$6~\AA), local models cannot distinguish neutral Cl$_2$ from hypothetical Cl$_2^{2-}$ based on local environment alone.
Left: Cl-Cl bond length (\AA) versus relaxation step. Non-electrostatic models (\MOMOL, \UMAS, \ORBMOL; red/pink/gray traces) drive unphysical bond dissociation, while electrostatic \MFP models (teal traces) preserve the covalent bond near its gas-phase equilibrium value ($\sim$2.0~\AA). Shaded regions indicate bonded (green, $<$2.5~\AA) and dissociated (pink, $>$3.5~\AA) regimes.
Right: Molecular snapshots showing the initial solvated Cl$_2$ configuration (top) and final relaxed structures for each model.}
    \label{fig:cl2_dissociation}
\end{figure*}
\subsubsection{Artificial dissociation of solvated Cl$_2$/Cl$^-$ pairs}
Figure~\ref{fig:cl2_dissociation} demonstrates the critical role of proper charge localisation in describing anionic species in aqueous solution. Starting from a Cl$_2$ molecule solvated in a water cluster alongside two distant Cl$^-$ ions in separate water clusters, the system should maintain the covalent Cl$_2$ bond, as the excess $-2$ charge resides on the separated chloride ions, as shown in the top panel of Fig.~\ref{fig:cl2_dissociation}. This configuration presents a fundamental challenge for machine learning potentials that treat total system charge through global charge embedding: because the two solvated clusters are separated by distances exceeding the message-passing network cutoff ($>$6 Å), the local atomic environment around the Cl$_2$ molecule contains no information about the presence of the distant Cl$^-$ ions. Consequently, models without explicit long-range electrostatics (\MOMOL, \UMAS and \ORBMOL) interpret the local Cl environment descriptors as if each atom carries a $-1$ charge in an unstable anionic geometry, rather than recognising the neutral covalently bonded Cl$_2$ configuration. This misidentification drives erroneous destabilisation of the Cl$_2$ bond, leading to unphysical dissociation with the Cl--Cl bond length increasing from $\sim$2.0 Å to $>$4.0 Å during geometry relaxation. Importantly, when the isolated Cl$_2$ cluster is simulated at net zero charge---where the global embedding correctly identifies the neutral
 state—all models preserve the intact Cl$_2$ bond, confirming that the failure arises specifically from the inability to communicate total charge information beyond the local cutoff radius. In contrast, \MFP models with explicit
 long-range electrostatics correctly compute the electrostatic potential at each atom from all charges in the system, properly localising the $-2$ charge on the distant chloride ions while maintaining the Cl$_2$ molecule at its
 equilibrium bond length. 

The hybrid DFT \wb{} has very low self-interaction error, and therefore we expect near-exact localisation of two additional electrons on the cluster containing the two Cl$^-$ ions and zero electrons on the cluster with Cl$_2$. We verify this by computing the same system with fewer water molecules and observe exact localisation for \wb{}. We also test localisation in the \MFP models by summing the charges obtained by the models on each cluster. We observe substantial localisation, with $1.952\,|e|$ on the two Cl$^-$ ions and $0.048\,|e|$ on the Cl$_2$ cluster.

\subsubsection{$\text{Fe}^{3+}/\text{Fe}^{2+}$ redox pair in solution}

  \begin{figure*}[t!]
  \centering
  \includegraphics[width=\linewidth]{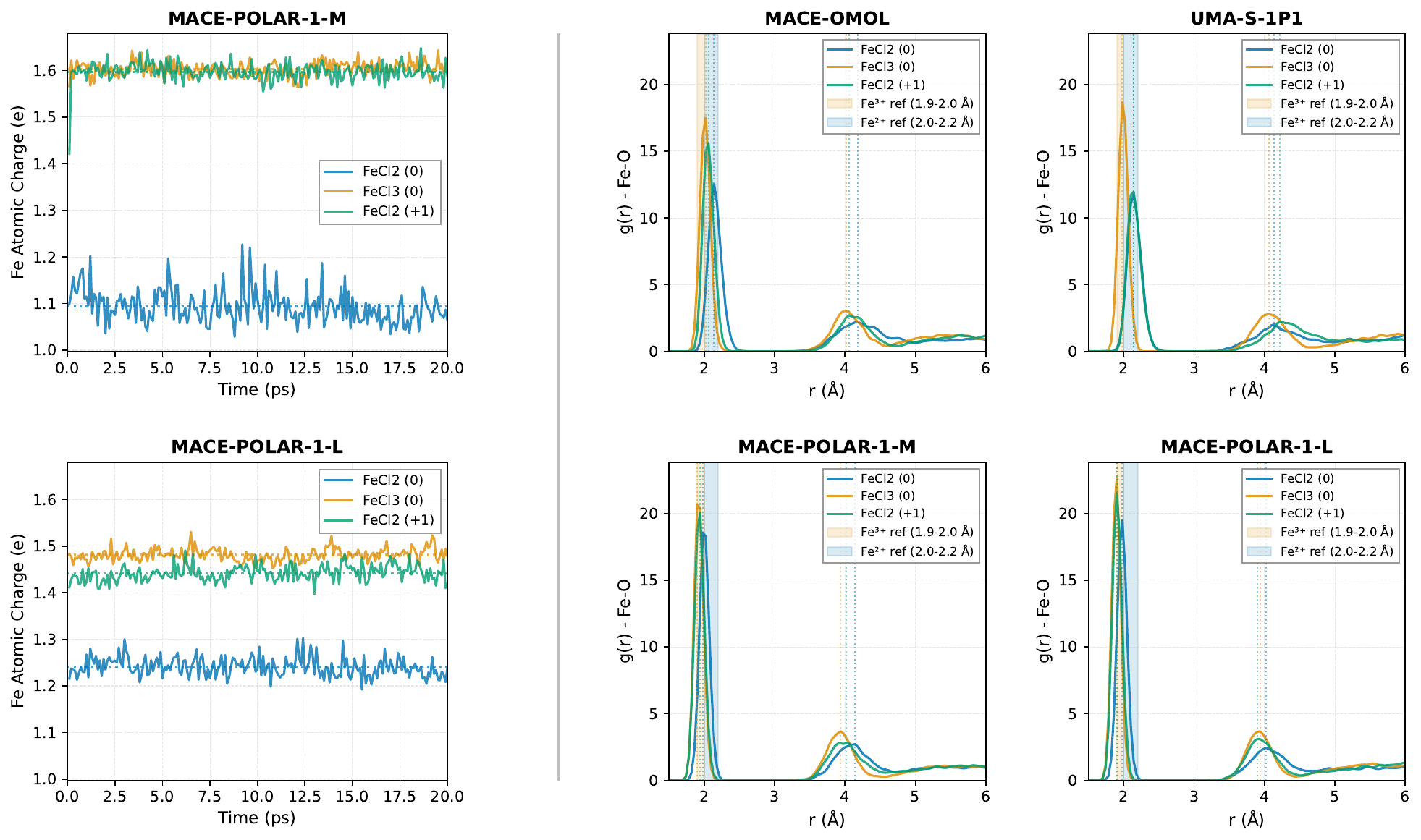}
  \caption{\textbf{Solvation structure and charge dynamics of aqueous iron chloride systems.}
  Left column: Fe atomic partial charge (in charge atomic units, $|e|$) as a function of simulation time over 20~ps of NVT molecular dynamics at 300~K. Results are shown for \MFPM (top) and \MFPL (bottom), comparing three systems: FeCl$_2$ with net charge 0 (blue), FeCl$_3$ with net charge 0 (red), and FeCl$_2$ with net charge $+1$ (teal). Dotted horizontal lines indicate time-averaged Fe charges. The similar charge values for FeCl$_3$ (0) and FeCl$_2$ (+1) indicate correct localisation of charge to Fe$^{3+}$ in both cases.
  Right column: Fe-O radial distribution functions $g(r)$ showing the probability of finding water oxygen atoms at distance $r$ from the Fe centre. All three systems are overlaid in each panel for \MOMOL, \UMAS, \MFPM, and \MFPL. Vertical dotted lines mark first and second coordination shell peaks. The first-shell contraction from Fe$^{2+}$ to Fe$^{3+}$ reflects the smaller ionic radius at higher oxidation state.}
  \label{fig:fe_solvation}
  \end{figure*}
Iron in a chloride--water solution is a prototypical redox system in solution. This system tests whether the models correctly localise charge into well-defined oxidation states~\cite{kocer2024machinelearningpotentialsredox}. We study a
  system involving either a high-spin Fe$^{2+}$ (d$^6$, $S=2$, multiplicity 5) or Fe$^{3+}$ (d$^5$, $S=5/2$, multiplicity 6) centres, solvated in water with Cl$^-$ ions as counter-ions. The box size is 12~\AA{}, so all the local models can see the full box in their receptive fields. We simulate the two systems at zero total charge, indicated as (0) in Figure~\ref{fig:fe_solvation}. The difference in oxidation states between the Fe$^{2+}$ and Fe$^{3+}$ affects the whole solvation shell, and therefore results in differences in the radial distribution functions. The right panels of Figure~\ref{fig:fe_solvation} plot the Fe--O radial distribution functions $g(r)$ for the different tested models along with shaded bands corresponding to the experimental values of the first peak, corresponding to the first solvation shell. The values are taken from~\cite{kocer2024machinelearningpotentialsredox}. From the experimental values, we expect the first solvation shell to systematically contract from 2.0--2.2~\AA{} in FeCl$_2$ to 1.9--2.0~\AA{} in
  FeCl$_3$, reflecting the stronger electrostatic attraction and reduced ionic radius of the iron ion in the highest oxidation state. This solvation structure results from a complex interplay of multiple factors: the
  d-orbital occupancy and spin state determine the metal's ionic radius and ligand field stabilisation energy; the oxidation state controls the electrostatic attraction between the metal centre and water oxygen atoms; and the presence of
  chloride counter-ions modulates the electrostatic environment through charge screening and hydrogen-bonding networks with the solvent. \MFP models capture this complex physics through explicit treatment of
  long-range electrostatics and produce distinct and well-resolved first-shell peaks for each oxidation state. To further confirm charge localisation on the iron centre, we evaluated FeCl$_2$ at a total charge of +1, which should produce an oxidation state of +3. We observe that FeCl$_3$ (0) and FeCl$_2$ (+1) show nearly superimposable $g(r)$ profiles consistent with their equivalent Fe$^{3+}$ oxidation states. In contrast, \UMAS shows structural differentiation between FeCl$_3$ (0) and FeCl$_2$ (0); however, it incorrectly predicts FeCl$_2$ (+1) as the same as FeCl$_2$ (0), showing that it delocalises the extra charge on all the solvent and Cl atoms instead of localising it in the Fe centre. \MOMOL shows mixed results, with clear separation of FeCl$_3$ (0) and FeCl$_2$ (0) peaks, but an intermediate value for FeCl$_2$ (+1) showing moderate spurious charge delocalisation.

  The Fe atomic charges as a function of time (left panels) provide further evidence of the proper description of oxidation states by the \MFP models. Indeed, \MFP models predict well-defined, nearly constant Fe partial charges across the trajectory, revealing the presence of well-defined oxidation states. These partial charges
  are not integers due to the charge partitioning between the metal centre and its coordination environment: covalent donation from water oxygen lone pairs into empty Fe d-orbitals,
  polarisation of the metal's d-electron density toward the electronegative ligands, and delocalisation of the oxidation-state charge across the entire [Fe(H$_2$O)$_6$Cl$_n$]$^{q}$ complex result in the formal +2 or +3 charge being
  distributed over the Fe atom, coordinating water molecules, and chloride ligands. Consequently, the Fe partial charge reflects only the fraction of the total oxidation-state charge localised on the metal nucleus, with the remainder
  residing on the solvation shell. The FeCl$_2$ (+1) case provides further evidence of the correctness of charge transfer physics: with two Cl$^-$ ligands and a net +1 system charge, the iron centre should adopt an effective +3 oxidation state
  identical to FeCl$_3$ (0), yielding comparable Fe partial charges and solvation structures. \MFP models correctly reproduce this electronic equivalence, with FeCl$_3$ (0) and FeCl$_2$ (+1) exhibiting Fe charges differing by only
  $\sim$0.1$e$ for \MFPL and identical charges for \MFPM. 

\subsubsection{Micro-solvated ionisation potentials of transition metals in water}

\begin{figure*}
    \centering
\includegraphics[width=\linewidth]{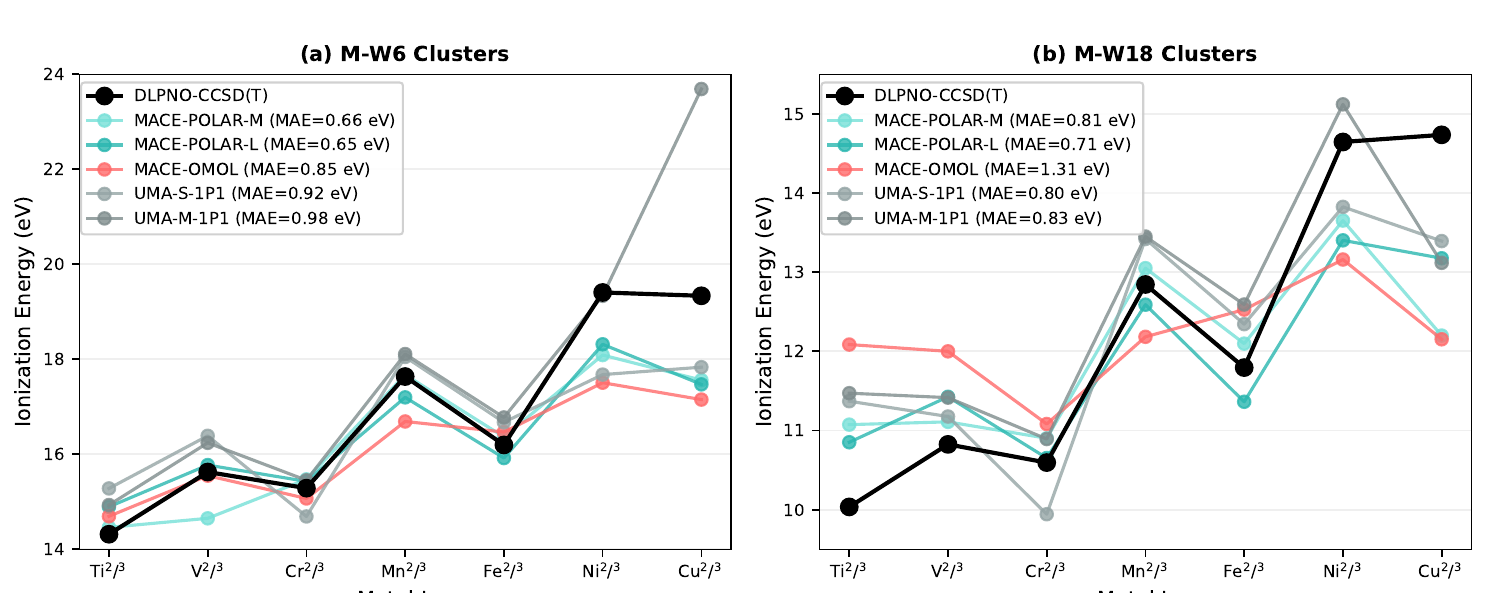}
    \caption{\textbf{Vertical ionisation energies of hydrated first-row transition metal ions.}
Line plots show the computed ionisation energy in eV for removing one electron from M$^{2+}$ to form M$^{3+}$ at fixed geometry, for metals Ti, V, Cr, Mn, Fe, Ni, and Cu.
(a) M-W6 clusters: metal ion coordinated by six water molecules in octahedral geometry.
(b) M-W18 clusters: same ions with an explicit second solvation shell of 12 additional water molecules.
Black circles show DLPNO-CCSD(T) reference values from Bhattacharjee \textit{et al.}~\cite{bhattacharjee2022dlpno}. Coloured lines show ML potential predictions: \MFPM (cyan), \MFPL (teal), \MOMOL (red), \UMAS (light gray), and \UMAM (dark gray). Mean absolute errors relative to the reference are given in the legend. The 4--5~eV reduction from M-W6 to M-W18 reflects electrostatic screening by the second solvation shell.}
    \label{fig:ionization_energies}
\end{figure*}

We evaluate vertical ionisation energies on the same hydrated cluster geometries and reference data reported by Bhattacharjee \textit{et al.}~\cite{bhattacharjee2022dlpno}. The benchmark considers first-row transition metal ions in the 2+/3+ oxidation states (Ti, V, Cr, Mn, Fe, Ni, Cu) using two explicit hydration models: M-W6 ($\mathrm{[M(H_2O)_6]^{2+/3+}}$) and M-W18 ($\mathrm{[M(H_2O)_6\cdot(H_2O)_{12}]^{2+/3+}}$). The reference values are computed at the DLPNO-CCSD(T) level of theory, and the geometry is relaxed at the BP86/def2-TZVP level of theory. For each cluster, we compute the vertical ionisation energy as the difference in electronic energy between the two oxidation states at fixed nuclei:
\begin{equation}
\text{IE}(\mathbf{R}_i) = E_{\text{M}^{3+}}(\mathbf{R}_i) - E_{\text{M}^{2+}}(\mathbf{R}_i)
\label{eq:ie_vertical}
\end{equation}
where $\mathbf{R}_i$ denotes the fixed cluster geometry. When evaluating the different oxidation states, we ensure the use of high-spin multiplicities. All spin multiplicities are tabulated in Table~\ref{tab:dual_redox_potentials_she}. The electrostatic \MFP models provide the closest agreement to the DLPNO references, with MAEs of 0.65--0.66~eV on M-W6 and 0.71--0.81~eV on M-W18. The local baseline \MOMOL deviates more strongly, especially for the larger M-W18 clusters (MAE 1.31~eV), while \UMAS/\UMAM are intermediate. Across the metal series, all models reproduce the strong reduction in ionisation energies when moving from M-W6 to M-W18, consistent with enhanced electrostatic screening by the second solvation shell.

\subsubsection{Redox potentials of transition metals in water}
\begin{table*}[tb]
    \centering
    \resizebox{\linewidth}{!}{%
\begin{tabular}{lcccccccccc}
	\toprule
	& & & \multicolumn{4}{c}{Self-distributed ($E$, V vs SHE)} &
	  \multicolumn{4}{c}{On the \MFPL trajectories ($E$, V vs SHE)} \\
	\cmidrule(r){4-7}\cmidrule(l){8-11}
	Ion & Spin multiplicities & $E_{\mathrm{exp}}$ &
	\MFPM & \MFPL & \MOMOL & \UMAS &
	\MFPM & \MFPL & \MOMOL & \UMAS \\
	\midrule
	Co$^{2+}$/Co$^{3+}$ & $(4, 1)$ & 1.92 &
	\cellcolor{cF}     0.309 &
	\cellcolor{cF}     0.663 &
	\cellcolor{cFail}  Crashed. &
	\cellcolor{cFail}  Crashed. &
	\cellcolor{cF}     0.383 &
	\cellcolor{cF}     0.664 &
	\cellcolor{cH}    10.412 &
	\cellcolor{cH}    14.394 \\
	Fe$^{2+}$/Fe$^{3+}$ & $(5, 6)$ & 0.77 &
	\cellcolor{cF}     1.561 &
	\cellcolor{cE}     1.380 &
	\cellcolor{cFail}  Crashed. &
	\cellcolor{cFail}  Crashed. &
	\cellcolor{cF}     1.557 &
	\cellcolor{cE}     1.380 &
	\cellcolor{cG}     3.313 &
	\cellcolor{cF}     2.223 \\
	Mn$^{2+}$/Mn$^{3+}$ & $(6, 5)$ & 1.50 &
	\cellcolor{cC}     1.734 &
	\cellcolor{cC}     1.280 &
	\cellcolor{cFail}  Crashed. &
	\cellcolor{cFail}  Crashed. &
	\cellcolor{cA}     1.590 &
	\cellcolor{cC}     1.280 &
	\cellcolor{cF}     2.892 &
	\cellcolor{cH}    -1.664 \\
	Ti$^{2+}$/Ti$^{3+}$ & $(3, 2)$ & -0.37 &
	\cellcolor{cA}    -0.367 &
	\cellcolor{cA}    -0.273 &
	\cellcolor{cH}    10.498 &
	\cellcolor{cFail}  Crashed. &
	\cellcolor{cA}    -0.350 &
	\cellcolor{cA}    -0.273 &
	\cellcolor{cG}    -3.248 &
	\cellcolor{cH}   -11.807 \\
	V$^{2+}$/V$^{3+}$ & $(4, 3)$ & -0.255 &
	\cellcolor{cE}     0.329 &
	\cellcolor{cF}     0.514 &
	\cellcolor{cFail}  Crashed. &
	\cellcolor{cFail}  Crashed. &
	\cellcolor{cE}     0.385 &
	\cellcolor{cF}     0.514 &
	\cellcolor{cH}    -9.804 &
	\cellcolor{cE}     0.419 \\
	\midrule
	MAE (V) &  &  &
	\cellcolor{cE}     0.679 &
	\cellcolor{cE}     0.588 &
	\cellcolor{cFail}  - &
	\cellcolor{cFail}  - &
	\cellcolor{cE}     0.673 &
	\cellcolor{cE}     0.676 &
	\cellcolor{cH}     6.534 &
	\cellcolor{cH}     6.902 \\
	\bottomrule
\end{tabular}%
    }
    \caption{
        Redox potentials $E$ in volts for M$^{2+}$/M$^{3+}$ aqueous couples computed from Marcus-theory free energy.
        Model values have been shifted by a constant estimated from a linear regression fit to account for differences in reference potential.
        Experimental potentials $E_{\mathrm{exp}}$ (V vs SHE) are taken from the same dataset as in the main text.
        Cell colours indicate the absolute deviation $|E - E_{\mathrm{exp}}|$, from green (small error) to red (large error).
    }
    \label{tab:dual_redox_potentials_she}
\end{table*}

Redox potentials of solvated transition metal ions test the model's ability to describe charge transfer, charge localisation and oxidation-state-dependent solvation. The accurate prediction of redox thermodynamics requires capturing the complex interactions between metal d-orbital energetics, solvent reorganisation, and long-range electrostatic interactions, which makes it a challenging frontier test for MLIPs. We evaluated the models on aqueous M$^{2+}$/M$^{3+}$ redox couples (M = Ti, V, Mn, Fe, Co) following the Marcus theory computational framework established in~\cite{Blumberger2005, Mandal2022}.

For each metal ion, we constructed systems containing a single M$^{2+}$ or M$^{3+}$ ion with 64 water molecules in a cubic periodic cell (12.4~Å). The initial structures were generated with Packmol \cite{packmol} and then relaxed with each model separately. Both oxidation states were simulated in their high-spin configurations: M$^{2+}$ with spin multiplicities of 3 (Ti), 4 (V), 5 (Fe), 6 (Mn), and 4 (Co); M$^{3+}$ with multiplicities of 2 (Ti), 3 (V), 6 (Fe), 5 (Mn), and 1 (Co).

Following previous protocols~\cite{Mandal2022}, we performed canonical (NVT) Langevin molecular dynamics simulations at 300~K for each oxidation state independently. After 50~ps equilibration, we collected 150~ps production trajectories, sampling configurations every 20~fs. For each sampled configuration $\mathbf{R}$ from the M$^{2+}$ ensemble, we computed the vertical energy gap:
\begin{equation}
\Delta E(\mathbf{R}) = E_{\text{M}^{3+}}(\mathbf{R}) - E_{\text{M}^{2+}}(\mathbf{R})
\end{equation}
via single-point energy calculations at both oxidation states. The same procedure was applied to configurations from the M$^{3+}$ ensemble. The ensemble averages $\langle \Delta E \rangle_{\text{M}^{2+}}$ and $\langle \Delta E \rangle_{\text{M}^{3+}}$ directly yield the reorganisation free energy and redox free energy through linear response relations:
\begin{align}
\Delta F &= \frac{1}{2}\left(\langle \Delta E \rangle_{\text{M}^{2+}} + \langle \Delta E \rangle_{\text{M}^{3+}}\right) \label{eq:deltaF}
\end{align}

Crucially, $\Delta F$ computed via Eq.~\ref{eq:deltaF} approximates the free energy difference of the different oxidation states. To compare with the experimental standard reduction potentials $E^{\circ}_{\text{exp}}$ measured vs.\ the standard hydrogen electrode (SHE), the arbitrary model-dependent shift in the free energy must be taken into account. Since our simulations do not include explicit calculations for proton insertion~\cite{Blumberger2005} or vacuum alignment to establish the absolute reference, we apply a constant shift per-model $C_{\text{model}}$ determined by least-squares fitting for each model to experimental data:
\begin{equation}
E^{\circ}_{\text{pred}} = \frac{\Delta F}{F} + C_{\text{model}}
\end{equation}
where $F$ is the Faraday constant. This approach enables the evaluation of relative redox trends and model performance while acknowledging the absence of an ab initio absolute reference.

Table~\ref{tab:dual_redox_potentials_she} presents the predicted redox potentials after applying the optimal shift for each model. We report results from two protocols: using per-model MD trajectories (left columns, called "Self-distributed") and using shared \MFPL trajectories evaluated with all models (right columns). The shared-trajectory protocol enables evaluation of the accuracy of the unstable models \MOMOL and \UMAS.

\MFPM and \MFPL achieve mean absolute errors (MAEs) of 0.60--0.64~V across the five redox couples, representing a dramatic improvement over local models. Critically, \MOMOL and \UMAS fail catastrophically on their own trajectories, with most simulations crashing during the equilibration. When forced to evaluate on stable \MFPL trajectories, these local models produce errors exceeding 10~V, demonstrating that they fundamentally misrepresent the energetics of solvated transition metal ions.

The relatively modest accuracy (MAE $\sim$0.6~V) despite correct qualitative physics reflects the challenge of this benchmark: transition metal redox couples involve multi-reference character, strong electron correlation, and subtle spin-state energetics that push the limits of the current foundation models. Notably, it is difficult to evaluate the accuracy of the hybrid DFT reference itself. The \MFP models' ability to approach this accuracy while maintaining computational efficiency demonstrates that explicit long-range electrostatics successfully transfers the qualitative physics of charge localisation and solvation response to this challenging domain, far outside the training distribution. These results confirm that \MFP models capture qualitatively the essential physics of aqueous redox chemistry—charge localisation, polarisation, and solvent reorganisation—even though quantitative accuracy will require further fine-tuning.

\subsection{Effect of external fields}

\begin{figure}
    \centering
\includegraphics[width=0.8\columnwidth]{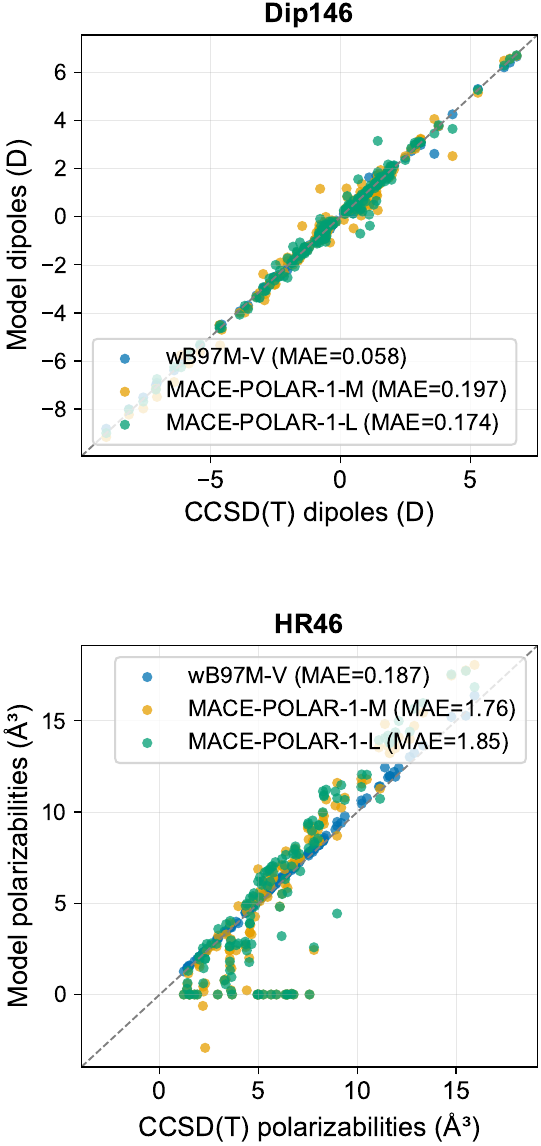}
    \caption{\textbf{External electric field response properties.}
Scatter plots compare model predictions (y-axis) against $\omega$B97M-V reference values (x-axis) for molecular response properties.
Left: Dip146 dataset of molecular dipole magnitudes (Debye).
Right: HR46 dataset of static isotropic polarisabilities (\AA$^3$).
The diagonal line indicates perfect agreement; points closer to the diagonal represent more accurate predictions. Notably, the \MFP models were trained only on ground-state energies and forces, yet predict these response properties through the learned electrostatic physics without explicit supervision.}
    \label{fig:external_field}
\end{figure}

The \MFP models are trained only on energies and forces, but can naturally extrapolate to external fields by using Equation~\ref{eq:field-features} and adding the potential generated by the external field to the potential from the charge density. Computing the response to external fields is a challenging extrapolation test of the learned electrostatic response.
Figure~\ref{fig:external_field} presents the results for external-field response properties: molecular dipoles (Dip146 dataset) and polarisabilities (HR46 dataset). The dipoles and polarisabilities are computed by finite differences of the total energy with respect to a uniform applied field: dipole moments are obtained from a first-order finite difference of the total energy $(\mu_i = -[E(+\Delta \mathbf{e})-E(-\Delta \mathbf{e})]/(2\Delta \mathbf{e}))$ with $\Delta \mathbf{e}=10^{-3}$ V/Å, and polarisabilities are obtained from the second derivative $(\alpha_{ii} = -[E(+\Delta \mathbf{e})-2E(0)+E(-\Delta \mathbf{e})]/ \Delta \mathbf{e}^2)$ with $\Delta \mathbf{e}=5\times10^{-4}$ V/Å. The geometries are re-centred to the molecular centroid in order to avoid translation problems. Critically, \MFP models were trained exclusively on ground-state
energies and forces without any explicit supervision on response properties, making this a pure test of whether the machine learning potentials have learned physically correct induction physics from the underlying quantum mechanical
training data. External field response properties represent a hierarchical series of increasingly challenging extrapolation tasks: molecular dipoles probe the first-order linear response of the electron density to an applied electric field, while polarisabilities characterise the second-order nonlinear response, requiring accurate prediction of field-induced redistribution of charge density. For the Dip146 benchmark, wB97M-V (the reference method used in training data generation) achieves MAE = 0.058 D, demonstrating internal consistency, while \MFPM and \MFPL attain MAE = 0.197 D and 0.174 D, respectively—representing $\sim$3$\times$ degradation relative to the reference but still maintaining quantitative accuracy for most molecules. The polarisability benchmark (HR46) presents a substantially more demanding test of second-order response, where wB97M-V achieves MAE = 0.187 Å$^3$, while \MFPM and \MFPL show MAE = 1.76 Å$^3$ and 1.85 Å$^3$—an order of magnitude degradation. Notably, the \MFP models exhibit systematic errors for low-polarisability molecules (cluster of predictions near zero in the HR46 plot), suggesting that while the models capture the dominant electrostatic induction effects through explicit long-range interactions, the many-body polarisation tensor requires higher-order correlation effects not fully encoded in the short-range message-passing architecture. Nevertheless, the ability of models trained solely on energies and forces to predict dipoles with $\sim$0.2 D accuracy and polarisabilities with $\sim$2 Å$^3$ accuracy demonstrates that physically meaningful electronic response emerges naturally from learning accurate potential energy surfaces.

\subsection{Lanthanides}

\begin{figure}
    \centering
    \includegraphics[width=\columnwidth]{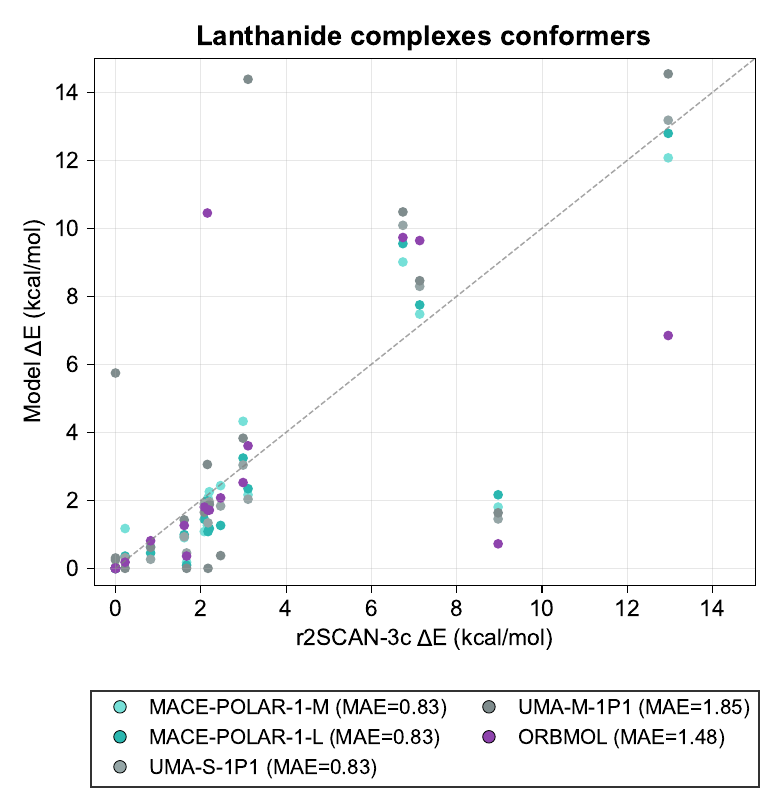}
    \caption{\textbf{Lanthanide isomerisation energies.} Relative energies (kcal/mol) for seven Ln complexes (La, Ce, Nd, Sm, Eu, Lu; 18 isomers total) from the GFN-FF study~\cite{Rose2024} compared against r2SCAN-3c references (Table S4). Points show model $\Delta E$ versus reference; dashed line is perfect agreement. Models: \MFPM/\MFPL (teal), \UMAS/\UMAM (gray), \ORBMOL (purple). \MOMOL is excluded due to unstable topologies on this set.}
    \label{fig:lanthanide_isomers}
\end{figure}

We evaluate the lanthanide benchmark of Rose \textit{et al.}~\cite{Rose2024} on seven published complexes containing isomers of lanthanum, cerium, neodymium, samarium, europium, and lutetium. We use the provided ORCA geometries, charges, and spins with
  r2SCAN-3c reference energies. Lanthanide isomerisation energies probe heavy‑element bonding and relative stability in f‑block complexes; small
  splittings and open‑shell f‑electron configurations make the ordering sensitive to the electronic description. All MLIPs match the reference ordering to within $\sim$1 kcal/mol on average except \ORBMOL, which shows several outliers; the
  tightest splittings (La, Eu) drive the visible offsets. Local \MOMOL failed to converge for several topologies and is excluded.

\subsection{Charge Transfer Tests}

\begin{figure*}[hbt!]
    \centering
\includegraphics[width=\linewidth]{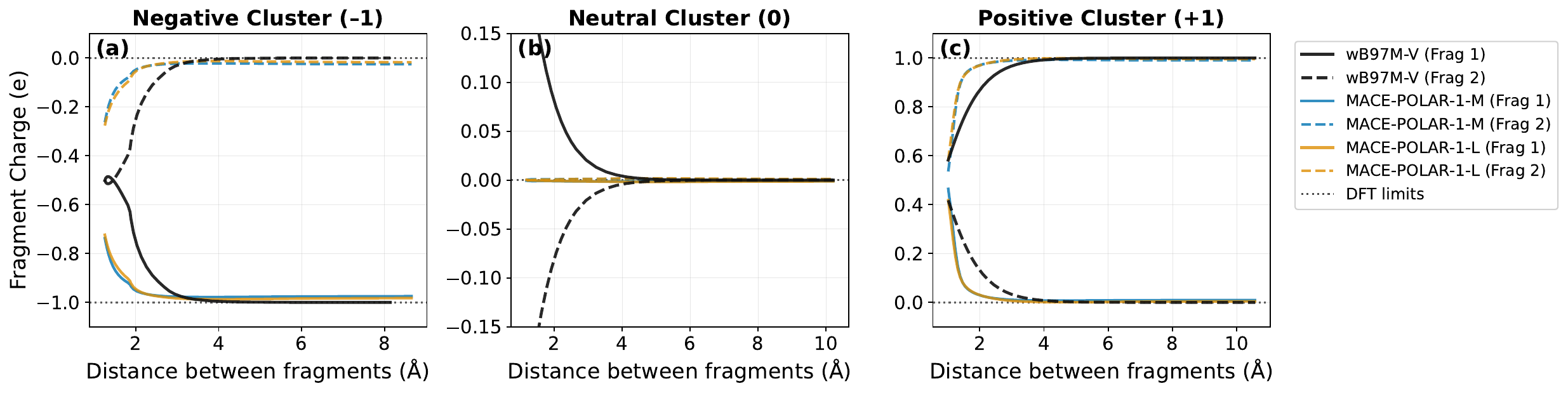}
     \caption{\textbf{Charge localisation during water cluster dissociation.}
  Line plots show the partial charge (in charge atomic units, $|e|$) on each of two water cluster fragments as a function of interfragment separation (\AA), for systems with total charge (a) $-1$, (b) $0$, and (c) $+1$ (all in $|e|$). Solid and dashed lines show charges on fragments 1 and 2, respectively, as predicted by \MFPM (blue) and \MFPL (orange). Black solid lines show $\omega$B97M-V Hirshfeld reference charges; dotted horizontal lines indicate the expected integer charges at infinite separation. At short distances (1--2~\AA), polarisation in the contact regime distributes charge across both fragments. As fragments separate beyond 4~\AA, excess charge localises onto one fragment, converging to the physical $-1/0$ or $+1/0$ limits.}
    \label{fig:water_cluster_sep}
\end{figure*}

To validate charge transfer and localisation during dissociation, we analysed the separation of charged water clusters into two fragments (Fig.~\ref{fig:water_cluster_sep}). We track the evolution of fragment charges as a function of interfragment distance for systems with total charges of $-1$, $0$, and $+1$ (in $|e|$). This test probes how the model partitions and localises charge between separating fragments—a critical capability for describing bond breaking in charged systems. For the neutral cluster
 (Fig.~\ref{fig:water_cluster_sep}b), both \MFPM and \MFPL correctly maintain near-zero charges on both fragments throughout separation. For charged clusters (Fig.~\ref{fig:water_cluster_sep}a,c), the models demonstrate physically correct charge localisation: as fragments separate beyond 4 Å, the excess charge localises quantitatively on one fragment, smoothly converging to the DFT-predicted limits of $-1/0$ and $+1/0$ (in $|e|$) for negative and positive systems, respectively. Notably, both fragments carry comparable partial charges at short distances (1--2~\AA), reflecting polarisation in the contact regime, before the charge localises as electrostatic coupling weakens. This smooth evolution validates that explicit long-range electrostatics enables physically correct charge redistribution during dissociation without spurious charge delocalisation between distant fragments.

\section{Conclusions and Outlook}

MFP models represent a significant advance in molecular modelling, demonstrating that explicit treatment of electrostatics within machine learning potentials yields substantial improvements over state-of-the-art local MLIPs. Following extensive testing, we confirmed the excellent quantitative accuracy of the models across diverse chemical systems, including small organic molecules, large protein-ligand complexes, redox ions in solution, transition metal complexes, and molecular crystals. The broad quantitative accuracy of the \MFP models on molecular chemistry positions them as powerful tools for computational chemistry.
The key architectural innovation of the polarisable long-range update provides a blueprint for next-generation foundation MLIPs. By capturing both short-range quantum effects and long-range electrostatics including polarisation effects and charge transfer, the model bridges the gap between the accurate description of short-range quantum effects and the physical description of long-range electrostatic interactions.

Several directions for future development emerge from this work. The incorporation of machine-learned dispersion corrections could further enhance accuracy for van der Waals complexes. The liquid densities, including water density, follow the correct qualitative trends but remain systematically off by about 5--10\% compared to experiment. This likely reflects limitations of the reference DFT functional used in OMol25, although model-architecture and simulation-protocol effects may also contribute. These results suggest that moving beyond hybrid DFT references, towards coupled-cluster-quality data, could improve quantitative liquid-property accuracy. Furthermore, the development of specialised variants for specific domains (proteins, materials, catalysis) could push accuracy even further while allowing higher computational efficiency by, e.g., distillation strategies.
More broadly, this work demonstrates that physics-informed machine learning—combining the flexibility of neural networks with rigorous physical principles—offers a path toward the simulation of molecular systems with \textit{ab-initio} quantum accuracy at scale.

\section*{Data Availability}

The OMol25 dataset is publicly available at \url{https://huggingface.co/facebook/OMol25}. All benchmark datasets are available from their respective publications. Processed data and analysis scripts to reproduce the benchmarks are provided via the ML Performance Guide (ML-PEG):  \url{https://ml-peg.stfc.ac.uk}.

\section*{Code Availability}

The MACE code is available at \url{https://github.com/ACEsuit/mace}.
The training scripts and models are available at \url{https://github.com/ACEsuit/mace-foundations/mace-polar-1}. ML-PEG code is available at \url{https://github.com/ddmms/ml-peg}.

\section*{Acknowledgements}
 
We would like to thank NVIDIA Lepton Compute for providing compute
to run experiments.
The authors acknowledge the use of resources provided by the Isambard-AI National AI Research Resource (AIRR)~\cite{bristol-ai}. Isambard-AI is operated by the University of Bristol and is funded by the UK Government’s Department for Science, Innovation and Technology (DSIT) via UK Research and Innovation; and the Science and Technology Facilities Council [ST/AIRR/I-A-I/1023].
We would like to thank UK Sovereign AI for providing compute on Isambard-AI.

E.K. and A.M.E. were supported by the Ada Lovelace Centre at the Science and Technology Facilities Council (https://adalovelacecentre.ac.uk/), the Physical Sciences Data Infrastructure (https://psdi.ac.uk; jointly STFC and the University of Southampton) under grants EP/X032663/1 and EP/X032701/1, and EPSRC under grants EP/W026775/1 and EP/V028537/1.
We are grateful for computational
support from the UK national high-performance computing service, ARCHER2, for which access was obtained via the UKCP consortium and funded by EPSRC
grant reference EP/P022065/1 and EP/X035891/1.
I.B. was supported by the Harding Distinguished Postgraduate Scholarship. J. H. was supported by The LennardJones Centre Ruth Lynden-Bell Scholarship in Scientific Computing.

\section*{Competing Interests}

GC is a partner in Symmetric Group LLP that licenses force fields commercially. GC and JHM have equity interests in Ångström AI.

\clearpage

\bibliography{references}

@misc{batatia2024foundationmodelatomisticmaterials,
      title={A foundation model for atomistic materials chemistry}, 
      author={Ilyes Batatia and Philipp Benner and Yuan Chiang and Alin M. Elena and Dávid P. Kovács and Janosh Riebesell and Xavier R. Advincula and Mark Asta and Matthew Avaylon and William J. Baldwin and Fabian Berger and Noam Bernstein and Arghya Bhowmik and Samuel M. Blau and Vlad Cărare and James P. Darby and Sandip De and Flaviano Della Pia and Volker L. Deringer and Rokas Elijošius and Zakariya El-Machachi and Fabio Falcioni and Edvin Fako and Andrea C. Ferrari and Annalena Genreith-Schriever and Janine George and Rhys E. A. Goodall and Clare P. Grey and Petr Grigorev and Shuang Han and Will Handley and Hendrik H. Heenen and Kersti Hermansson and Christian Holm and Jad Jaafar and Stephan Hofmann and Konstantin S. Jakob and Hyunwook Jung and Venkat Kapil and Aaron D. Kaplan and Nima Karimitari and James R. Kermode and Namu Kroupa and Jolla Kullgren and Matthew C. Kuner and Domantas Kuryla and Guoda Liepuoniute and Johannes T. Margraf and Ioan-Bogdan Magdău and Angelos Michaelides and J. Harry Moore and Aakash A. Naik and Samuel P. Niblett and Sam Walton Norwood and Niamh O'Neill and Christoph Ortner and Kristin A. Persson and Karsten Reuter and Andrew S. Rosen and Lars L. Schaaf and Christoph Schran and Benjamin X. Shi and Eric Sivonxay and Tamás K. Stenczel and Viktor Svahn and Christopher Sutton and Thomas D. Swinburne and Jules Tilly and Cas van der Oord and Eszter Varga-Umbrich and Tejs Vegge and Martin Vondrák and Yangshuai Wang and William C. Witt and Fabian Zills and Gábor Csányi},
      year={2024},
      eprint={2401.00096},
      archivePrefix={arXiv},
      primaryClass={physics.chem-ph},
      url={https://arxiv.org/abs/2401.00096}, 
}

@article{batatia_mace_2023,
  title={MACE: Higher order equivariant message passing neural networks for fast and accurate force fields},
  author={Batatia, Ilyes and Kovacs, David P and Simm, Gregor and Ortner, Christoph and Cs{\'a}nyi, G{\'a}bor},
  journal={Advances in Neural Information Processing Systems},
  volume={35},
  pages={11423--11436},
  year={2022}
}

@misc{levine2025openmolecules2025omol25,
      title={The Open Molecules 2025 (OMol25) Dataset, Evaluations, and Models}, 
      author={Daniel S. Levine and Muhammed Shuaibi and Evan Walter Clark Spotte-Smith and Michael G. Taylor and Muhammad R. Hasyim and Kyle Michel and Ilyes Batatia and Gábor Csányi and Misko Dzamba and Peter Eastman and Nathan C. Frey and Xiang Fu and Vahe Gharakhanyan and Aditi S. Krishnapriyan and Joshua A. Rackers and Sanjeev Raja and Ammar Rizvi and Andrew S. Rosen and Zachary Ulissi and Santiago Vargas and C. Lawrence Zitnick and Samuel M. Blau and Brandon M. Wood},
      year={2025},
      eprint={2505.08762},
      archivePrefix={arXiv},
      primaryClass={physics.chem-ph},
      url={https://arxiv.org/abs/2505.08762}, 
}

@article{Zhang2024,
  title = {DPA-2: a large atomic model as a multi-task learner},
  volume = {10},
  ISSN = {2057-3960},
  url = {http://dx.doi.org/10.1038/s41524-024-01493-2},
  DOI = {10.1038/s41524-024-01493-2},
  number = {1},
  journal = {npj Computational Materials},
  publisher = {Springer Science and Business Media LLC},
  author = {Zhang,  Duo and Liu,  Xinzijian and Zhang,  Xiangyu and Zhang,  Chengqian and Cai,  Chun and Bi,  Hangrui and Du,  Yiming and Qin,  Xuejian and Peng,  Anyang and Huang,  Jiameng and Li,  Bowen and Shan,  Yifan and Zeng,  Jinzhe and Zhang,  Yuzhi and Liu,  Siyuan and Li,  Yifan and Chang,  Junhan and Wang,  Xinyan and Zhou,  Shuo and Liu,  Jianchuan and Luo,  Xiaoshan and Wang,  Zhenyu and Jiang,  Wanrun and Wu,  Jing and Yang,  Yudi and Yang,  Jiyuan and Yang,  Manyi and Gong,  Fu-Qiang and Zhang,  Linshuang and Shi,  Mengchao and Dai,  Fu-Zhi and York,  Darrin M. and Liu,  Shi and Zhu,  Tong and Zhong,  Zhicheng and Lv,  Jian and Cheng,  Jun and Jia,  Weile and Chen,  Mohan and Ke,  Guolin and E,  Weinan and Zhang,  Linfeng and Wang,  Han},
  year = {2024},
  month = dec 
}

@article{Kovcs2023,
  title = {Evaluation of the MACE force field architecture: From medicinal chemistry to materials science},
  volume = {159},
  ISSN = {1089-7690},
  url = {http://dx.doi.org/10.1063/5.0155322},
  DOI = {10.1063/5.0155322},
  number = {4},
  journal = {The Journal of Chemical Physics},
  publisher = {AIP Publishing},
  author = {Kovács,  Dávid Péter and Batatia,  Ilyes and Arany,  Eszter Sára and Csányi,  Gábor},
  year = {2023},
  month = jul 
}

@article{kriz2020benchmarking,
  title={Benchmarking of semiempirical quantum-mechanical methods on systems relevant to computer-aided drug design},
  author={Kriz, Kristian and Rezac, Jan},
  journal={Journal of Chemical Information and Modeling},
  volume={60},
  number={3},
  pages={1453--1460},
  year={2020},
  publisher={ACS Publications}
}

@article{puleva2025quid,
  title={Extending quantum-mechanical benchmark accuracy to biological ligand-pocket interactions},
  author={Puleva, Mirela and Medrano Sandonas, Leonardo and L{\"o}rincz, Bal{\'a}zs D. and Charry, Jorge and Rogers, David M. and Nagy, P{\'e}ter R. and Tkatchenko, Alexandre},
  journal={Nature Communications},
  volume={16},
  number={1},
  year={2025},
  doi={10.1038/s41467-025-63587-9},
  publisher={Springer Nature}
}

@misc{wood2025umafamilyuniversalmodels,
      title={UMA: A Family of Universal Models for Atoms}, 
      author={Brandon M. Wood and Misko Dzamba and Xiang Fu and Meng Gao and Muhammed Shuaibi and Luis Barroso-Luque and Kareem Abdelmaqsoud and Vahe Gharakhanyan and John R. Kitchin and Daniel S. Levine and Kyle Michel and Anuroop Sriram and Taco Cohen and Abhishek Das and Ammar Rizvi and Sushree Jagriti Sahoo and Zachary W. Ulissi and C. Lawrence Zitnick},
      year={2025},
      eprint={2506.23971},
      archivePrefix={arXiv},
      primaryClass={cs.LG},
      url={https://arxiv.org/abs/2506.23971}, 
}

@article{batzner20223NequIP,
  title={E (3)-equivariant graph neural networks for data-efficient and accurate interatomic potentials},
  author={Batzner, Simon and Musaelian, Albert and Sun, Lixin and Geiger, Mario and Mailoa, Jonathan P and Kornbluth, Mordechai and Molinari, Nicola and Smidt, Tess E and Kozinsky, Boris},
  journal={Nature communications},
  volume={13},
  number={1},
  pages={2453},
  year={2022},
  publisher={Nature Publishing Group UK London}
}

@article{aimnet2019,
   author = {Roman Zubatyuk and Justin S Smith and Jerzy Leszczynski and Olexandr Isayev},
   doi = {10.1126/sciadv.aav6490},
   issue = {8},
   journal = {Science Advances},
   pages = {eaav6490},
   title = {Accurate and transferable multitask prediction of chemical properties with an atoms-in-molecules neural network},
   volume = {5},
   url = {https://www.science.org/doi/abs/10.1126/sciadv.aav6490},
   year = {2019},
}

@article{behler_4gnn_review_2021,
   author = {Jörg Behler},
   doi = {10.1021/acs.chemrev.0c00868},
   issue = {16},
   journal = {Chemical Reviews},
   note = {PMID: 33779150},
   pages = {10037-10072},
   title = {Four Generations of High-Dimensional Neural Network Potentials},
   volume = {121},
   url = {https://doi.org/10.1021/acs.chemrev.0c00868},
   year = {2021},
}

@article{cent2015,
   author = {S Alireza Ghasemi and Albert Hofstetter and Santanu Saha and Stefan Goedecker},
   doi = {10.1103/PhysRevB.92.045131},
   issue = {4},
   journal = {Phys. Rev. B},
   month = {7},
   pages = {45131},
   publisher = {American Physical Society},
   title = {Interatomic potentials for ionic systems with density functional accuracy based on charge densities obtained by a neural network},
   volume = {92},
   url = {https://link.aps.org/doi/10.1103/PhysRevB.92.045131},
   year = {2015},
}

@article{cent2017,
   author = {Somayeh Faraji and S Alireza Ghasemi and Samare Rostami and Robabe Rasoulkhani and Bastian Schaefer and Stefan Goedecker and Maximilian Amsler},
   doi = {10.1103/PhysRevB.95.104105},
   issue = {10},
   journal = {Phys. Rev. B},
   month = {3},
   pages = {104105},
   publisher = {American Physical Society},
   title = {High accuracy and transferability of a neural network potential through charge equilibration for calcium fluoride},
   volume = {95},
   url = {https://link.aps.org/doi/10.1103/PhysRevB.95.104105},
   year = {2017},
}

@article{cent2019,
   author = {Somayeh Faraji and S Alireza Ghasemi and Behnam Parsaeifard and Stefan Goedecker},
   doi = {10.1039/C9CP02213A},
   issue = {29},
   journal = {Phys. Chem. Chem. Phys.},
   pages = {16270-16281},
   publisher = {The Royal Society of Chemistry},
   title = {Surface reconstructions and premelting of the (100) CaF2 surface},
   volume = {21},
   url = {http://dx.doi.org/10.1039/C9CP02213A},
   year = {2019},
}

@misc{fu2025learningsmoothexpressiveinteratomic,
      title={Learning Smooth and Expressive Interatomic Potentials for Physical Property Prediction}, 
      author={Xiang Fu and Brandon M. Wood and Luis Barroso-Luque and Daniel S. Levine and Meng Gao and Misko Dzamba and C. Lawrence Zitnick},
      year={2025},
      eprint={2502.12147},
      archivePrefix={arXiv},
      primaryClass={physics.comp-ph},
      url={https://arxiv.org/abs/2502.12147}, 
}

@misc{rhodes2025orbv3atomisticsimulationscale,
      title={Orb-v3: atomistic simulation at scale}, 
      author={Benjamin Rhodes and Sander Vandenhaute and Vaidotas Šimkus and James Gin and Jonathan Godwin and Tim Duignan and Mark Neumann},
      year={2025},
      eprint={2504.06231},
      archivePrefix={arXiv},
      primaryClass={cond-mat.mtrl-sci},
      url={https://arxiv.org/abs/2504.06231}, 
}

@article{aimnetnse,
  title = {Teaching a neural network to attach and detach electrons from molecules},
  volume = {12},
  ISSN = {2041-1723},
  url = {http://dx.doi.org/10.1038/s41467-021-24904-0},
  DOI = {10.1038/s41467-021-24904-0},
  number = {1},
  journal = {Nature Communications},
  publisher = {Springer Science and Business Media LLC},
  author = {Zubatyuk,  Roman and Smith,  Justin S. and Nebgen,  Benjamin T. and Tretiak,  Sergei and Isayev,  Olexandr},
  year = {2021},
  month = aug 
}

@misc{gxtb2025,
  title = {g-xTB: A General-Purpose Extended Tight-Binding Electronic Structure Method For the Elements H to Lr (Z=1–103)},
  url = {http://dx.doi.org/10.26434/chemrxiv-2025-bjxvt},
  DOI = {10.26434/chemrxiv-2025-bjxvt},
  publisher = {American Chemical Society (ACS)},
  author = {Froitzheim,  Thomas and M\"{u}ller,  Marcel and Hansen,  Andreas and Grimme,  Stefan},
  year = {2025},
  month = jun 
}

@article{DellaPia2024,
  title = {How Accurate Are Simulations and Experiments for the Lattice Energies of Molecular Crystals?},
  volume = {133},
  ISSN = {1079-7114},
  url = {http://dx.doi.org/10.1103/PhysRevLett.133.046401},
  DOI = {10.1103/physrevlett.133.046401},
  number = {4},
  journal = {Physical Review Letters},
  publisher = {American Physical Society (APS)},
  author = {Della Pia,  Flaviano and Zen,  Andrea and Alfè,  Dario and Michaelides,  Angelos},
  year = {2024},
  month = jul 
}

@article{Goerigk2017,
  title = {A look at the density functional theory zoo with the advanced GMTKN55 database for general main group thermochemistry,  kinetics and noncovalent interactions},
  volume = {19},
  ISSN = {1463-9084},
  url = {http://dx.doi.org/10.1039/C7CP04913G},
  DOI = {10.1039/c7cp04913g},
  number = {48},
  journal = {Physical Chemistry Chemical Physics},
  publisher = {Royal Society of Chemistry (RSC)},
  author = {Goerigk,  Lars and Hansen,  Andreas and Bauer,  Christoph and Ehrlich,  Stephan and Najibi,  Asim and Grimme,  Stefan},
  year = {2017},
  pages = {32184–32215}
}

@article{sharapa2019robust,
  title={A robust and cost-efficient scheme for accurate conformational energies of organic molecules},
  author={Sharapa, Dmitry I and Genaev, Alexander and Cavallo, Luigi and Minenkov, Yury},
  journal={ChemPhysChem},
  volume={20},
  number={1},
  pages={92--102},
  year={2019},
  publisher={Wiley Online Library}
}

@article{ehlert2022conformational,
  title={Conformational energy benchmark for longer n-alkane chains},
  author={Ehlert, Sebastian and Grimme, Stefan and Hansen, Andreas},
  journal={The Journal of Physical Chemistry A},
  volume={126},
  number={22},
  pages={3521--3535},
  year={2022},
  publisher={ACS Publications}
}

@article{werner2023accurate,
  title={Accurate calculation of isomerization and conformational energies of larger molecules using explicitly correlated local coupled cluster methods in Molpro and ORCA},
  author={Werner, Hans-Joachim and Hansen, Andreas},
  journal={Journal of Chemical Theory and Computation},
  volume={19},
  number={20},
  pages={7007--7030},
  year={2023},
  publisher={ACS Publications}
}

@article{marianski2016assessing,
  title={Assessing the accuracy of across-the-scale methods for predicting carbohydrate conformational energies for the examples of glucose and $\alpha$-maltose},
  author={Marianski, Mateusz and Supady, Adriana and Ingram, Teresa and Schneider, Markus and Baldauf, Carsten},
  journal={Journal of chemical theory and computation},
  volume={12},
  number={12},
  pages={6157--6168},
  year={2016},
  publisher={ACS Publications}
}

@article{rezac2018mpconf196,
  title={Toward Accurate Conformational Energies of Smaller Peptides and Medium-Sized Macrocycles: MPCONF196 Benchmark Energy Data Set},
  author={Řezáč, Jan and Bím, Daniel and Gutten, Ondrej and Rulíšek, Lubomír},
  journal={Journal of Chemical Theory and Computation},
  volume={14},
  pages={1254--1266},
  year={2018},
  doi={10.1021/acs.jctc.7b01074}
}

@article{plett2023mpconf196water,
  title={Conformational energies of biomolecules in solution: Extending the MPCONF196 benchmark with explicit water molecules},
  author={Plett, Christoph and Grimme, Stefan and Hansen, Andreas},
  journal={Journal of Computational Chemistry},
  volume={45},
  pages={419--429},
  year={2023},
  doi={10.1002/jcc.27248}
}

@article{behara2024openff,
  title={Benchmarking Quantum Mechanical Levels of Theory for Valence Parametrization in Force Fields},
  author={Behara, Pavan Kumar and Jang, Hyesu and Horton, Joshua T. and Gokey, Trevor and Dotson, David L. and Boothroyd, Simon and Bayly, Christopher I. and Cole, Daniel J. and Wang, Lee-Ping and Mobley, David L.},
  journal={The Journal of Physical Chemistry B},
  volume={128},
  pages={7888--7902},
  year={2024},
  doi={10.1021/acs.jpcb.4c03167}
}

@article{wiktor2017scanrvv10,
  title={Note: Assessment of the SCAN+rVV10 functional for the structure of liquid water},
  author={Wiktor, Julia and Ambrosio, Francesco and Pasquarello, Alfredo},
  journal={The Journal of Chemical Physics},
  volume={147},
  number={21},
  pages={216101},
  year={2017},
  doi={10.1063/1.5006146}
}

@article{pruitt2016threebodywater,
  title={Importance of Three-Body Interactions in Molecular Dynamics Simulations of Water Demonstrated with the Fragment Molecular Orbital Method},
  author={Pruitt, Samuel R. and Nakata, Haruki and Nagata, Tetsuya and Mayes, Eric R. and Alexeev, Yuri and Fletcher, Geoffrey D. and Fedorov, Dmitri G. and Kitaura, Kazuo and Gordon, Mark S.},
  journal={Journal of Chemical Theory and Computation},
  volume={12},
  number={4},
  pages={1423--1435},
  year={2016},
  doi={10.1021/acs.jctc.5b01208}
}

@article{manna2017water27,
  title={Conventional and Explicitly Correlated ab Initio Benchmark Study on Water Clusters: Revision of the BEGDB and WATER27 Data Sets},
  author={Manna, Debashree and Kesharwani, Manoj K. and Sylvetsky, Nitai and Martin, Jan M. L.},
  journal={Journal of Chemical Theory and Computation},
  volume={13},
  number={7},
  pages={3136--3152},
  year={2017},
  doi={10.1021/acs.jctc.6b01046}
}

@article{kruse2015quantum,
  title={Quantum chemical benchmark study on 46 RNA backbone families using a dinucleotide unit},
  author={Kruse, Holger and Mladek, Arnost and Gkionis, Konstantinos and Hansen, Andreas and Grimme, Stefan and Sponer, Jiri},
  journal={Journal of Chemical Theory and Computation},
  volume={11},
  number={10},
  pages={4972--4991},
  year={2015},
  publisher={ACS Publications}
}

@article{ez2020-1,
  title = {Non-Covalent Interactions Atlas Benchmark Data Sets: Hydrogen Bonding},
  volume = {16},
  ISSN = {1549-9626},
  url = {http://dx.doi.org/10.1021/acs.jctc.9b01265},
  DOI = {10.1021/acs.jctc.9b01265},
  number = {4},
  journal = {Journal of Chemical Theory and Computation},
  publisher = {American Chemical Society (ACS)},
  author = {Řezáč,  Jan},
  year = {2020},
  month = mar,
  pages = {2355–2368}
}

@article{plett2024toward,
  title={Toward Reliable Conformational Energies of Amino Acids and Dipeptides - The DipCONFS Benchmark and DipCONL Datasets},
  author={Plett, Christoph and Grimme, Stefan and Hansen, Andreas},
  journal={Journal of Chemical Theory and Computation},
  volume={20},
  number={18},
  pages={8329--8339},
  year={2024},
  publisher={ACS Publications}
}

@article{Sure2015,
  title = {Comprehensive Benchmark of Association (Free) Energies of Realistic Host–Guest Complexes},
  volume = {11},
  ISSN = {1549-9626},
  url = {http://dx.doi.org/10.1021/acs.jctc.5b00296},
  DOI = {10.1021/acs.jctc.5b00296},
  number = {8},
  journal = {Journal of Chemical Theory and Computation},
  publisher = {American Chemical Society (ACS)},
  author = {Sure,  Rebecca and Grimme,  Stefan},
  year = {2015},
  month = jul,
  pages = {3785–3801}
}

@misc{gharakhanyan2025openmolecularcrystals2025,
      title={Open Molecular Crystals 2025 (OMC25) Dataset and Models}, 
      author={Vahe Gharakhanyan and Luis Barroso-Luque and Yi Yang and Muhammed Shuaibi and Kyle Michel and Daniel S. Levine and Misko Dzamba and Xiang Fu and Meng Gao and Xingyu Liu and Haoran Ni and Keian Noori and Brandon M. Wood and Matt Uyttendaele and Arman Boromand and C. Lawrence Zitnick and Noa Marom and Zachary W. Ulissi and Anuroop Sriram},
      year={2025},
      eprint={2508.02651},
      archivePrefix={arXiv},
      primaryClass={physics.chem-ph},
      url={https://arxiv.org/abs/2508.02651}, 
}

@article{Neugebauer2023,
  title = {Toward Benchmark-Quality Ab Initio Predictions for 3d Transition Metal Electrocatalysts: A Comparison of CCSD(T) and ph-AFQMC},
  volume = {19},
  ISSN = {1549-9626},
  url = {http://dx.doi.org/10.1021/acs.jctc.3c00617},
  DOI = {10.1021/acs.jctc.3c00617},
  number = {18},
  journal = {Journal of Chemical Theory and Computation},
  publisher = {American Chemical Society (ACS)},
  author = {Neugebauer,  Hagen and Vuong,  Hung T. and Weber,  John L. and Friesner,  Richard A. and Shee,  James and Hansen,  Andreas},
  year = {2023},
  month = sep,
  pages = {6208–6225}
}

@article{Wappett2023,
  title = {Benchmarking Density Functional Theory Methods for Metalloenzyme Reactions: The Introduction of the MME55 Set},
  volume = {19},
  ISSN = {1549-9626},
  url = {http://dx.doi.org/10.1021/acs.jctc.3c00558},
  DOI = {10.1021/acs.jctc.3c00558},
  number = {22},
  journal = {Journal of Chemical Theory and Computation},
  publisher = {American Chemical Society (ACS)},
  author = {Wappett,  Dominique A. and Goerigk,  Lars},
  year = {2023},
  month = nov,
  pages = {8365–8383}
}

@article{maurer2021assessing,
  title={Assessing density functional theory for chemically relevant open-shell transition metal reactions},
  author={Maurer, Leonard R and Bursch, Markus and Grimme, Stefan and Hansen, Andreas},
  journal={Journal of Chemical Theory and Computation},
  volume={17},
  number={10},
  pages={6134--6151},
  year={2021},
  publisher={ACS Publications}
}

@article{Gong2025,
  title = {A predictive machine learning force-field framework for liquid electrolyte development},
  volume = {7},
  ISSN = {2522-5839},
  url = {http://dx.doi.org/10.1038/s42256-025-01009-7},
  DOI = {10.1038/s42256-025-01009-7},
  number = {4},
  journal = {Nature Machine Intelligence},
  publisher = {Springer Science and Business Media LLC},
  author = {Gong,  Sheng and Zhang,  Yumin and Mu,  Zhenliang and Pu,  Zhichen and Wang,  Hongyi and Han,  Xu and Yu,  Zhiao and Chen,  Mengyi and Zheng,  Tianze and Wang,  Zhi and Chen,  Lifei and Yang,  Zhenze and Wu,  Xiaojie and Shi,  Shaochen and Gao,  Weihao and Yan,  Wen and Xiang,  Liang},
  year = {2025},
  month = apr,
  pages = {543–552}
}

@article{Parthiban2001,
  title = {Assessment of {W1} and {W2} Theories for the Computation of Electron Affinities, Ionization Potentials, Heats of Formation, and Proton Affinities},
  author = {Parthiban, S. and Martin, Jan M. L.},
  journal = {J. Chem. Phys.},
  volume = {114},
  pages = {6014--6029},
  year = {2001}
}

@article{Cheng2007,
  title = {Vertical Ionization Potentials of Small Molecules: A Benchmark Study},
  author = {Cheng, Lei and others},
  journal = {J. Chem. Phys.},
  volume = {127},
  pages = {214105},
  year = {2007}
}

@article{Luo2012,
  title = {How Evenly Can Approximate Density Functionals Treat the Different Multiplicities and Ionization States of {4d} Transition Metal Atoms?},
  author = {Luo, Sijie and Truhlar, Donald G.},
  journal = {J. Chem. Theory Comput.},
  volume = {8},
  pages = {4112--4126},
  year = {2012}
}

@article{Ermis2021,
  title = {State-Of-The-Art Computations of Vertical Electron Affinities with the Extended {Koopmans}' Theorem Integrated with the {CCSD(T)} Method},
  author = {Ermi{\c{s}}, B{\"u}lent and Ekinci, Emine and Bozkaya, U{\u{g}}ur},
  journal = {J. Chem. Theory Comput.},
  volume = {17},
  pages = {7648--7656},
  year = {2021}
}

@article{Chan2017,
  title = {Assessment of {DFT} Methods for Transition Metals with the {TMC151} Compilation of Data Sets and Comparison with Accuracies for Main-Group Chemistry},
  author = {Chan, Bun and others},
  journal = {J. Chem. Theory Comput.},
  volume = {13},
  pages = {3385--3398},
  year = {2017}
}

@article{Chan2018,
  title = {Barriometry---An Enhanced Database of Accurate Barrier Heights for Gas-Phase Reactions},
  author = {Chan, Bun and Simmie, John M.},
  journal = {Phys. Chem. Chem. Phys.},
  volume = {20},
  pages = {10732--10740},
  year = {2018}
}

@article{Zhao2012,
  title = {Benchmark Database for Ylidic Bond Dissociation Energies and Its Use for Assessments of Electronic Structure Methods},
  author = {Zhao, Yan and Ng, Hoi Ting and Peverati, Roberto and Truhlar, Donald G.},
  journal = {J. Chem. Theory Comput.},
  volume = {8},
  pages = {2824--2834},
  year = {2012}
}

@article{Karton2017HAT,
  title = {Heavy Atom Transfer Reactions: {HAT707} Benchmark},
  author = {Karton, Amir and others},
  journal = {J. Phys. Chem. A},
  year = {2017},
  note = {From W4-11 dataset}
}

@article{Zhao2012RSE,
  title = {Radical Stabilization Energies: {RSE43} Benchmark},
  author = {Zhao, Yan and others},
  journal = {J. Chem. Theory Comput.},
  volume = {8},
  year = {2012}
}

@article{Chan2016BSR,
  title = {Bond Separation Reactions: {BSR36} Benchmark},
  author = {Chan, Bun and others},
  journal = {J. Chem. Theory Comput.},
  year = {2016}
}

@article{Prasad2021,
  title = {{BH9}, a New Comprehensive Benchmark Data Set for Barrier Heights and Reaction Energies: Assessment of Density Functional Approximations and Basis Set Incompleteness Potentials},
  author = {Prasad, Viki Kumar and Pei, Zhen and Edelmann, Steven and Otero-de-la Roza, A. and DiLabio, Gino A.},
  journal = {J. Chem. Theory Comput.},
  volume = {18},
  pages = {151--166},
  year = {2022}
}

@article{Zhao2008BH76,
  title = {Design of Density Functionals by Combining the Method of Constraint Satisfaction with Parametrization for Thermochemistry, Thermochemical Kinetics, and Noncovalent Interactions},
  author = {Zhao, Yan and Truhlar, Donald G.},
  journal = {J. Chem. Theory Comput.},
  volume = {4},
  pages = {1849--1868},
  year = {2008}
}

@article{Yu2015CRBH,
  title = {Reaction Barrier Heights for Cycloreversion of Heterocyclic Rings: An {Achilles}' Heel for {DFT} and Standard Ab Initio Procedures},
  author = {Yu, Li-Juan and Sarrami, Fatemeh and O'Reilly, Robert J. and Karton, Amir},
  journal = {Chem. Phys.},
  volume = {458},
  pages = {1--8},
  year = {2015}
}

@article{Karton2008DBH,
  title = {Highly Accurate First-Principles Benchmark Data Sets for the Parametrization and Validation of Density Functional and Other Approximate Methods. {Derivation} of a Robust, Generally Applicable, Double-Hybrid Functional for Thermochemistry and Thermochemical Kinetics},
  author = {Karton, Amir and Tarnopolsky, Alex and Lam{\`e}re, Jean-Fran{\c{c}}ois and Schatz, George C. and Martin, Jan M. L.},
  journal = {J. Phys. Chem. A},
  volume = {112},
  pages = {12868--12886},
  year = {2008}
}

@article{Karton2012PX,
  title = {Determination of Barrier Heights for Proton Exchange in Small Water, Ammonia, and Hydrogen Fluoride Clusters with {G4(MP2)}-Type, {MPn}, and {SCS-MPn} Procedures---A Caveat},
  author = {Karton, Amir and O'Reilly, Robert J. and Chan, Bun and Radom, Leo},
  journal = {J. Chem. Theory Comput.},
  volume = {8},
  pages = {3128--3136},
  year = {2012}
}

@article{Karton2012WCPT,
  title = {Assessment of Theoretical Procedures for Calculating Barrier Heights for a Diverse Set of Water-Catalyzed Proton-Transfer Reactions},
  author = {Karton, Amir and O'Reilly, Robert J. and Radom, Leo},
  journal = {J. Phys. Chem. A},
  volume = {116},
  pages = {4211--4221},
  year = {2012}
}

@article{Mardirossian2017,
  title = {Thirty years of density functional theory in computational chemistry: an overview and extensive assessment of 200 density functionals},
  volume = {115},
  ISSN = {1362-3028},
  url = {http://dx.doi.org/10.1080/00268976.2017.1333644},
  DOI = {10.1080/00268976.2017.1333644},
  number = {19},
  journal = {Molecular Physics},
  publisher = {Informa UK Limited},
  author = {Mardirossian,  Narbe and Head-Gordon,  Martin},
  year = {2017},
  month = jun,
  pages = {2315–2372}
}

@article{soper_radial_2013,
	title = {The {Radial} {Distribution} {Functions} of {Water} as {Derived} from {Radiation} {Total} {Scattering} {Experiments}: {Is} {There} {Anything} {We} {Can} {Say} for {Sure}?},
	volume = {2013},
	copyright = {http://creativecommons.org/licenses/by/3.0/},
	issn = {2090-7761},
	shorttitle = {The {Radial} {Distribution} {Functions} of {Water} as {Derived} from {Radiation} {Total} {Scattering} {Experiments}},
	url = {https://www.hindawi.com/journals/isrn/2013/279463/},
	doi = {10.1155/2013/279463},
	urldate = {2025-12-03},
	journal = {ISRN Physical Chemistry},
	author = {Soper, A. K.},
	month = feb,
	year = {2013},
	pages = {1--67},
}

@article{Chan_2021,
    author = {Chan, Bun},
    doi = {10.1021/acs.jctc.1c00598},
    issn = {1549-9626},
    journal = {J. Chem. Theory Comput.},
    number = {9},
    pages = {5704-5714},
    publisher = {American Chemical Society (ACS)},
    title = {Accurate Thermochemistry for Main-Group Elements up to Xenon with the Wn-P34 Series of Composite Methods},
    url = {http://dx.doi.org/10.1021/acs.jctc.1c00598},
    volume = {17},
    year = {2021}
}

@article{Friedrich_2015,
    author = {Friedrich, Joachim},
    doi = {10.1021/acs.jctc.5b00087},
    issn = {1549-9626},
    journal = {J. Chem. Theory Comput.},
    number = {8},
    pages = {3596--3609},
    publisher = {American Chemical Society (ACS)},
    title = {Efficient Calculation of Accurate Reaction Energies-Assessment of Different Models in Electronic Structure Theory},
    url = {http://dx.doi.org/10.1021/acs.jctc.5b00087},
    volume = {11},
    year = {2015}
}

@article{Gruzman_2009,
    author = {Gruzman, David and Karton, Amir and Martin, Jan M. L.},
    doi = {10.1021/jp903640h},
    issn = {1520-5215},
    journal = {J. Phys. Chem. A},
    number = {43},
    pages = {11974-11983},
    publisher = {American Chemical Society (ACS)},
    title = {Performance of Ab Initio and Density Functional Methods for Conformational Equilibria of CnH2n+2 Alkane Isomers (n = 4-8)},
    url = {http://dx.doi.org/10.1021/jp903640h},
    volume = {113},
    year = {2009}
}

@article{Karton_2009,
    author = {Karton, Amir and Gruzman, David and Martin, Jan M. L.},
    doi = {10.1021/jp904369h},
    issn = {1520-5215},
    journal = {J. Phys. Chem. A},
    number = {29},
    pages = {8434-8447},
    publisher = {American Chemical Society (ACS)},
    title = {Benchmark Thermochemistry of the CnH2n+2 Alkane Isomers (n = 2-8) and Performance of DFT and Composite Ab Initio Methods for Dispersion-Driven Isomeric Equilibria},
    url = {http://dx.doi.org/10.1021/jp904369h},
    volume = {113},
    year = {2009}
}

@article{Karton_2016Heats,
    author = {Karton, Amir and Schreiner, Peter R. and Martin, Jan M. L.},
    doi = {10.1002/jcc.23963},
    issn = {1096-987X},
    journal = {J. Comput. Chem.},
    number = {1},
    pages = {49-58},
    publisher = {Wiley},
    title = {Heats of formation of platonic hydrocarbon cages by means of high‐level thermochemical procedures},
    url = {http://dx.doi.org/10.1002/jcc.23963},
    volume = {37},
    year = {2016}
}

@article{Yu_2015An,
    author = {Yu, Li-Juan and Sarrami, Farzaneh and Karton, Amir and O’Reilly, Robert J.},
    doi = {10.1080/00268976.2014.986238},
    issn = {1362-3028},
    journal = {Mol. Phys.},
    number = {11},
    pages = {1284-1296},
    publisher = {Informa UK Limited},
    title = {An assessment of theoretical procedures for $\pi$-conjugation stabilisation energies in enones},
    url = {http://dx.doi.org/10.1080/00268976.2014.986238},
    volume = {113},
    year = {2015}
}

@article{curtiss1991gaussian,
  title={Gaussian-2 theory for molecular energies of first-and second-row compounds},
  author={Curtiss, Larry A and Raghavachari, Krishnan and Trucks, Gary W and Pople, John A},
  journal = {J. Chem. Phys.},
  volume={94},
  number={11},
  pages={7221--7230},
  year={1991},
  publisher={American Institute of Physics}
}

@article{goerigk2010general,
  title={A general database for main group thermochemistry, kinetics, and noncovalent interactions- assessment of common and reparameterized (meta-) GGA density functionals},
  author={Goerigk, Lars and Grimme, Stefan},
  journal = {J. Chem. Theory Comput.},
  volume={6},
  number={1},
  pages={107--126},
  year={2010},
  publisher={ACS Publications}
}

@article{hait2018dipole,
 author = {Hait, Diptarka and Head-Gordon, Martin},
 title = {How Accurate Is Density Functional Theory at Predicting Dipole Moments? An Assessment Using a New Database of 200 Benchmark Values},
 journal = {J. Chem. Theory Comput.},
 volume = {14},
 number = {4},
 pages = {1969-1981},
 year = {2018},
 publisher = {American Chemical Society (ACS)},
 doi = {10.1021/acs.jctc.7b01252},
 source = {Crossref},
 url = {https://doi.org/10.1021/acs.jctc.7b01252},
 issn = {1549-9618, 1549-9626},
}

@article{korth2009mindless,
  title={“Mindless” DFT benchmarking},
  author={Korth, Martin and Grimme, Stefan},
  journal = {J. Chem. Theory Comput.},
  volume={5},
  number={4},
  pages={993--1003},
  year={2009},
  publisher={ACS Publications}
}

@article{yu2016can,
  title={Can DFT and ab initio methods describe all aspects of the potential energy surface of cycloreversion reactions?},
  author={Yu, Li-Juan and Sarrami, Farzaneh and O'Reilly, Robert J and Karton, Amir},
  journal = {Mol. Phys.},
  volume={114},
  number={1},
  pages={21--33},
  year={2016},
  publisher={Taylor \& Francis}
}

@misc{Weber2025MPNICE,
      title={Efficient Long-Range Machine Learning Force Fields for Liquid and Materials Properties}, 
      author={John L. Weber and Rishabh D. Guha and Garvit Agarwal and Yujing Wei and Aidan A. Fike and Xiaowei Xie and James Stevenson and Karl Leswing and Mathew D. Halls and Robert Abel and Leif D. Jacobson},
      year={2025},
      eprint={2505.06462},
      archivePrefix={arXiv},
      primaryClass={physics.chem-ph},
      url={https://arxiv.org/abs/2505.06462}, 
}

@article{packmol,
author = {Martínez, L. and Andrade, R. and Birgin, E. G. and Martínez, J. M.},
title = {PACKMOL: A package for building initial configurations for molecular dynamics simulations},
journal = {Journal of Computational Chemistry},
volume = {30},
number = {13},
pages = {2157-2164},
doi = {https://doi.org/10.1002/jcc.21224},
url = {https://onlinelibrary.wiley.com/doi/abs/10.1002/jcc.21224},
eprint = {https://onlinelibrary.wiley.com/doi/pdf/10.1002/jcc.21224},
year = {2009}
}

@article{mtkbarostat,
    author = {Martyna, Glenn J. and Tobias, Douglas J. and Klein, Michael L.},
    title = {Constant pressure molecular dynamics algorithms},
    journal = {The Journal of Chemical Physics},
    volume = {101},
    number = {5},
    pages = {4177-4189},
    year = {1994},
    month = {09},
    issn = {0021-9606},
    doi = {10.1063/1.467468},
    url = {https://doi.org/10.1063/1.467468}
}

@article{ASE,
doi = {10.1088/1361-648X/aa680e},
url = {https://doi.org/10.1088/1361-648X/aa680e},
year = {2017},
month = {jun},
publisher = {IOP Publishing},
volume = {29},
number = {27},
pages = {273002},
author = {Hjorth Larsen, Ask and Jørgen Mortensen, Jens and Blomqvist, Jakob and Castelli, Ivano E and Christensen, Rune and Dułak, Marcin and Friis, Jesper and Groves, Michael N and Hammer, Bjørk and Hargus, Cory and Hermes, Eric D and Jennings, Paul C and Bjerre Jensen, Peter and Kermode, James and Kitchin, John R and Leonhard Kolsbjerg, Esben and Kubal, Joseph and Kaasbjerg, Kristen and Lysgaard, Steen and Bergmann Maronsson, Jón and Maxson, Tristan and Olsen, Thomas and Pastewka, Lars and Peterson, Andrew and Rostgaard, Carsten and Schiøtz, Jakob and Schütt, Ole and Strange, Mikkel and Thygesen, Kristian S and Vegge, Tejs and Vilhelmsen, Lasse and Walter, Michael and Zeng, Zhenhua and Jacobsen, Karsten W},
title = {The atomic simulation environment—a Python library for working with atoms},
journal = {Journal of Physics: Condensed Matter}
}

@misc{batatia2025crosslearningelectronicstructure,
      title={Cross Learning between Electronic Structure Theories for Unifying Molecular, Surface, and Inorganic Crystal Foundation Force Fields}, 
      author={Ilyes Batatia and Chen Lin and Joseph Hart and Elliott Kasoar and Alin M. Elena and Sam Walton Norwood and Thomas Wolf and Gábor Csányi},
      year={2025},
      eprint={2510.25380},
      archivePrefix={arXiv},
      primaryClass={physics.chem-ph},
      url={https://arxiv.org/abs/2510.25380}, 
}

@article{bhattacharjee2022dlpno,
    author = {Bhattacharjee, Sinjini and Isegawa, Miho and Garcia-Rat{\'e}s, Miquel and Neese, Frank and Pantazis, Dimitrios A.},
    title = {Ionization Energies and Redox Potentials of Hydrated Transition Metal Ions: Evaluation of Domain-Based Local Pair Natural Orbital Coupled Cluster Approaches},
    journal = {Journal of Chemical Theory and Computation},
    volume = {18},
    number = {3},
    pages = {1619--1632},
    year = {2022},
    doi = {10.1021/acs.jctc.1c01267},
    url = {https://doi.org/10.1021/acs.jctc.1c01267}
}

@article{cposs209,
author = {Price, Louise S. and Paloni, Matteo and Salvalaglio, Matteo and Price, Sarah L.},
title = {One Size Fits All? Development of the CPOSS209 Data Set of Experimental and Hypothetical Polymorphs for Testing Computational Modeling Methods},
journal = {Crystal Growth \& Design},
volume = {25},
number = {9},
pages = {3186-3209},
year = {2025},
doi = {10.1021/acs.cgd.5c00255},
URL = {https://doi.org/10.1021/acs.cgd.5c00255},
eprint = {https://doi.org/10.1021/acs.cgd.5c00255}
}

@article{Liang2025,
  title = {Gold-Standard Chemical Database 137 (GSCDB137): A Diverse Set of Accurate Energy Differences for Assessing and Developing Density Functionals},
  volume = {21},
  ISSN = {1549-9626},
  url = {http://dx.doi.org/10.1021/acs.jctc.5c01380},
  DOI = {10.1021/acs.jctc.5c01380},
  number = {24},
  journal = {Journal of Chemical Theory and Computation},
  publisher = {American Chemical Society (ACS)},
  author = {Liang,  Jiashu and Head-Gordon,  Martin},
  year = {2025},
  month = dec,
  pages = {12601–12621}
}

@article{zhao2005multi,
  title={Multi-coefficient extrapolated density functional theory for thermochemistry and thermochemical kinetics},
  author={Zhao, Yan and Lynch, Benjamin J and Truhlar, Donald G},
  journal = {Phys. Chem. Chem. Phys.},
  volume={7},
  number={1},
  pages={43--52},
  year={2005},
  publisher={Royal Society of Chemistry}
}

@article{zhao2005benchmark,
  title={Benchmark database of barrier heights for heavy atom transfer, nucleophilic substitution, association, and unimolecular reactions and its use to test theoretical methods},
  author={Zhao, Yan and Gonz{\'a}lez-Garc{\'\i}a, N{\'u}ria and Truhlar, Donald G},
  journal = {J. Phys. Chem. A},
  volume={109},
  number={9},
  pages={2012--2018},
  year={2005},
  publisher={ACS Publications}
}

@article{Zheng_2007,
    author = {Zheng, Jingjing and Zhao, Yan and Truhlar, Donald G.},
    doi = {10.1021/ct600281g},
    issn = {1549-9626},
    journal = {J. Chem. Theory Comput.},
    number = {2},
    pages = {569-582},
    publisher = {American Chemical Society (ACS)},
    title = {Representative Benchmark Suites for Barrier Heights of Diverse Reaction Types and Assessment of Electronic Structure Methods for Thermochemical Kinetics},
    url = {http://dx.doi.org/10.1021/ct600281g},
    volume = {3},
    year = {2007}
}

@article{hickey2014benchmarking,
  title={Benchmarking quantum chemical methods for the calculation of molecular dipole moments and polarizabilities},
  author={Hickey, A Leif and Rowley, Christopher N},
  journal = {J. Phys. Chem. A},
  volume={118},
  number={20},
  pages={3678--3687},
  year={2014},
  publisher={ACS Publications}
}

@article{grimme2007compute,
  title={How to compute isomerization energies of organic molecules with quantum chemical methods},
  author={Grimme, Stefan and Steinmetz, Marc and Korth, Martin},
  journal = {J. Org. Chem.},
  volume={72},
  number={6},
  pages={2118--2126},
  year={2007},
  publisher={ACS Publications}
}

@article{huenerbein2010effects,
  title={Effects of London dispersion on the isomerization reactions of large organic molecules: a density functional benchmark study},
  author={Huenerbein, Robert and Schirmer, Birgitta and Moellmann, Jonas and Grimme, Stefan},
  journal = {Phys. Chem. Chem. Phys.},
  volume={12},
  number={26},
  pages={6940--6948},
  year={2010},
  publisher={Royal Society of Chemistry}
}

@article{kozuch2014conformational,
  title={Conformational equilibria in butane-1, 4-diol: a benchmark of a prototypical system with strong intramolecular H-bonds},
  author={Kozuch, Sebastian and Bachrach, Steven M and Martin, Jan ML},
  journal = {J. Phys. Chem. A},
  volume={118},
  number={1},
  pages={293-303},
  year={2014},
  publisher={ACS Publications}
}

@article{fogueri2013melatonin,
  title={The melatonin conformer space: Benchmark and assessment of wave function and DFT methods for a paradigmatic biological and pharmacological molecule},
  author={Fogueri, Uma R and Kozuch, Sebastian and Karton, Amir and Martin, Jan ML},
  journal = {J. Phys. Chem. A},
  volume={117},
  number={10},
  pages={2269--2277},
  year={2013},
  publisher={ACS Publications}
}

@article{vreha2005structure,
  title={Structure and IR Spectrum of Phenylalanyl--Glycyl--Glycine Tripetide in the Gas-Phase: IR/UV Experiments, Ab Initio Quantum Chemical Calculations, and Molecular Dynamic Simulations},
  author={{\v{R}}eha, D and Valdes, H and Vondr{\'a}{\v{s}}ek, J and Hobza, P and Abu-Riziq, Ali and Crews, Bridgit and De Vries, Mattanjah S},
  journal = {Chem. Eur. J.},
  volume={11},
  number={23},
  pages={6803--6817},
  year={2005},
  publisher={Wiley Online Library}
}

@article{goerigk2013accurate,
  title={Accurate quantum chemical energies for tetrapeptide conformations: why MP2 data with an insufficient basis set should be handled with caution},
  author={Goerigk, Lars and Karton, Amir and Martin, Jan ML and Radom, Leo},
  journal = {Phys. Chem. Chem. Phys.},
  volume={15},
  number={19},
  pages={7028--7031},
  year={2013},
  publisher={Royal Society of Chemistry}
}

@article{csonka2009evaluation,
  title={Evaluation of density functionals and basis sets for carbohydrates},
  author={Csonka, G{\'a}bor I and French, Alfred D and Johnson, Glenn P and Stortz, Carlos A},
  journal = {J. Chem. Theory Comput.},
  volume={5},
  number={4},
  pages={679--692},
  year={2009},
  publisher={ACS Publications}
}

@article{Rezac2013Describing,
    author = {Řezáč, Jan and Hobza, Pavel},
    doi = {10.1021/ct400057w},
    issn = {1549-9626},
    journal = {J. Chem. Theory Comput.},
    number = {5},
    pages = {2151-2155},
    publisher = {American Chemical Society (ACS)},
    title = {Describing Noncovalent Interactions beyond the Common Approximations: How Accurate Is the “Gold Standard,” CCSD(T) at the Complete Basis Set Limit?},
    url = {http://dx.doi.org/10.1021/ct400057w},
    volume = {9},
    year = {2013}
}

@article{grimme2010consistent,
    author = {Grimme, Stefan and Antony, Jens and Ehrlich, Stephan and Krieg, Helge},
    title = "{A consistent and accurate ab initio parametrization of density functional dispersion correction (DFT-D) for the 94 elements H-Pu}",
    journal = {J. Chem. Phys.},
    volume = {132},
    number = {15},
    pages = {154104},
    year = {2010},
    issn = {0021-9606},
    doi = {10.1063/1.3382344},
    url = {https://doi.org/10.1063/1.3382344},
}

@article{Bauza_2013,
    author = {Bauzá, Antonio and Alkorta, Ibon and Frontera, Antonio and Elguero, José},
    doi = {10.1021/ct400818v},
    issn = {1549-9626},
    journal = {J. Chem. Theory Comput.},
    number = {11},
    pages = {5201-5210},
    publisher = {American Chemical Society (ACS)},
    title = {On the Reliability of Pure and Hybrid DFT Methods for the Evaluation of Halogen, Chalcogen, and Pnicogen Bonds Involving Anionic and Neutral Electron Donors},
    url = {http://dx.doi.org/10.1021/ct400818v},
    volume = {9},
    year = {2013}
}

@article{lao2013improved,
  title={An improved treatment of empirical dispersion and a many-body energy decomposition scheme for the explicit polarization plus symmetry-adapted perturbation theory (XSAPT) method},
  author={Lao, Ka Un and Herbert, John M},
  journal = {J. Chem. Phys.},
  volume={139},
  number={3},
    pages = {034107},
  year={2013},
  publisher={AIP Publishing}
}

@article{kazimirski2003search,
  title={Search for low energy structures of water clusters (H2O) n, n= 20- 22, 48, 123, and 293},
  author={Kazimirski, Jan K and Buch, Victoria},
  journal = {J. Phys. Chem. A},
  volume={107},
  number={46},
  pages={9762--9775},
  year={2003},
  publisher={ACS Publications}
}

@article{faver2011formal,
  title={Formal estimation of errors in computed absolute interaction energies of protein- ligand complexes},
  author={Faver, John C and Benson, Mark L and He, Xiao and Roberts, Benjamin P and Wang, Bing and Marshall, Michael S and Kennedy, Matthew R and Sherrill, C David and Merz Jr, Kenneth M},
  journal = {J. Chem. Theory Comput.},
  volume={7},
  number={3},
  pages={790--797},
  year={2011},
  publisher={ACS Publications}
}

@article{takatani2007performance,
  title={Performance of spin-component-scaled M{\o}ller--Plesset theory (SCS-MP2) for potential energy curves of noncovalent interactions},
  author={Takatani, Tait and Sherrill, C David},
  journal = {Phys. Chem. Chem. Phys.},
  volume={9},
  number={46},
  pages={6106--6114},
  year={2007},
  publisher={Royal Society of Chemistry}
}

@article{hohenstein2009effects,
  title={Effects of heteroatoms on aromatic $\pi$- $\pi$ interactions: benzene- pyridine and pyridine dimer},
  author={Hohenstein, Edward G and Sherrill, C David},
  journal = {J. Phys. Chem. A},
  volume={113},
  number={5},
  pages={878--886},
  year={2009},
  publisher={ACS Publications}
}

@article{setiawan2015strength,
  title={Strength of the pnicogen bond in complexes involving group Va elements N, P, and As},
  author={Setiawan, Dani and Kraka, Elfi and Cremer, Dieter},
  journal = {J. Phys. Chem. A},
  volume={119},
  number={9},
  pages={1642--1656},
  year={2015},
  publisher={ACS Publications}
}

@article{jurevcka2006benchmark,
  title={Benchmark database of accurate (MP2 and CCSD (T) complete basis set limit) interaction energies of small model complexes, DNA base pairs, and amino acid pairs},
  author={Jure{\v{c}}ka, Petr and {\v{S}}poner, Ji{\v{r}}{\'\i} and {\v{C}}ern{\`y}, Ji{\v{r}}{\'\i} and Hobza, Pavel},
  journal = {Phys. Chem. Chem. Phys.},
  volume={8},
  number={17},
  pages={1985--1993},
  year={2006},
  publisher={Royal Society of Chemistry}
}

@article{Rezac2011_1,
author = {Řezáč, Jan and Riley, Kevin E. and Hobza, Pavel},
title = {S66: A Well-balanced Database of Benchmark Interaction Energies Relevant to Biomolecular Structures},
journal = {J. Chem. Theory Comput.},
volume = {7},
number = {8},
pages = {2427-2438},
year = {2011},
doi = {10.1021/ct2002946}
}

@article{Temelso_2011,
    author = {Temelso, Berhane and Archer, Kaye A. and Shields, George C.},
    doi = {10.1021/jp2069489},
    issn = {1520-5215},
    journal = {J. Phys. Chem. A},
    number = {43},
    pages = {12034-12046},
    publisher = {American Chemical Society (ACS)},
    title = {Benchmark Structures and Binding Energies of Small Water Clusters with Anharmonicity Corrections},
    url = {http://dx.doi.org/10.1021/jp2069489},
    volume = {115},
    year = {2011}
}

@article{Bryantsev_2009,
    author = {Bryantsev, Vyacheslav S. and Diallo, Mamadou S. and van Duin, Adri C. T. and Goddard, William A.},
    doi = {10.1021/ct800549f},
    issn = {1549-9626},
    journal = {J. Chem. Theory Comput.},
    number = {4},
    pages = {1016-1026},
    publisher = {American Chemical Society (ACS)},
    title = {Evaluation of B3LYP, X3LYP, and M06-Class Density Functionals for Predicting the Binding Energies of Neutral, Protonated, and Deprotonated Water Clusters},
    url = {http://dx.doi.org/10.1021/ct800549f},
    volume = {5},
    year = {2009}
}

@article{Rezac2012,
    author = {Řezáč, Jan and Riley, Kevin E. and Hobza, Pavel},
    doi = {10.1021/ct300647k},
    issn = {1549-9626},
    journal = {J. Chem. Theory Comput.},
    number = {11},
    pages = {4285-4292},
    publisher = {American Chemical Society (ACS)},
    title = {Benchmark Calculations of Noncovalent Interactions of Halogenated Molecules},
    url = {http://dx.doi.org/10.1021/ct300647k},
    volume = {8},
    year = {2012}
}

@article{johnson2008delocalization,
  title={Delocalization errors in density functionals and implications for main-group thermochemistry},
  author={Johnson, Erin R and Mori-S{\'a}nchez, Paula and Cohen, Aron J and Yang, Weitao},
  journal = {J. Chem. Phys.},
  volume={129},
  number={20},
 pages = {204112},
  year={2008},
  publisher={AIP Publishing}
}

@article{steinmann2009unified,
  title={Unified inter-and intramolecular dispersion correction formula for generalized gradient approximation density functional theory},
  author={Steinmann, Stephan N and Csonka, Gabor and Corminboeuf, Clemence},
  journal = {J. Chem. Theory Comput.},
  volume={5},
  number={11},
  pages={2950--2958},
  year={2009},
  publisher={ACS Publications}
}

@article{friedrich2013incremental,
  title={Incremental CCSD (T)(F12*)| MP2: A black box method to obtain highly accurate reaction energies},
  author={Friedrich, Joachim and Ha\"{a}nchen, Julia},
  journal = {J. Chem. Theory Comput.},
  volume={9},
  number={12},
  pages={5381--5394},
  year={2013},
  publisher={ACS Publications}
}

@article{goerigk2011efficient,
  title={Efficient and Accurate Double-Hybrid-Meta-GGA Density Functionals--Evaluation with the Extended GMTKN30 Database for General Main Group Thermochemistry, Kinetics, and Noncovalent Interactions},
  author={Goerigk, Lars and Grimme, Stefan},
  journal = {J. Chem. Theory Comput.},
  volume={7},
  number={2},
  pages={291--309},
  year={2011},
  publisher={ACS Publications}
}

@article{grimme2013towards,
  title={Towards first principles calculation of electron impact mass spectra of molecules.},
  author={Grimme, Stefan},
  journal = {Angew. Chem. Int. Ed.},
  volume={52},
  number={24},
  year={2013}
}

@article{neese2009assessment,
  title={Assessment of orbital-optimized, spin-component scaled second-order many-body perturbation theory for thermochemistry and kinetics},
  author={Neese, Frank and Schwabe, Tobias and Kossmann, Simone and Schirmer, Birgitta and Grimme, Stefan},
  journal = {J. Chem. Theory Comput.},
  volume={5},
  number={11},
  pages={3060--3073},
  year={2009},
  publisher={ACS Publications}
}

@article{iron2019evaluating,
  title={Evaluating transition metal barrier heights with the latest density functional theory exchange--correlation functionals: The MOBH35 benchmark database},
  author={Iron, Mark A and Janes, Trevor},
  journal = {J. Phys. Chem. A},
  volume={123},
  number={17},
  pages={3761--3781},
  year={2019},
  publisher={ACS Publications}
}

@article{chan2019assessment,
 author = {Chan, Bun and Gill, Peter M. W. and Kimura, Masanari},
 title = {Assessment of DFT Methods for Transition Metals with the TMC151 Compilation of Data Sets and Comparison with Accuracies for Main-Group Chemistry},
 journal = {J. Chem. Theory Comput.},
 volume = {15},
 number = {6},
 pages = {3610-3622},
 year = {2019},
 publisher = {American Chemical Society (ACS)},
 doi = {10.1021/acs.jctc.9b00239},
 source = {Crossref},
 url = {https://doi.org/10.1021/acs.jctc.9b00239},
 issn = {1549-9618, 1549-9626},
}

@article{Blumberger2005,
  title = {Ab Initio Molecular Dynamics Simulation of the Aqueous Ru2+/Ru3+ Redox Reaction: The Marcus Perspective},
  volume = {109},
  ISSN = {1520-5207},
  url = {http://dx.doi.org/10.1021/jp0455879},
  DOI = {10.1021/jp0455879},
  number = {14},
  journal = {The Journal of Physical Chemistry B},
  publisher = {American Chemical Society (ACS)},
  author = {Blumberger,  Jochen and Sprik,  Michiel},
  year = {2005},
  month = jan,
  pages = {6793–6804}
}

@article{Mandal2022,
  title = {Hybrid Functional and Plane Waves based Ab Initio Molecular Dynamics Study of the Aqueous Fe2+/Fe3+ Redox Reaction**},
  volume = {24},
  ISSN = {1439-7641},
  url = {http://dx.doi.org/10.1002/cphc.202200617},
  DOI = {10.1002/cphc.202200617},
  number = {3},
  journal = {ChemPhysChem},
  publisher = {Wiley},
  author = {Mandal,  Sagarmoy and Kar,  Ritama and Meyer,  Bernd and Nair,  Nisanth N.},
  year = {2022},
  month = nov 
}

@article{Pl__2023,
   title={Force-field-enhanced neural network interactions: from local equivariant embedding to atom-in-molecule properties and long-range effects},
   volume={14},
   ISSN={2041-6539},
   url={http://dx.doi.org/10.1039/D3SC02581K},
   DOI={10.1039/d3sc02581k},
   number={44},
   journal={Chemical Science},
   publisher={Royal Society of Chemistry (RSC)},
   author={Plé, Thomas and Lagardère, Louis and Piquemal, Jean-Philip},
   year={2023},
   pages={12554–12569} }

@article{PL_2025,
author = {Plé, Thomas and Adjoua, Olivier and Benali, Anouar and Posenitskiy, Evgeny and Villot, Corentin and Lagardère, Louis and Piquemal, Jean-Philip},
year = {2025},
month = {05},
pages = {},
journal = {ChemRxiv},
title = {A Foundation Model for Accurate Atomistic Simulations in Drug Design},
doi = {10.26434/chemrxiv-2025-f1hgn-v3}
}

@article{Cheng2025,
  title = {Latent Ewald summation for machine learning of long-range interactions},
  volume = {11},
  ISSN = {2057-3960},
  url = {http://dx.doi.org/10.1038/s41524-025-01577-7},
  DOI = {10.1038/s41524-025-01577-7},
  number = {1},
  journal = {npj Computational Materials},
  publisher = {Springer Science and Business Media LLC},
  author = {Cheng,  Bingqing},
  year = {2025},
  month = mar 
}

@article{Mazitov2025,
  title = {PET-MAD as a lightweight universal interatomic potential for advanced materials modeling},
  volume = {16},
  ISSN = {2041-1723},
  url = {http://dx.doi.org/10.1038/s41467-025-65662-7},
  DOI = {10.1038/s41467-025-65662-7},
  number = {1},
  journal = {Nature Communications},
  publisher = {Springer Science and Business Media LLC},
  author = {Mazitov,  Arslan and Bigi,  Filippo and Kellner,  Matthias and Pegolo,  Paolo and Tisi,  Davide and Fraux,  Guillaume and Pozdnyakov,  Sergey and Loche,  Philip and Ceriotti,  Michele},
  year = {2025},
  month = nov 
}

@article{Rose2024,
  title = {Fast and Robust Modeling of Lanthanide and Actinide Complexes,  Biomolecules,  and Molecular Crystals with the Extended GFN-FF Model},
  volume = {63},
  ISSN = {1520-510X},
  url = {http://dx.doi.org/10.1021/acs.inorgchem.4c03215},
  DOI = {10.1021/acs.inorgchem.4c03215},
  number = {41},
  journal = {Inorganic Chemistry},
  publisher = {American Chemical Society (ACS)},
  author = {Rose,  Thomas and Bursch,  Markus and Mewes,  Jan-Michael and Grimme,  Stefan},
  year = {2024},
  month = sep,
  pages = {19364–19374}
}

@unpublished{Baldwin2026SCF,
    title = {Design Space of Self-Consistent Electrostatic Machine Learning Force Fields},
    author = {Baldwin, William J. and Batatia, Ilyes and Vondr\'ak, Martin and Margraf, Johannes T. and Cs\'anyi, G\'abor},
    year = {2026},
    note = {Manuscript in preparation}
  }

@article{Olexandr_lr_review,
  author = {Anstine, Dylan M. and Isayev, Olexandr},
  title = {Machine Learning Interatomic Potentials and Long-Range Physics},
  journal = {The Journal of Physical Chemistry A},
  volume = {127},
  number = {11},
  pages = {2417-2431},
  year = {2023},
  doi = {10.1021/acs.jpca.2c06778},
  note = {PMID: 36802360},
  url = {https://doi.org/10.1021/acs.jpca.2c06778}
}

@article{Morawietz2012ACharges,
  title = {{A neural network potential-energy surface for the water dimer based on environment-dependent atomic energies and charges}},
  year = {2012},
  journal = {The Journal of Chemical Physics},
  author = {Morawietz, Tobias and Sharma, Vikas and Behler, Jörg},
  number = {6},
  month = {2},
  volume = {136},
  doi = {10.1063/1.3682557},
  issn = {0021-9606}
}

@article{Artrith2011,
  title = {High-dimensional neural-network potentials for multicomponent systems: Applications to zinc oxide},
  author = {Artrith, Nongnuch and Morawietz, Tobias and Behler, J\"org},
  journal = {Phys. Rev. B},
  volume = {83},
  issue = {15},
  pages = {153101},
  numpages = {4},
  year = {2011},
  month = {Apr},
  publisher = {American Physical Society},
  doi = {10.1103/PhysRevB.83.153101},
  url = {https://link.aps.org/doi/10.1103/PhysRevB.83.153101}
}

@article{physnet2019,
  author = {Unke, Oliver T and Meuwly, Markus},
  doi = {10.1021/acs.jctc.9b00181},
  issue = {6},
  journal = {Journal of Chemical Theory and Computation},
  note = {PMID: 31042390},
  pages = {3678-3693},
  title = {PhysNet: A Neural Network for Predicting Energies, Forces, Dipole Moments, and Partial Charges},
  volume = {15},
  url = {https://doi.org/10.1021/acs.jctc.9b00181},
  year = {2019}
}

@article{Zhang2022AInteractions,
  title = {{A deep potential model with long-range electrostatic interactions}},
  year = {2022},
  journal = {The Journal of Chemical Physics},
  author = {Zhang, Linfeng and Wang, Han and Muniz, Maria Carolina and Panagiotopoulos, Athanassios Z. and Car, Roberto and E, Weinan},
  number = {12},
  month = {3},
  pages = {124107},
  volume = {156},
  doi = {10.1063/5.0083669},
  issn = {0021-9606}
}

@misc{kocer2024machinelearningpotentialsredox,
  title = {Machine learning potentials for redox chemistry in solution}, 
  author = {Kocer, Emir and El Haouari, Redouan and Dellago, Christoph and Behler, Jörg},
  year = {2024},
  eprint = {2410.03299},
  archivePrefix = {arXiv},
  primaryClass = {physics.chem-ph},
  url = {https://arxiv.org/abs/2410.03299}
}

@article{qeq1985,
  author = {Mortier, Wilfried J and Van Genechten, Karin and Gasteiger, Johann},
  doi = {10.1021/ja00290a017},
  issue = {4},
  journal = {Journal of the American Chemical Society},
  pages = {829-835},
  title = {Electronegativity equalization: application and parametrization},
  volume = {107},
  url = {https://doi.org/10.1021/ja00290a017},
  year = {1985}
}

@article{qeq1986,
  author = {Mortier, Wilfried J and Ghosh, Swapan K and Shankar, S},
  doi = {10.1021/ja00275a013},
  journal = {Journal of the American Chemical Society},
  number = {15},
  pages = {4315--4320},
  title = {{Electronegativity-equalization method for the calculation of atomic charges in molecules}},
  volume = {108},
  year = {1986}
}

@article{Rappe1991ChargeSimulations,
  title = {{Charge equilibration for molecular dynamics simulations}},
  year = {1991},
  journal = {The Journal of Physical Chemistry},
  author = {Rappe, Anthony K. and Goddard, William A.},
  number = {8},
  month = {4},
  pages = {3358--3363},
  volume = {95},
  doi = {10.1021/j100161a070},
  issn = {0022-3654}
}

@article{4gnn_ko_2020,
  author = {Ko, Tsz Wai and Finkler, Jonas A and Goedecker, Stefan and Behler, Jörg},
  doi = {10.1038/s41467-020-20427-2},
  issn = {2041-1723},
  issue = {1},
  journal = {Nature Communications},
  pages = {398},
  title = {A fourth-generation high-dimensional neural network potential with accurate electrostatics including non-local charge transfer},
  volume = {12},
  url = {https://doi.org/10.1038/s41467-020-20427-2},
  year = {2021}
}

@article{kqeq_og2022,
  author = {Staacke, Carsten G and Wengert, Simon and Kunkel, Christian and Csányi, Gábor and Reuter, Karsten and Margraf, Johannes T},
  doi = {10.1088/2632-2153/ac568d},
  issue = {1},
  journal = {Machine Learning: Science and Technology},
  month = {3},
  pages = {15032},
  publisher = {IOP Publishing},
  title = {Kernel charge equilibration: efficient and accurate prediction of molecular dipole moments with a machine-learning enhanced electron density model},
  volume = {3},
  url = {https://dx.doi.org/10.1088/2632-2153/ac568d},
  year = {2022}
}

@article{Jensen2023UnifyingModels,
  title = {{Unifying Charge-Flow Polarization Models}},
  year = {2023},
  journal = {Journal of Chemical Theory and Computation},
  author = {Jensen, Frank},
  number = {13},
  month = {6},
  pages = {4047--4073},
  volume = {19},
  doi = {10.1021/acs.jctc.3c00341},
  issn = {1549-9618}
}

@article{Perdew1982Density-FunctionalEnergy,
  title = {{Density-Functional Theory for Fractional Particle Number: Derivative Discontinuities of the Energy}},
  year = {1982},
  journal = {Physical Review Letters},
  author = {Perdew, John P. and Parr, Robert G. and Levy, Mel and Balduz, Jose L.},
  number = {23},
  month = {12},
  pages = {1691--1694},
  volume = {49},
  doi = {10.1103/PhysRevLett.49.1691},
  issn = {0031-9007}
}

@misc{Vondrak2025PushingLimits,
  title = {{Pushing Charge Equilibration Based Machine Learning Potentials to their Limits}},
  year = {2025},
  author = {Vondr{\'{a}}k, Martin and Reuter, Karsten and Margraf, Johannes T.},
  month = {6},
  doi = {10.26434/chemrxiv-2025-6wv52}
}

@article{LeeWarren2008OriginMethods,
  title = {{Origin and control of superlinear polarizability scaling in chemical potential equalization methods}},
  year = {2008},
  journal = {The Journal of Chemical Physics},
  author = {Lee Warren, G. and Davis, Joseph E. and Patel, Sandeep},
  number = {14},
  month = {4},
  volume = {128},
  doi = {10.1063/1.2872603},
  issn = {0021-9606}
}

@article{nonlinear_pol_fq,
  author = {Chelli, Riccardo and Procacci, Piero and Righini, Roberto and Califano, Salvatore},
  title = {Electrical response in chemical potential equalization schemes},
  journal = {The Journal of Chemical Physics},
  volume = {111},
  number = {18},
  pages = {8569-8575},
  year = {1999},
  month = {11},
  issn = {0021-9606},
  doi = {10.1063/1.480198},
  url = {https://doi.org/10.1063/1.480198}
}

@article{conducting_molecules,
  author = {Verstraelen, Toon and Pauwels, Ewald and De Proft, Frank and Van Speybroeck, Veronique and Geerlings, Paul and Waroquier, Michel},
  title = {Assessment of Atomic Charge Models for Gas-Phase Computations on Polypeptides},
  journal = {Journal of Chemical Theory and Computation},
  volume = {8},
  number = {2},
  pages = {661-676},
  year = {2012},
  doi = {10.1021/ct200512e},
  note = {PMID: 26596614},
  url = {https://doi.org/10.1021/ct200512e}
}

@article{scfnn,
  author = {Gao, Ang and Remsing, Richard C},
  doi = {10.1038/s41467-022-29243-2},
  issn = {2041-1723},
  issue = {1},
  journal = {Nature Communications},
  pages = {1572},
  title = {Self-consistent determination of long-range electrostatics in neural network potentials},
  volume = {13},
  url = {https://doi.org/10.1038/s41467-022-29243-2},
  year = {2022}
}

@article{eMLP,
  author = {Cools-Ceuppens, Maarten and Dambre, Joni and Verstraelen, Toon},
  doi = {10.1021/acs.jctc.1c00978},
  issue = {3},
  journal = {Journal of Chemical Theory and Computation},
  note = {PMID: 35171606},
  pages = {1672-1691},
  title = {Modeling Electronic Response Properties with an Explicit-Electron Machine Learning Potential},
  volume = {18},
  url = {https://doi.org/10.1021/acs.jctc.1c00978},
  year = {2022}
}

@article{bpopnn,
  author = {Xie, Xiaowei and Persson, Kristin A. and Small, David W.},
  title = {Incorporating Electronic Information into Machine Learning Potential Energy Surfaces via Approaching the Ground-State Electronic Energy as a Function of Atom-Based Electronic Populations},
  journal = {Journal of Chemical Theory and Computation},
  volume = {16},
  number = {7},
  pages = {4256-4270},
  year = {2020},
  doi = {10.1021/acs.jctc.0c00217},
  note = {PMID: 32502350},
  url = {https://doi.org/10.1021/acs.jctc.0c00217}
}

@article{Grisafi2019,
  title = {Incorporating long-range physics in atomic-scale machine
          learning},
  volume = {151},
  ISSN = {1089-7690},
  url = {http://dx.doi.org/10.1063/1.5128375},
  DOI = {10.1063/1.5128375},
  number = {20},
  journal = {The Journal of Chemical Physics},
  publisher = {AIP Publishing},
  author = {Grisafi,  Andrea and Ceriotti,  Michele},
  year = {2019},
  month = nov 
}

@article{parr1983,
  title = {Absolute hardness: companion parameter to absolute electronegativity},
  author = {Parr, Robert G. and Pearson, Ralph G.},
  journal = {J. Am. Chem. Soc.},
  volume = {105},
  number = {26},
  pages = {7512--7516},
  year = {1983},
  doi = {10.1021/ja00364a005}
}

@article{geerlings2003conceptual,
  title = {Conceptual density functional theory},
  author = {Geerlings, Paul and De Proft, Frank and Langenaeker, Wilfried},
  journal = {Chem. Rev.},
  volume = {103},
  number = {5},
  pages = {1793--1874},
  year = {2003},
  doi = {10.1021/cr990029p}
}

@article{mortier1986electronegativity,
  title = {Electronegativity-equalization method for the calculation of atomic charges in molecules},
  author = {Mortier, Wilfried J. and Ghosh, Swapan K. and Shankar, S.},
  journal = {J. Am. Chem. Soc.},
  volume = {108},
  number = {15},
  pages = {4315--4320},
  year = {1986},
  doi = {10.1021/ja00275a013}
}

@article{rappe1991charge,
  title = {Charge equilibration for molecular dynamics simulations},
  author = {Rapp{\'e}, Anthony K. and Goddard III, William A.},
  journal = {J. Phys. Chem.},
  volume = {95},
  number = {8},
  pages = {3358--3363},
  year = {1991},
  doi = {10.1021/j100161a070}
}

@article{Kabylda2025,
  title = {Molecular Simulations with a Pretrained Neural Network and Universal Pairwise Force Fields},
  volume = {147},
  ISSN = {1520-5126},
  url = {http://dx.doi.org/10.1021/jacs.5c09558},
  DOI = {10.1021/jacs.5c09558},
  number = {37},
  journal = {Journal of the American Chemical Society},
  publisher = {American Chemical Society (ACS)},
  author = {Kabylda,  Adil and Frank,  J. Thorben and Suárez-Dou,  Sergio and Khabibrakhmanov,  Almaz and Medrano Sandonas,  Leonardo and Unke,  Oliver T. and Chmiela,  Stefan and M\"{u}ller,  Klaus-Robert and Tkatchenko,  Alexandre},
  year = {2025},
  month = aug,
  pages = {33723–33734}
}

@article{Mills2011,
  title = {Intramolecular polarisable multipolar electrostatics from the machine learning method Kriging},
  volume = {975},
  ISSN = {2210-271X},
  url = {http://dx.doi.org/10.1016/j.comptc.2011.04.004},
  DOI = {10.1016/j.comptc.2011.04.004},
  number = {1–3},
  journal = {Computational and Theoretical Chemistry},
  publisher = {Elsevier BV},
  author = {Mills,  Matthew J.L. and Popelier,  Paul L.A.},
  year = {2011},
  month = nov,
  pages = {42–51}
}

@article{smith2017ani1,
  title = {ANI-1: an extensible neural network potential with DFT accuracy at force field computational cost},
  author = {Smith, Justin S. and Isayev, Olexandr and Roitberg, Adrian E.},
  journal = {Chemical Science},
  year = {2017},
  volume = {8},
  pages = {3192--3203},
  doi = {10.1039/C6SC05720A},
  url = {https://doi.org/10.1039/C6SC05720A}
}

@article{devereux2020ani2x,
  title = {Extending the Applicability of the ANI Deep Learning Molecular Potential to Sulfur and Halogens},
  author = {Devereux, Christian and Smith, Justin S. and Huddleston, Kate K. and Barros, Kipton and Zubatyuk, Roman and Isayev, Olexandr and Roitberg, Adrian E.},
  journal = {Journal of Chemical Theory and Computation},
  year = {2020},
  volume = {16},
  number = {7},
  pages = {4192--4202},
  doi = {10.1021/acs.jctc.0c00121},
  url = {https://doi.org/10.1021/acs.jctc.0c00121}
}

@article{smith2020ani1x,
  title = {The ANI-1ccx and ANI-1x data sets, coupled-cluster and density functional theory properties for molecules},
  author = {Smith, Justin S. and Zubatyuk, Roman and Nebgen, Benjamin and Lubbers, Nicholas and Barros, Kipton and Roitberg, Adrian E. and Isayev, Olexandr and Tretiak, Sergei},
  journal = {Scientific Data},
  year = {2020},
  volume = {7},
  number = {1},
  pages = {134},
  doi = {10.1038/s41597-020-0473-z},
  url = {https://doi.org/10.1038/s41597-020-0473-z}
}

@article{anstine2025aimnet2,
  title = {AIMNet2: a neural network potential to meet your neutral, charged, organic, and elemental-organic needs},
  author = {Anstine, Dylan M. and Zubatyuk, Roman and Isayev, Olexandr},
  journal = {Chemical Science},
  year = {2025},
  volume = {16},
  pages = {10228--10244},
  doi = {10.1039/D4SC08572H},
  url = {https://doi.org/10.1039/D4SC08572H}
}

@article{kovacs2025maceoff,
  title = {MACE-OFF: Short-Range Transferable Machine Learning Force Fields for Organic Molecules},
  author = {Kov{\'a}cs, D{\'a}vid P{\'e}ter and Moore, J. Harry and Browning, Nicholas J. and Batatia, Ilyes and Horton, Joshua T. and Pu, Yixuan and Kapil, Venkat and Witt, William C. and Magd{\'a}u, Ioan-Bogdan and Cole, Daniel J. and Cs{\'a}nyi, G{\'a}bor},
  journal = {Journal of the American Chemical Society},
  year = {2025},
  volume = {147},
  number = {21},
  pages = {17598--17611},
  doi = {10.1021/jacs.4c07099},
  url = {https://doi.org/10.1021/jacs.4c07099}
}

@article{kabylda2025so3lr,
  title = {Molecular Simulations with a Pretrained Neural Network and Universal Pairwise Force Fields},
  author = {Kabylda, Adil and Frank, J. Thorben and Su{\'a}rez-Dou, Sergio and Khabibrakhmanov, Almaz and Medrano Sandonas, Leonardo and Unke, Oliver T. and Chmiela, Stefan and M{\"u}ller, Klaus-Robert and Tkatchenko, Alexandre},
  journal = {Journal of the American Chemical Society},
  year = {2025},
  volume = {147},
  number = {37},
  pages = {33723--33734},
  doi = {10.1021/jacs.5c09558},
  url = {https://doi.org/10.1021/jacs.5c09558}
}

@article{magduau2023machine,
  title={Machine learning force fields for molecular liquids: Ethylene Carbonate/Ethyl Methyl Carbonate binary solvent},
  author={Magd{\u{a}}u, Ioan-Bogdan and Arismendi-Arrieta, Daniel J and Smith, Holly E and Grey, Clare P and Hermansson, Kersti and Cs{\'a}nyi, G{\'a}bor},
  journal={npj Computational Materials},
  volume={9},
  number={1},
  pages={146},
  year={2023},
  publisher={Nature Publishing Group UK London}
}

@article{MakovPayneChargedCellCorrections,
  title = {Periodic boundary conditions in ab initio calculations},
  author = {Makov, G. and Payne, M. C.},
  journal = {Phys. Rev. B},
  volume = {51},
  issue = {7},
  pages = {4014--4022},
  numpages = {0},
  year = {1995},
  month = {Feb},
  publisher = {American Physical Society},
  doi = {10.1103/PhysRevB.51.4014},
  url = {https://link.aps.org/doi/10.1103/PhysRevB.51.4014}
}

@article{Deng2023,
  title = {CHGNet as a pretrained universal neural network potential for charge-informed atomistic modelling},
  volume = {5},
  ISSN = {2522-5839},
  url = {http://dx.doi.org/10.1038/s42256-023-00716-3},
  DOI = {10.1038/s42256-023-00716-3},
  number = {9},
  journal = {Nature Machine Intelligence},
  publisher = {Springer Science and Business Media LLC},
  author = {Deng,  Bowen and Zhong,  Peichen and Jun,  KyuJung and Riebesell,  Janosh and Han,  Kevin and Bartel,  Christopher J. and Ceder,  Gerbrand},
  year = {2023},
  month = sep,
  pages = {1031–1041}
}

@misc{eastman2022spicedatasetdruglikemolecules,
      title={SPICE, A Dataset of Drug-like Molecules and Peptides for Training Machine Learning Potentials}, 
      author={Peter Eastman and Pavan Kumar Behara and David L. Dotson and Raimondas Galvelis and John E. Herr and Josh T. Horton and Yuezhi Mao and John D. Chodera and Benjamin P. Pritchard and Yuanqing Wang and Gianni De Fabritiis and Thomas E. Markland},
      year={2022},
      eprint={2209.10702},
      archivePrefix={arXiv},
      primaryClass={physics.chem-ph},
      url={https://arxiv.org/abs/2209.10702}, 
}

@article{Kalita2025,
  title = {AIMNet2‐NSE: A Transferable Reactive Neural Network Potential for Open‐Shell Chemistry},
  volume = {65},
  ISSN = {1521-3773},
  url = {http://dx.doi.org/10.1002/anie.202516763},
  DOI = {10.1002/anie.202516763},
  number = {5},
  journal = {Angewandte Chemie International Edition},
  publisher = {Wiley},
  author = {Kalita,  Bhupalee and Zubatyuk,  Roman and Anstine,  Dylan M. and Bergeler,  Maike and Settels,  Volker and Stork,  Conrad and Spicher,  Sebastian and Isayev,  Olexandr},
  year = {2025},
  month = dec 
}

@article{Boese_2015,
    author = {Boese, A. Daniel},
    doi = {10.1080/00268976.2014.1001806},
    issn = {1362-3028},
    journal = {Mol. Phys.},
    number = {13-14},
    pages = {1618-1629},
    publisher = {Informa UK Limited},
    title = {Basis set limit coupled-cluster studies of hydrogen-bonded systems},
    url = {http://dx.doi.org/10.1080/00268976.2014.1001806},
    volume = {113},
    year = {2015}
}

@article{Chakravorty_1993,
    author = {Chakravorty, Subhas J. and Gwaltney, Steven R. and Davidson, Ernest R. and Parpia, Farid A. and p Fischer, Charlotte Froese},
    doi = {10.1103/physreva.47.3649},
    issn = {1094-1622},
    journal = {Phys. Rev. A},
    number = {5},
    pages = {3649-3670},
    publisher = {American Physical Society (APS)},
    title = {Ground-state correlation energies for atomic ions with 3 to 18 electrons},
    url = {http://dx.doi.org/10.1103/physreva.47.3649},
    volume = {47},
    year = {1993}
}

@article{Chan_2019,
    author = {Chan, Bun},
    doi = {10.1021/acs.jpca.9b03976},
    issn = {1520-5215},
    journal = {J. Phys. Chem. A},
    number = {27},
    pages = {5781-5788},
    publisher = {American Chemical Society (ACS)},
    title = {The CUAGAU Set of Coupled-Cluster Reference Data for Small Copper, Silver, and Gold Compounds and Assessment of DFT Methods},
    url = {http://dx.doi.org/10.1021/acs.jpca.9b03976},
    volume = {123},
    year = {2019}
}

@article{Chan_2023,
    author = {Chan, Bun},
    doi = {10.1021/acs.jpca.3c01880},
    issn = {1520-5215},
    journal = {J. Phys. Chem. A},
    number = {27},
    pages = {5652-5661},
    publisher = {American Chemical Society (ACS)},
    title = {Compilation of Ionic Clusters with the Rock Salt Structure: Accurate Benchmark Thermochemical Data, Assessment of Quantum Chemistry Methods, and the Convergence Behavior of Lattice Energies},
    url = {http://dx.doi.org/10.1021/acs.jpca.3c01880},
    volume = {127},
    year = {2023}
}

@article{Copeland_2012,
    author = {Copeland, Kari L. and Tschumper, Gregory S.},
    doi = {10.1021/ct300132e},
    issn = {1549-9626},
    journal = {J. Chem. Theory Comput.},
    number = {5},
    pages = {1646-1656},
    publisher = {American Chemical Society (ACS)},
    title = {Hydrocarbon/Water Interactions: Encouraging Energetics and Structures from DFT but Disconcerting Discrepancies for Hessian Indices},
    url = {http://dx.doi.org/10.1021/ct300132e},
    volume = {8},
    year = {2012}
}

@article{Crittenden_2009,
    author = {Crittenden, Deborah L.},
    doi = {10.1021/jp809106b},
    issn = {1520-5215},
    journal = {J. Phys. Chem. A},
    number = {8},
    pages = {1663-1669},
    publisher = {American Chemical Society (ACS)},
    title = {A Systematic CCSD(T) Study of Long-Range and Noncovalent Interactions between Benzene and a Series of First- and Second-Row Hydrides and Rare Gas Atoms},
    url = {http://dx.doi.org/10.1021/jp809106b},
    volume = {113},
    year = {2009}
}

@article{Goerigk_2016,
    author = {Goerigk, Lars and Sharma, Rahul},
    doi = {10.1139/cjc-2016-0290},
    issn = {1480-3291},
    journal = {Can. J. Chem.},
    number = {12},
    pages = {1133-1143},
    publisher = {Canadian Science Publishing},
    title = {The INV24 test set: how well do quantum-chemical methods describe inversion and racemization barriers?},
    url = {http://dx.doi.org/10.1139/cjc-2016-0290},
    volume = {94},
    year = {2016}
}

@article{Karton_2012Explicitly,
    author = {Karton, Amir and Martin, Jan M.L.},
    doi = {10.1080/00268976.2012.698316},
    issn = {1362-3028},
    journal = {Mol. Phys.},
    number = {19-20},
    pages = {2477-2491},
    publisher = {Informa UK Limited},
    title = {Explicitly correlated benchmark calculations on C8H8 isomer energy separations: how accurate are DFT, double-hybrid, and composite ab initio procedures?},
    url = {http://dx.doi.org/10.1080/00268976.2012.698316},
    volume = {110},
    year = {2012}
}

@article{Karton_2019,
    author = {Karton, Amir},
    doi = {10.1021/acs.jpca.9b04611},
    issn = {1520-5215},
    journal = {J. Phys. Chem. A},
    number = {31},
    pages = {6720-6732},
    publisher = {American Chemical Society (ACS)},
    title = {Highly Accurate CCSDT(Q)/CBS Reaction Barrier Heights for a Diverse Set of Transition Structures: Basis Set Convergence and Cost-Effective Approaches for Estimating Post-CCSD(T) Contributions},
    url = {http://dx.doi.org/10.1021/acs.jpca.9b04611},
    volume = {123},
    year = {2019}
}

@article{Kesharwani_2016,
    author = {Kesharwani, Manoj K. and Karton, Amir and Martin, Jan M. L.},
    doi = {10.1021/acs.jctc.5b01066},
    issn = {1549-9626},
    journal = {J. Chem. Theory Comput.},
    number = {1},
    pages = {444-454},
    publisher = {American Chemical Society (ACS)},
    title = {Benchmark ab Initio Conformational Energies for the Proteinogenic Amino Acids through Explicitly Correlated Methods. Assessment of Density Functional Methods},
    url = {http://dx.doi.org/10.1021/acs.jctc.5b01066},
    volume = {12},
    year = {2016}
}

@article{Kozuch_2013,
    author = {Kozuch, Sebastian and Martin, Jan M. L.},
    doi = {10.1021/ct301064t},
    issn = {1549-9626},
    journal = {J. Chem. Theory Comput.},
    number = {4},
    pages = {1918-1931},
    publisher = {American Chemical Society (ACS)},
    title = {Halogen Bonds: Benchmarks and Theoretical Analysis},
    url = {http://dx.doi.org/10.1021/ct301064t},
    volume = {9},
    year = {2013}
}

@article{Kruse_2015,
    author = {Kruse, Holger and Mladek, Arnost and Gkionis, Konstantinos and Hansen, Andreas and Grimme, Stefan and Sponer, Jiri},
    doi = {10.1021/acs.jctc.5b00515},
    issn = {1549-9626},
    journal = {J. Chem. Theory Comput.},
    number = {10},
    pages = {4972-4991},
    publisher = {American Chemical Society (ACS)},
    title = {Quantum Chemical Benchmark Study on 46 RNA Backbone Families Using a Dinucleotide Unit},
    url = {http://dx.doi.org/10.1021/acs.jctc.5b00515},
    volume = {11},
    year = {2015}
}

@article{Lao_2015,
    author = {Lao, Ka Un and Schäffer, Rainer and Jansen, Georg and Herbert, John M.},
    doi = {10.1021/ct5010593},
    issn = {1549-9626},
    journal = {J. Chem. Theory Comput.},
    number = {6},
    pages = {2473-2486},
    publisher = {American Chemical Society (ACS)},
    title = {Accurate Description of Intermolecular Interactions Involving Ions Using Symmetry-Adapted Perturbation Theory},
    url = {http://dx.doi.org/10.1021/ct5010593},
    volume = {11},
    year = {2015}
}

@article{Manna_2016,
    author = {Manna, Debashree and Martin, Jan M. L.},
    doi = {10.1021/acs.jpca.5b10266},
    issn = {1520-5215},
    journal = {J. Phys. Chem. A},
    number = {1},
    pages = {153-160},
    publisher = {American Chemical Society (ACS)},
    title = {What Are the Ground State Structures of C20 and C24? An Explicitly Correlated Ab Initio Approach},
    url = {http://dx.doi.org/10.1021/acs.jpca.5b10266},
    volume = {120},
    year = {2016}
}

@article{Mardirossian_2013,
    author = {Mardirossian, Narbe and Lambrecht, Daniel S. and McCaslin, Laura and Xantheas, Sotiris S. and Head-Gordon, Martin},
    doi = {10.1021/ct4000235},
    issn = {1549-9626},
    journal = {J. Chem. Theory Comput.},
    number = {3},
    pages = {1368-1380},
    publisher = {American Chemical Society (ACS)},
    title = {The Performance of Density Functionals for Sulfate-Water Clusters},
    url = {http://dx.doi.org/10.1021/ct4000235},
    volume = {9},
    year = {2013}
}

@article{Martin_2013,
    author = {Martin, Jan M. L.},
    doi = {10.1021/jp401429u},
    issn = {1520-5215},
    journal = {J. Phys. Chem. A},
    number = {14},
    pages = {3118-3132},
    publisher = {American Chemical Society (ACS)},
    title = {What Can We Learn about Dispersion from the Conformer Surface of n-Pentane?},
    url = {http://dx.doi.org/10.1021/jp401429u},
    volume = {117},
    year = {2013}
}

@article{Maurer_2021,
    author = {Maurer, Leonard R. and Bursch, Markus and Grimme, Stefan and Hansen, Andreas},
    doi = {10.1021/acs.jctc.1c00659},
    issn = {1549-9626},
    journal = {J. Chem. Theory Comput.},
    number = {10},
    pages = {6134-6151},
    publisher = {American Chemical Society (ACS)},
    title = {Assessing Density Functional Theory for Chemically Relevant Open-Shell Transition Metal Reactions},
    url = {http://dx.doi.org/10.1021/acs.jctc.1c00659},
    volume = {17},
    year = {2021}
}

@article{Mintz2012,
author = {Mintz, Benjamin J. and Parks, Jerry M.},
title = {Benchmark Interaction Energies for Biologically Relevant Noncovalent Complexes Containing Divalent Sulfur},
journal = {J. Phys. Chem. A},
volume = {116},
number = {3},
pages = {1086-1092},
year = {2012},
doi = {10.1021/jp209536e}
}

@article{Ochieng_2023,
    author = {Ochieng, Sharon A. and Patkowski, Konrad},
    doi = {10.1039/d3cp03938b},
    issn = {1463-9084},
    journal = {Phys. Chem. Chem. Phys.},
    number = {42},
    pages = {28621-28637},
    publisher = {Royal Society of Chemistry (RSC)},
    title = {Accurate three-body noncovalent interactions: the insights from energy decomposition},
    url = {http://dx.doi.org/10.1039/d3cp03938b},
    volume = {25},
    year = {2023}
}

@article{Oliveira_2017,
    author = {Oliveira, Vytor and Kraka, Elfi},
    doi = {10.1021/acs.jpca.7b10196},
    issn = {1520-5215},
    journal = {J. Phys. Chem. A},
    number = {49},
    pages = {9544-9556},
    publisher = {American Chemical Society (ACS)},
    title = {Systematic Coupled Cluster Study of Noncovalent Interactions Involving Halogens, Chalcogens, and Pnicogens},
    url = {http://dx.doi.org/10.1021/acs.jpca.7b10196},
    volume = {121},
    year = {2017}
}

@article{Rezac2015Benchmark,
    author = {Řezáč, Jan and Huang, Yuanhang and Hobza, Pavel and Beran, Gregory J. O.},
    doi = {10.1021/acs.jctc.5b00281},
    issn = {1549-9626},
    journal = {J. Chem. Theory Comput.},
    number = {7},
    pages = {3065-3079},
    publisher = {American Chemical Society (ACS)},
    title = {Benchmark Calculations of Three-Body Intermolecular Interactions and the Performance of Low-Cost Electronic Structure Methods},
    url = {http://dx.doi.org/10.1021/acs.jctc.5b00281},
    volume = {11},
    year = {2015}
}

@article{Rezac2020,
    author = {Řezáč, Jan},
    doi = {10.1021/acs.jctc.9b01265},
    issn = {1549-9626},
    journal = {J. Chem. Theory Comput.},
    number = {4},
    pages = {2355-2368},
    publisher = {American Chemical Society (ACS)},
    title = {Non-Covalent Interactions Atlas Benchmark Data Sets: Hydrogen Bonding},
    url = {http://dx.doi.org/10.1021/acs.jctc.9b01265},
    volume = {16},
    year = {2020}
}

@article{Smith_2014,
    author = {Smith, Daniel G. A. and Jankowski, Piotr and Slawik, Michał and Witek, Henryk A. and Patkowski, Konrad},
    doi = {10.1021/ct500347q},
    issn = {1549-9626},
    journal = {J. Chem. Theory Comput.},
    number = {8},
    pages = {3140-3150},
    publisher = {American Chemical Society (ACS)},
    title = {Basis Set Convergence of the Post-CCSD(T) Contribution to Noncovalent Interaction Energies},
    url = {http://dx.doi.org/10.1021/ct500347q},
    volume = {10},
    year = {2014}
}

@article{Steinmann_2012,
    author = {Steinmann, Stephan N. and Piemontesi, Cyril and Delachat, Aurore and Corminboeuf, Clemence},
    doi = {10.1021/ct200930x},
    issn = {1549-9626},
    journal = {J. Chem. Theory Comput.},
    number = {5},
    pages = {1629-1640},
    publisher = {American Chemical Society (ACS)},
    title = {Why are the Interaction Energies of Charge-Transfer Complexes Challenging for DFT?},
    url = {http://dx.doi.org/10.1021/ct200930x},
    volume = {8},
    year = {2012}
}

@article{Sure_2017,
    author = {Sure, Rebecca and Hansen, Andreas and Schwerdtfeger, Peter and Grimme, Stefan},
    doi = {10.1039/c7cp00735c},
    issn = {1463-9084},
    journal = {Phys. Chem. Chem. Phys.},
    number = {22},
    pages = {14296-14305},
    publisher = {Royal Society of Chemistry (RSC)},
    title = {Comprehensive theoretical study of all 1812 C60 isomers},
    url = {http://dx.doi.org/10.1039/c7cp00735c},
    volume = {19},
    year = {2017}
}

@article{Tentscher_2013,
    author = {Tentscher, Peter R. and Arey, J. Samuel},
    doi = {10.1021/ct300846m},
    issn = {1549-9626},
    journal = {J. Chem. Theory Comput.},
    number = {3},
    pages = {1568-1579},
    publisher = {American Chemical Society (ACS)},
    title = {Binding in Radical-Solvent Binary Complexes: Benchmark Energies and Performance of Approximate Methods},
    url = {http://dx.doi.org/10.1021/ct300846m},
    volume = {9},
    year = {2013}
}

@article{Witte_2015,
    author = {Witte, Jonathon and Goldey, Matthew and Neaton, Jeffrey B. and Head-Gordon, Martin},
    doi = {10.1021/ct501050s},
    issn = {1549-9626},
    journal = {J. Chem. Theory Comput.},
    number = {4},
    pages = {1481-1492},
    publisher = {American Chemical Society (ACS)},
    title = {Beyond Energies: Geometries of Nonbonded Molecular Complexes as Metrics for Assessing Electronic Structure Approaches},
    url = {http://dx.doi.org/10.1021/ct501050s},
    volume = {11},
    year = {2015}
}

@article{Yu_2014,
    author = {Yu, Li-Juan and Karton, Amir},
    doi = {10.1016/j.chemphys.2014.07.015},
    issn = {0301-0104},
    journal = {Chem. Phys.},
    pages = {166-177},
    publisher = {Elsevier BV},
    title = {Assessment of theoretical procedures for a diverse set of isomerization reactions involving double-bond migration in conjugated dienes},
    url = {http://dx.doi.org/10.1016/j.chemphys.2014.07.015},
    volume = {441},
    year = {2014}
}

@article{Yu_2015Components,
    author = {Yu, Haoyu and Truhlar, Donald G.},
    doi = {10.1021/acs.jctc.5b00083},
    issn = {1549-9626},
    journal = {J. Chem. Theory Comput.},
    number = {7},
    pages = {2968-2983},
    publisher = {American Chemical Society (ACS)},
    title = {Components of the Bond Energy in Polar Diatomic Molecules, Radicals, and Ions Formed by Group-1 and Group-2 Metal Atoms},
    url = {http://dx.doi.org/10.1021/acs.jctc.5b00083},
    volume = {11},
    year = {2015}
}

@article{balabanov2006basis,
  title={Basis set limit electronic excitation energies, ionization potentials, and electron affinities for the 3d transition metal atoms: Coupled cluster and multireference methods},
  author={Balabanov, Nikolai B and Peterson, Kirk A},
  journal = {J. Chem. Phys.},
  volume={125},
  number={7},
    pages = {074110},
  year={2006},
  publisher={AIP Publishing}
}

@article{chan2023dapd,
  title={DAPD Set of Pd-Containing Diatomic Molecules: Accurate Molecular Properties and the Great Lengths to Obtain Them},
  author={Chan, Bun},
  journal = {J. Chem. Theory Comput.},
  volume={19},
  number={24},
  pages={9260--9268},
  year={2023},
  publisher={ACS Publications}
}

@article{karton2011w4,
  title={W4-11: A high-confidence benchmark dataset for computational thermochemistry derived from first-principles W4 data},
  author={Karton, Amir and Daon, Shauli and Martin, Jan ML},
  journal = {Chem. Phys. Lett.},
  volume={510},
  number={4-6},
  pages={165--178},
  year={2011},
  publisher={Elsevier}
}

@article{lang2023three,
  title={Three-body potential and third virial coefficients for helium including relativistic and nuclear-motion effects},
  author={Lang, Jakub and Garberoglio, Giovanni and Przybytek, Micha{\l} and Jeziorska, Ma{\l}gorzata and Jeziorski, Bogumi{\l}},
  journal = {Phys. Chem. Chem. Phys.},
  volume={25},
  number={35},
  pages={23395--23416},
  year={2023},
  publisher={Royal Society of Chemistry}
}

@article{madajczyk2021dataset,
  title={Dataset of noncovalent intermolecular interaction energy curves for 24 small high-spin open-shell dimers},
  author={Madajczyk, Katarzyna and {\.Z}uchowski, Piotr S and Brz\c{e}k, Filip and Rajchel, {\L}ukasz and K\c{e}dziera, Dariusz and Modrzejewski, Marcin and Hapka, Micha{\l}},
  journal = {J. Chem. Phys.},
  volume={154},
  number={13},
    pages = {134106},
  year={2021},
  publisher={AIP Publishing}
}

@article{mccarthy2011accurate,
  title={Accurate all-electron correlation energies for the closed-shell atoms from Ar to Rn and their relationship to the corresponding MP2 correlation energies},
  author={McCarthy, Shane P and Thakkar, Ajit J},
  journal = {J. Chem. Phys.},
  volume={134},
  number={4},
    pages = {044102},
  year={2011},
  publisher={AIP Publishing}
}

@article{przybytek2017pair,
  title={Pair potential with submillikelvin uncertainties and nonadiabatic treatment of the halo state of the helium dimer},
  author={Przybytek, Micha{\l} and Cencek, Wojciech and Jeziorski, Bogumi{\l} and Szalewicz, Krzysztof},
  journal = {Phys. Rev. Lett.},
  volume={119},
  number={12},
  pages={123401},
  year={2017},
  publisher={APS}
}

@article{yoo2010high,
  title={High-level ab initio electronic structure calculations of water clusters (H2O) 16 and (H2O) 17: A new global minimum for (H2O) 16},
  author={Yoo, Soohaeng and Apra, Edoardo and Zeng, Xiao Cheng and Xantheas, Sotiris S},
  journal = {J. Phys. Chem. Lett.},
  volume={1},
  number={20},
  pages={3122--3127},
  year={2010},
  publisher={ACS Publications}
}

@article{Thomas2025,
  title = {Self-consistent Coulomb interactions for machine learning interatomic potentials},
  volume = {38},
  ISSN = {1361-6544},
  url = {http://dx.doi.org/10.1088/1361-6544/ae0402},
  DOI = {10.1088/1361-6544/ae0402},
  number = {9},
  journal = {Nonlinearity},
  publisher = {IOP Publishing},
  author = {Thomas,  Jack and Baldwin,  Will and Csanyi,  Gabor and Ortner,  Christoph},
  year = {2025},
  month = sep,
  pages = {095024}
}

@misc{grasselli2026longrangeelectrostaticsatomisticmachine,
      title={Long-range electrostatics in atomistic machine learning: a physical perspective}, 
      author={Federico Grasselli and Kevin Rossi and Stefano de Gironcoli and Andrea Grisafi},
      year={2026},
      eprint={2602.11071},
      archivePrefix={arXiv},
      primaryClass={cond-mat.mtrl-sci},
      url={https://arxiv.org/abs/2602.11071}, 
}

@misc{bristol-ai,
  doi = {10.48550/ARXIV.2410.11199},
  url = {https://arxiv.org/abs/2410.11199},
  author = {McIntosh-Smith,  Simon and Alam,  Sadaf R and Woods,  Christopher},
  title = {Isambard-AI: a leadership class supercomputer optimised specifically for Artificial Intelligence},
  publisher = {arXiv},
  year = {2024},
  copyright = {Creative Commons Attribution 4.0 International}
}

\clearpage
\onecolumngrid

\section{Supplementary Information}

\subsection{WTMAD-2 aggregation}
\label{sec:wtmad2}
The weighted total mean absolute deviation (WTMAD-2) follows the GMTKN55 convention~\cite{Goerigk2017}. For a collection of benchmark subsets $k$ with $N_k$ entries and absolute errors $\Delta E_{k,i}$, the aggregate score is
\begin{equation}
\mathrm{WTMAD\text{-}2} = \frac{\sum_k w_k \, \frac{1}{N_k}\sum_{i=1}^{N_k} |\Delta E_{k,i}|}{\sum_k w_k},
\end{equation}
where $w_k = \frac{56.84~\mathrm{kcal/mol}}{\langle |E_k|\rangle}$ rescales each subset by the inverse of its mean absolute reference magnitude $\langle |E_k|\rangle$. This weighting balances datasets of very different energetic scales (e.g., small reaction energies vs.\ large atomisation energies) and allows a single number to compare broad benchmark suites. We report WTMAD-2 in kcal/mol.

\subsection{Outlier filtering and excluded structures}
\label{sec:exclusions}
A small number of benchmark structures are excluded per model. For aggregation, we report two exclusion categories: (i) single-atom configurations (\texttt{num\_atoms} $\leq 1$ in the model CSVs), and (ii) outliers with $|\Delta E| > 100$~kcal/mol (summary bar plots only). Full error distributions are reported without this outlier cap in the per-dataset SI figures. Table~\ref{tab:excluded_structures} reports the per-model breakdown for the thermochemistry and reaction-barrier subsets used in Fig.~\ref{fig:thermochem}. Lists of excluded identifiers are provided in the uploaded CSV files (column \texttt{exclusion\_type} indicates the reason): \texttt{excluded\_structures\_MACE-POLAR-1-M.csv}, \texttt{excluded\_structures\_MACE-POLAR-1-L.csv}, \texttt{excluded\_structures\_MACE-OMOL.csv}, \texttt{excluded\_structures\_UMA-S-1P1.csv}, \texttt{excluded\_structures\_UMA-M-1P1.csv}, \texttt{excluded\_structures\_ORBMOL.csv}, \texttt{excluded\_structures\_g-xTB.csv}, and the combined file \texttt{excluded\_structures\_all\_models.csv} in the accompanying data files.
\begin{table}[h]
\centering
\caption{Breakdown of excluded structures for the thermochemistry and reaction-barrier summary plots (Fig.~\ref{fig:thermochem}). Columns report single-atom exclusions (\texttt{num\_atoms} $\leq 1$) and outliers with $|\Delta E|>100$~kcal/mol (summary bars only). The final column gives the fraction of excluded structures relative to the total number of structures evaluated for each model on these subsets.}
\label{tab:excluded_structures}
\begin{tabular}{lrrrrr}
\toprule
Model & Single-atom & Outlier $>100$ & Total & \% excluded \\
\midrule
\MFPM & 86 & 9 & 95 & 3.08 \\
\MFPL & 86 & 11 & 97 & 3.14 \\
\MOMOL & 86 & 96 & 182 & 6.26 \\
\UMAS & 86 & 66 & 152 & 4.92 \\
\UMAM & 86 & 59 & 145 & 4.69 \\
\ORBMOL & 86 & 121 & 207 & 6.70 \\
g-xTB & 86 & 1 & 87 & 2.99 \\
\bottomrule
\end{tabular}
\end{table}

\subsection{Hyper-parameters}

\begin{table*}[h]
\centering
\caption{Hyper-parameter settings for the two \MFP variants used in this work.}
\label{tab:mace_hparams}
{
\setlength{\tabcolsep}{6pt}
\renewcommand{\arraystretch}{1.15}
\begin{tabular}{@{}lcc@{}}
\toprule
& \multicolumn{2}{c}{\textbf{Models}} \\
\cmidrule(lr){2-3}
\textbf{hyper-parameter} & \textbf{\MFPM} & \textbf{\MFPL} \\
\midrule
max\_ell                & 3   & 3   \\
correlation             & 3   & 3   \\
max\_L                  & 1   & 1   \\
num\_channels\_edge      & 128 & 128 \\
num\_channels\_node      & 512 & 512 \\
num\_interactions       & 2   & 3   \\
num\_radial\_basis       & 8   & 8   \\
r\_max (\AA)            & 6   & 6   \\
interactions\_class     & non-linear~\cite{batatia2025crosslearningelectronicstructure} & non-linear~\cite{batatia2025crosslearningelectronicstructure} \\
irreps                  & 16x0e & 16x0e \\
smearing\_sigma\_charge (\AA)        &   1.5   &   1.5  \\
num\_field\_features     &  2   &   2 \\
smearing\_sigma\_field (\AA)        &   1.5, 2.0   &   1.5, 2.0  \\
field\_features\_l\_max  &   1  &   1  \\
spin\_charge\_density\_l\_max &  1   &  1   \\
num\_update             & 2 & 2 \\
batch size              & 256 & 256 \\
energy coefficient      & 1   & 1   \\
force coefficient       & 10  & 10  \\
\bottomrule
\end{tabular}}
\end{table*}

\paragraph{Hyperparameter definitions.}
\noindent We provide here a brief summary of the meaning of each hyper-parameter in the table. More details on the hyper-parameter choices and their implementation can be found in the original MACE paper~\cite{batatia_mace_2023} and the MACE-MH-1 paper~\cite{batatia2025crosslearningelectronicstructure}.
\begin{itemize}
  \item \textbf{max\_ell} sets the highest angular momentum used in the spherical-harmonic expansion of local environments, controlling angular resolution.
  \item \textbf{correlation} sets the maximum correlation order retained in the MACE basis.
  \item \textbf{max\_L} sets the maximum total angular momentum of messages passed between interaction layers.
  \item \textbf{num\_channels\_edge} is the size of the edge features in the local MACE part; see~\cite{batatia2025crosslearningelectronicstructure}.
  \item \textbf{num\_channels\_node} is the size of the node features in the local MACE part; see~\cite{batatia2025crosslearningelectronicstructure}.
  \item \textbf{num\_interactions} is the number of MACE message-passing layers.
  \item \textbf{num\_radial\_basis} is the number of radial basis functions used to expand interatomic distances.
  \item \textbf{r\_max} is the neighbour cutoff radius that defines the local graph used by MACE.
  \item \textbf{interactions\_class} specifies whether interaction blocks are linear or non-linear; see~\cite{batatia2025crosslearningelectronicstructure}.
  \item \textbf{hidden\_irreps} gives the hidden layer dimension used in the readout layers.
  \item \textbf{smearing\_sigma\_charge} is the Gaussian width for charge-density smearing (Eq.~\ref{eq:gaussian-basis}).
  \item \textbf{smearing\_sigma\_field} is the Gaussian width used for field-feature smearing (Eq.~\ref{eq:gaussian-basis}).
  \item \textbf{num\_field\_features} is the number of learned field-feature channels per update step.
  \item \textbf{field\_features\_l\_max} sets the maximum angular momentum of the electrostatic features used in non-local updates.
  \item \textbf{spin\_charge\_density\_l\_max} sets the maximum multipole order predicted for the spin/charge density.
  \item \textbf{num\_update} is the number of non-local field-update iterations applied during inference.
  \item \textbf{batch size} is the effective global batch size used for training across all devices.
  \item \textbf{energy coefficient} is the weight applied to the energy term in the training loss.
  \item \textbf{force coefficient} is the weight applied to the force term in the training loss.
\end{itemize}
\subsection{Thermochemistry and barrier benchmark datasets}
\label{sec:tc_benchmarks}
\textbf{Thermochemistry subsets}  Ionisation potentials (G21IP adiabatic first-row organics~\cite{Parthiban2001}; IP23~\cite{Cheng2007} and IP30~\cite{Luo2012} vertical IPs of small organic/main-group systems); electron affinities (G21EA~\cite{Parthiban2001}, EA50~\cite{Ermis2021}); proton affinities (PA26, small organic bases~\cite{Parthiban2001}); mindless molecules (MB08-165~\cite{korth2009mindless}, MB16-43~\cite{Goerigk2017}); bond dissociation and HAT sets (BDE99MR and BDE99nonMR~\cite{Chan2017}, HNBrBDE18~\cite{Chan2018}, YBDE18~\cite{Zhao2012}, HAT707MR/HAT707nonMR~\cite{Karton2017HAT}, RSE43~\cite{Zhao2012RSE}, BSR36~\cite{Chan2016BSR}). All benchmark names and citations are collected in this SI section for reference. \\
\textbf{Reaction barrier subsets} Barrier heights (BH28~\cite{Liang2025}, BH46~\cite{Zhao2008BH76}, BH876~\cite{Prasad2021}, BHDIV7~\cite{Goerigk2017}, ORBH35~\cite{Chan2018}, CRBH14~\cite{Yu2015CRBH}); proton transfer (DBH22~\cite{Karton2008DBH}, PX9~\cite{Karton2012PX}, WCPT26~\cite{Karton2012WCPT}); general reaction energies (FH51~\cite{Friedrich_2015}, G2RC24~\cite{curtiss1991gaussian}, BH76RC~\cite{Goerigk2017}, CR20~\cite{yu2016can}, DARC~\cite{Goerigk2017}, NBPRC~\cite{goerigk2010general}, RC21~\cite{goerigk2010general}). All reference values are CCSD(T)/CBS or explicitly correlated F12 where available.

\paragraph{WTMAD table subsets.} Table~\ref{tab:wtmad_tc_bh_subgroups} reports WTMAD-2 on disjoint subsets to avoid overlap between rows. The thermochemistry rows are: TC: Bond Energies + HAT (BDE99MR, BDE99nonMR, BSR36, HAT707MR, HAT707nonMR, RSE43, YBDE18); TC: Reaction Energies (BH76RC, CR20, DARC, FH51, G2RC24, RC21); and TC: Thermochemistry (AlkIsod14, DC13, EA50, G21EA, HEAVYSB11, MB08-165, SN13, TAE\_W4-17MR, WCPT6). The barrier row is BH: Barrier + Proton (BH28, BH46, BH876, BHDIV7, BHROT27, CRBH14, ORBH35, DBH22, PX9, WCPT26). HAT denotes hydrogen-atom transfer.

\subsection{Real Space Electrostatic Energy Computation}
\label{sec:si_realspace_electrostatics}

\subsubsection{Real Space Energy}

As described in the main text, the electrostatic energy in open boundary conditions reduces to the following sum:
\begin{align}
    E_\text{Hartree} &= \frac{1}{2}\sum_{ilm,jl'm'} p_{i,lm}p_{j,l'm'} \mathcal{T}_{ilm,jl'm'} \label{eq:SI_def_E_Hartree} \\
    \mathcal{T}_{ilm,jl'm'} &:=\iint \frac{\phi_{lm}(\mathbf{r}-\mathbf{r}_i)\phi_{l'm'}(\mathbf{r}'-\mathbf{r}_j)}{|\mathbf{r}-\mathbf{r}'|} d\mathbf{r} d\mathbf{r}'
    \label{eq:SI_def_T_coeffs}
\end{align}
In the same open-boundary setting, the electrostatic potential generated by the multipoles is
\begin{equation}
v(\mathbf{r})=\sum_{jl'm'} p_{j,l'm'}\int \frac{\phi_{l'm'}(\mathbf{r}'-\mathbf{r}_j)}{|\mathbf{r}-\mathbf{r}'|}\,d\mathbf{r}'
\label{eq:si_realspace_potential_field}
\end{equation}
and the atom-centred projected potential feature entering Equation~\ref{eq:field-features} is
\begin{equation}
v_{i,nlm}=\frac{1}{\mathcal N_{nl}}\sum_{jl'm'} p_{j,l'm'}\iint
\frac{\phi_{nlm}(\mathbf{r}-\mathbf{r}_i)\phi_{l'm'}(\mathbf{r}'-\mathbf{r}_j)}{|\mathbf{r}-\mathbf{r}'|}
\,d\mathbf{r}\,d\mathbf{r}' .
\label{eq:si_realspace_projected_potential}
\end{equation}
For $l=l'=0$, with source width $\sigma_{\text{charge}}$ and receiver width $\sigma_{\text{field},n}$, this becomes
\begin{equation}
v_{i,n00}
=
\sum_j p_{j,00}
\frac{\mathrm{erf}\!\left(r_{ij}/(2\sigma_{\mathrm{tot},n})\right)}{r_{ij}},
\qquad
\sigma_{\mathrm{tot},n}=\sqrt{\frac{\sigma_{\text{charge}}^2+\sigma_{\text{field},n}^2}{2}}.
\label{eq:si_realspace_l0_projector}
\end{equation}
The electrostatic energy is recovered from the same contraction when source and receiver basis widths are identical.
One can easily evaluate the monopole coefficients for distinct $i,j$ as:
\begin{equation}
\mathcal{T}_{i00,j00} = \frac{\text{erf}(r_{ij}/\sqrt{2}\sigma)}{r_{ij}}
\label{eq:SI_monopole_energy}
\end{equation}
where $\sigma$ is the half-width of the Gaussian function $\phi_{n00}$. For the dipole interactions, one can approximate an $l=1$ Gaussian type orbital as a linear combination of $l=0$ orbitals which are slightly displaced from one another. This is valid not only for distant $i$ and $j$, but also when atoms $i$ and $j$ are close together so that the Gaussian functions on each atom overlap.

To show this, consider elements of $\phi_{1m}$ for $m=-1,0,1$ with a fixed half-width $\sigma$. These are Gaussian type orbitals (GTOs) with the following form:
\begin{align}
    \phi_{1m}(\mathbf{r}) = C_1 r^1 \exp\left(-\frac{|\mathbf{r} - \mathbf{r}_i|^2}{2\sigma^2}\right) Y_{1m}\!\left(\widehat{\mathbf{r} - \mathbf{r}_i}\right)
    \label{eq:SI_phi_1m}
\end{align}
where $C_1$ is the normalisation constant. Using the Condon-Shortley phase convention for the spherical harmonics, these three GTOs ($m=-1,0,1$) are aligned with the $y, z$ and $x$ axes, in that order. Let $\phi_{1\alpha}$ for $\alpha=x,y,z$ be a permutation of $\phi_{1 m}$ which swaps from the Condon-Shortley phase convention ($y,z,x$) back to the Cartesian ordering $x,y,z$, so that $\phi_{1x}$ is aligned with the $x$ axis, and so on.

We want to approximate $\phi_{1x}(\mathbf{r})$ using $\phi_{00}(\mathbf{r})$ and $\phi_{00}(\mathbf{r} + a \hat{\mathbf{x}})$, where $\hat{\mathbf{x}}$ is a unit vector in the $x$ direction, and $a$ is a small constant. This can be done as follows. Using the expression \eqref{eq:SI_phi_1m}, one can show that
\begin{align}
    \phi_{1 x}(\mathbf{r}) = \sqrt{3}\sigma^2\frac{\partial}{\partial x} \phi_{0 0}(\mathbf{r}) \times \frac{C_{1}}{C_{0}}
\end{align}
where $C_0$ is the normalisation constant for the $l=0$ GTO and $C_1$ is the normalisation constant for the $l=1$ GTO. Then, since
\begin{align}
    \frac{\partial}{\partial x} f(x) = \lim_{a\rightarrow0}\frac{f(x+a) - f(x)}{a},
\end{align}
we can write
\begin{align}
    \phi_{1 x}(\mathbf{r}) = \frac{C_1}{C_0} \sqrt{3}\sigma^2 \lim_{a\rightarrow 0} \left( \frac{\phi_{0 0}(\mathbf{r}+a\hat{\mathbf{x}}) - \phi_{0 0}(\mathbf{r})}{a} \right) 
\end{align}
In other words, with a few normalisation terms, one can approximate $l=1$ orbitals by using two $l=0$ orbitals displaced by a distance $a$, and the approximation is exact for all $\mathbf{r}$ as $a$ goes to zero. In practice, using $a=0.02$ \AA \space is sufficient to achieve sub-meV total energy convergence.

In order to compute the interaction energy of a charge $p_{i,00}$ on atom $i$ and a dipole $p_{j,1m}$ on atom $j$, we therefore proceed as follows. For the $x$-component of the dipole, approximate
\begin{align}
    p_{j,1x} \phi_{1x}(\mathbf{r}-\mathbf{r}_j) &= q \phi_{0 0}(\mathbf{r}-\mathbf{r}_j+a\hat{\mathbf{x}}) - q \phi_{0 0}(\mathbf{r}-\mathbf{r}_j) \\
    q &= p_{j,1x} \frac{C_1\sqrt{3}\sigma^2}{a C_0}
\end{align}
Repeat for the $y$ and $z$ components, and then simply compute the interaction energy of these $6$ Gaussian charges with other charges and dipoles using \eqref{eq:SI_monopole_energy}.
The same displaced-charge construction is used for projected potential features. For feature evaluation we use $a=0.1$~\AA{}, which improves numerical stability of the finite-difference projection while preserving the accuracy of the long-range feature map.

\subsubsection{Self-Interaction Energy}

We found it beneficial to add the energy of the self-interaction of the Gaussian charge distribution interacting with itself to the electrostatic energy. The self-energy for a single Gaussian of multipole order $l$ is:
\begin{equation}
E_{\text{self},i}^{(l)} = \frac{1}{2(2l+1)} \int_0^\infty r^{2l+2} \phi_l^2(r) V_l(r) dr
\end{equation}
where $\phi_l(r) = C_l r^l e^{-r^2/(2\sigma^2)}$ and $V_l(r)$ is the potential generated by $\phi_l$. In the implementation, we define:
\begin{align}
I_1(r,l,\sigma) &= 2^{l+1/2}\sigma^{2l+3}r^{-(l+1)}\,P\!\left(\frac{2l+3}{2}, \frac{r^2}{2\sigma^2}\right)\Gamma\!\left(\frac{2l+3}{2}\right) \\
I_2(r,l,\sigma) &= \sigma^2 r^l e^{-r^2/(2\sigma^2)}
\end{align}
where $P(a,x)$ is the regularised lower incomplete gamma function. The potential contribution is then evaluated as $V_l(r)=I_1(r,l,\sigma)+I_2(r,l,\sigma)$ and integrated numerically once during model initialisation for each $(l, \sigma)$ pair, then stored as constant tensors. The total self-interaction correction is:
\begin{equation}
E_{\text{self}} = \frac{1}{2} \sum_i \sum_{lm} \sum_{l'm'} p_{i,lm} \Gamma_{ll'}^{mm'}(\sigma) p_{i,l'm'} = \frac{1}{2} \sum_i \mathbf{p}_i^T \mathbf{\Gamma}(\sigma) \mathbf{p}_i
\end{equation}
where the overlap matrix $\mathbf{\Gamma}$ is diagonal in $(l,m)$ indices due to angular momentum conservation.

\subsection{Periodic Electrostatic Energy Computation}
\label{sec:si_periodic_electrostatics}

\subsubsection{Reciprocal Space Grid Construction}

For a periodic system with lattice vectors forming the matrix $\mathbf{L} = [\mathbf{a}_1, \mathbf{a}_2, \mathbf{a}_3]$, the reciprocal lattice vectors are defined through $\mathbf{L}^* = 2\pi(\mathbf{L}^T)^{-1}$. The k-space grid is constructed as:
\begin{equation}
\mathbf{k}_{n_1,n_2,n_3} = n_1 \mathbf{b}_1 + n_2 \mathbf{b}_2 + n_3 \mathbf{b}_3
\end{equation}
where $\mathbf{b}_i$ are the reciprocal lattice vectors (columns of $\mathbf{L}^*$) and $(n_1, n_2, n_3) \in \mathbb{Z}^3$ are integer indices chosen such that $|\mathbf{k}_{n_1,n_2,n_3}| \leq k_{\text{cutoff}}$. While one could use all these vectors, we make use of the fact that the density and potential are scalar fields and hence have $\tilde{f}(\mathbf{k}) = \tilde{f}(-\mathbf{k})^*$. We therefore store only a values on a half-grid and adjust the formulae below to make use of the conjugate symmetry. 

The k-space cutoff is determined adaptively from the Gaussian basis parameters using:
\begin{equation}
k_{\text{cutoff}} = \kappa \cdot \frac{2.25}{\sigma_{\text{min}}} (l_{\text{max}} + 1)^{0.3}
\end{equation}
where $\kappa$ is a user-specified factor (typically 1.5), $\sigma_{\text{min}} = \min(\sigma_{\text{charge}}, \sigma_{\text{field},1}, \ldots)$ is the minimum smearing width across all basis functions, and $l_{\text{max}}$ is the maximum angular momentum. This heuristic ensures that the Fourier transform of the most localised basis function is adequately resolved.

\subsubsection{Fourier Transform of Gaussian Multipoles}

The Fourier transform of a Gaussian spherical harmonic basis function $\phi_{nlm}(\mathbf{r} - \mathbf{r}_i)$ involves separating angular and radial components. Using the plane wave expansion $e^{-i\mathbf{k}\cdot\mathbf{r}} = 4\pi \sum_{lm} (-i)^l j_l(kr) Y_{lm}^*(\hat{\mathbf{k}}) Y_{lm}(\hat{\mathbf{r}})$ where $j_l$ is the spherical Bessel function, and integrating the Gaussian radial function $r^l e^{-r^2/(2\sigma^2)}$ against $j_l(kr)$, we obtain:
\begin{equation}
\tilde{\phi}_{nlm}(\mathbf{k}) = \int \phi_{nlm}(\mathbf{r}) e^{-i\mathbf{k}\cdot\mathbf{r}} d\mathbf{r} = C_l \cdot (-i)^l Y_{lm}(\hat{\mathbf{k}}) \cdot I_{l,\sigma}(k)
\end{equation}
where the radial integral $I_{l,\sigma}(k)$ evaluates to:
\begin{equation}
I_{l,\sigma}(k) = 4\pi \sqrt{\frac{\pi}{2}} \, \sigma^{2l+3} \, k^l \, e^{-k^2\sigma^2/2}
\end{equation}
The fourier series of the full charge density is then computed as:
\begin{equation}
\tilde{\rho}(\mathbf{k}) = \frac{(2\pi)^3}{\Omega} \sum_{i=1}^{N_{\text{atoms}}} \sum_{l=0}^{l_{\text{max}}} \sum_{m=-l}^{l} p_{i,lm} \tilde{\phi}_{lm}(\mathbf{k})e^{-\mathbf{k}\cdot \mathbf{r}_i}
\end{equation}
This is evaluated by first computing the real and imaginary parts separately:
\begin{align}
&\text{Re}[\tilde{\rho}(\mathbf{k})] \\
&= \frac{(2\pi)^3}{\Omega} \sum_{i,lm} p_{i,lm} [\text{Re}[\tilde{\phi}_{lm}(\mathbf{k})] \cos(\mathbf{k}\cdot\mathbf{r}_i) + \text{Im}[\tilde{\phi}_{lm}(\mathbf{k})] \sin(\mathbf{k}\cdot\mathbf{r}_i)] \\
&\text{Im}[\tilde{\rho}(\mathbf{k})] \\
&= \frac{(2\pi)^3}{\Omega} \sum_{i,lm} p_{i,lm} [\text{Im}[\tilde{\phi}_{lm}(\mathbf{k})] \cos(\mathbf{k}\cdot\mathbf{r}_i) - \text{Re}[\tilde{\phi}_{lm}(\mathbf{k})] \sin(\mathbf{k}\cdot\mathbf{r}_i)]
\end{align}
The phases $(-i)^l$ from the spherical harmonics expansion lead to real parts for even $l$ and imaginary parts for odd $l$:
\begin{equation}
\text{Re}[\tilde{\phi}_{lm}(\mathbf{k})] = \begin{cases}
C_l (-1)^{l/2} Y_{lm}(\hat{\mathbf{k}}) I_{l,\sigma}(k) & \text{if } l \text{ even} \\
0 & \text{if } l \text{ odd}
\end{cases}
\end{equation}
\begin{equation}
\text{Im}[\tilde{\phi}_{lm}(\mathbf{k})] = \begin{cases}
0 & \text{if } l \text{ even} \\
-C_l (-1)^{(l-1)/2} Y_{lm}(\hat{\mathbf{k}}) I_{l,\sigma}(k) & \text{if } l \text{ odd}
\end{cases}
\end{equation}

\subsubsection{Coulomb Operator in Reciprocal Space}

The Coulomb potential corresponding to density $\tilde{\rho}(\mathbf{k})$ is obtained by applying the bare Coulomb kernel:
\begin{equation}
\tilde{V}(\mathbf{k}) = \frac{4\pi}{k^2} \tilde{\rho}(\mathbf{k})
\end{equation}
The electrostatic energy is the inner product of density and potential in reciprocal space:
\begin{equation}
E_{\text{elec}}^{\text{k-space}} = \frac{\Omega}{2} \int_{\mathbb{R}^3} \frac{d^3k}{(2\pi)^3} \tilde{\rho}^*(\mathbf{k}) \tilde{V}(\mathbf{k})
\end{equation}
Discretising over the k-point grid and noting that $|\tilde{\rho}(\mathbf{k})|^2 = \text{Re}[\tilde{\rho}]^2 + \text{Im}[\tilde{\rho}]^2$, we obtain:
\begin{equation}
E_{\text{elec}}^{\text{k-space}} = \frac{\Omega}{(2\pi)^6} \sum_{\substack{\mathbf{k}\neq \mathbf{0}\\ \mathbf{k}\in \text{half-grid}}} \frac{4\pi}{k^2} \left( \text{Re}[\tilde{\rho}(\mathbf{k})]^2 + \text{Im}[\tilde{\rho}(\mathbf{k})]^2 \right)
\end{equation}
The factor of $(2\pi)^6$ arises from the double Fourier transform in the convolution. The $\mathbf{k}=\mathbf{0}$ term is excluded because it corresponds to the interaction of the uniform background charge density and vanishes for neutral systems. For systems evaluated in a sufficiently large cell, the k-space sum converges exponentially with $k_{\text{cutoff}}$.

\subsubsection{Projected Potential Features in Reciprocal Space}

The projected potential features are evaluated with the same reciprocal-space representation. Given $\tilde{V}(\mathbf{k})$, the projection on receiver basis function $(n,l,m)$ centred at atom $i$ is
\begin{equation}
v_{i,nlm}
=
\frac{1}{(2\pi)^3}\left(
\Re\!\left[\tilde{V}^*(\mathbf{0})\,\tilde{\phi}_{nlm}(\mathbf{0})\right]
+2\sum_{\substack{\mathbf{k}\neq \mathbf{0}\\ \mathbf{k}\in \text{half-grid}}}
\Re\!\left[
\tilde{V}^*(\mathbf{k})\,
\tilde{\phi}_{nlm}(\mathbf{k})\,
e^{-i\mathbf{k}\cdot\mathbf{r}_i}
\right]\right),
\label{eq:si_periodic_projected_potential}
\end{equation}
where ``half-grid'' denotes the same open-half reciprocal set used throughout the reciprocal-space evaluation; the $\mathbf{k}=\mathbf{0}$ term is treated separately to avoid double counting.

When self-interaction features are not included, the on-site overlap contribution is subtracted from the projected coefficients. For molecular and slab systems evaluated with the reciprocal-space solver, we then add the same finite-size correction fields (Makov--Payne--Dabo for molecules, slab dipole correction for 2D-periodic cells), projected onto the receiver basis, before combining them with the reciprocal-space contribution.

\subsubsection{Finite-Size Corrections: Makov-Payne-Dabo Formalism}

When computing energies of isolated molecules or clusters within periodic boundary conditions, artificial interactions with periodic images must be corrected. Following Makov and Payne [Phys. Rev. B 51, 4014 (1995)] and Dabo et al. [Phys. Rev. B 77, 115139 (2008)], we apply multipole-based corrections.

The leading monopole correction accounts for the spurious self-interaction of the net charge $Q$ with its periodic images:
\begin{equation}
\Delta E_{\text{mono}} = \frac{\alpha_{\text{M}}}{2L} Q^2
\end{equation}
where $L = \Omega^{1/3}$ and $\alpha_{\text{M}} = 2.837297$ is the Madelung constant for a simple cubic lattice, approximating the long-range electrostatic potential experienced by a point charge in a periodic cubic array.

The dipole correction arises from the quadrupole interaction energy of the system with the field generated by its periodic dipole images:
\begin{equation}
\Delta E_{\text{dip}} = \frac{2\pi}{3\Omega} |\mathbf{p}|^2
\end{equation}
where $\mathbf{p} = \sum_i q_i \mathbf{r}_i + \sum_i \mathbf{p}_i^{\text{local}}$ is the total dipole moment, including both charge-position contributions and intrinsic atomic dipoles (for $l_{\text{max}} \geq 1$). This correction depends on the choice of origin; we define $\mathbf{r}_i$ relative to the centre of mass of the atoms.

The isotropic quadrupole correction accounts for the finite spatial extent of the charge distribution:
\begin{equation}
\Delta E_{\text{quad}} = -\frac{2\pi Q}{3\Omega} \text{Tr}[\mathbf{Q}]
\end{equation}
where the trace of the quadrupole tensor is $\text{Tr}[\mathbf{Q}] = \sum_i q_i r_i^2 + 2 \sum_i \mathbf{r}_i \cdot \mathbf{p}_i^{\text{local}}$.

These corrections are applied selectively based on the periodicity tensor \texttt{pbc}: for fully non-periodic systems (\texttt{pbc} = [False, False, False]), all three corrections are applied; for periodic systems (\texttt{pbc} = [True, True, True]), none are applied; for slab geometries (\texttt{pbc} = [True, True, False]), a dipole correction along the non-periodic direction is applied using the slab-specific formalism of Bengtsson [Phys. Rev. B 59, 12301 (1999)].

\clearpage 
\begin{figure*}[t]
\centering
\includegraphics[width=\linewidth]{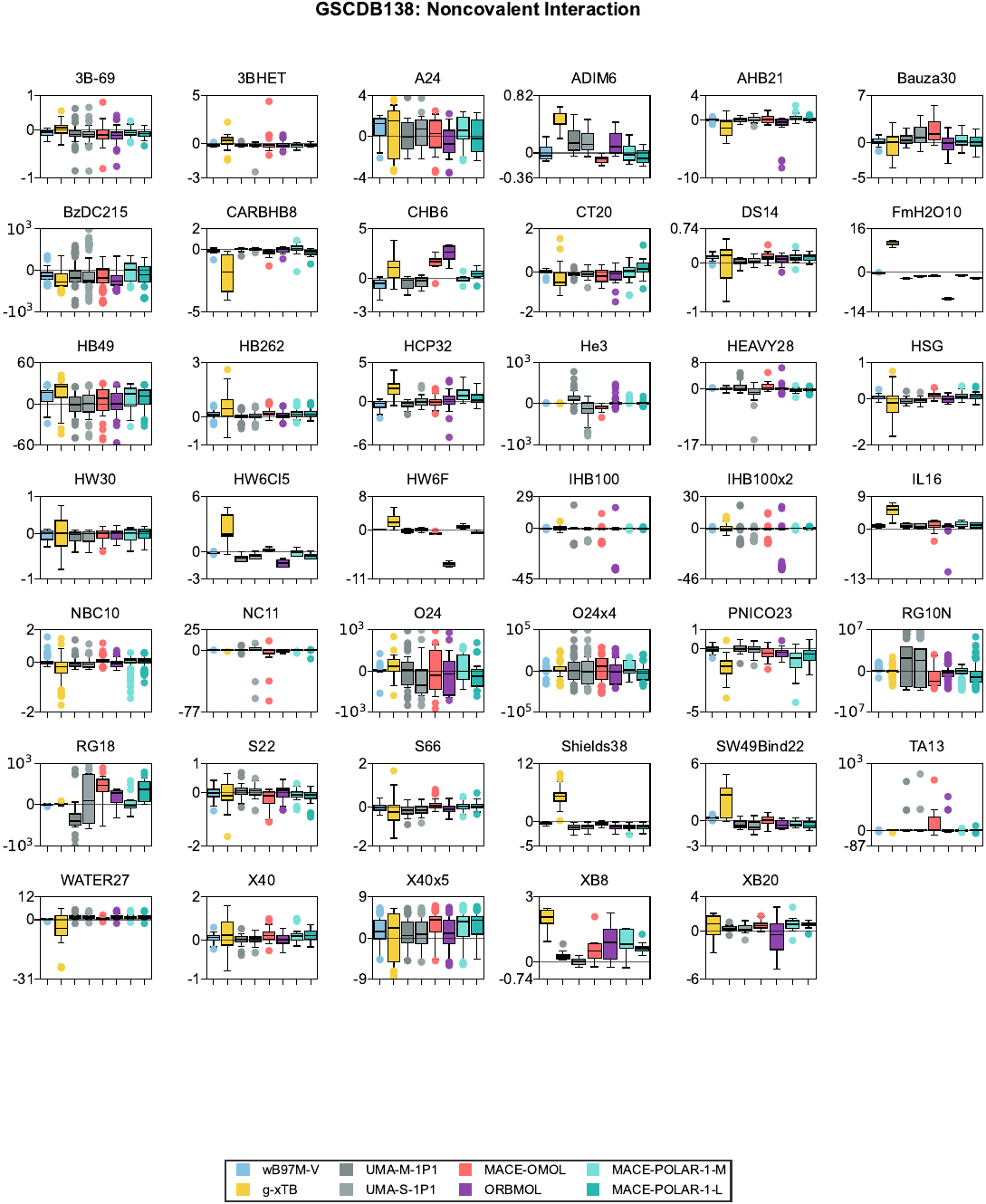}
\caption{Distribution of signed interaction‑energy errors (kcal mol$^{-1}$) across non-covalent benchmarks. Boxplots compare $\omega$B97M‑V, r2SCAN, \UMAM, \UMAS, \MFPL, and \MFPM. Panels use symmetric log scaling only
when outliers require it; y‑axis ticks are placed at the lower bound, 0 (if within range), and the upper bound.}
\label{fig:gscdb138_nc}
\end{figure*}

\begin{figure*}[t]
\centering
\includegraphics[width=\linewidth]{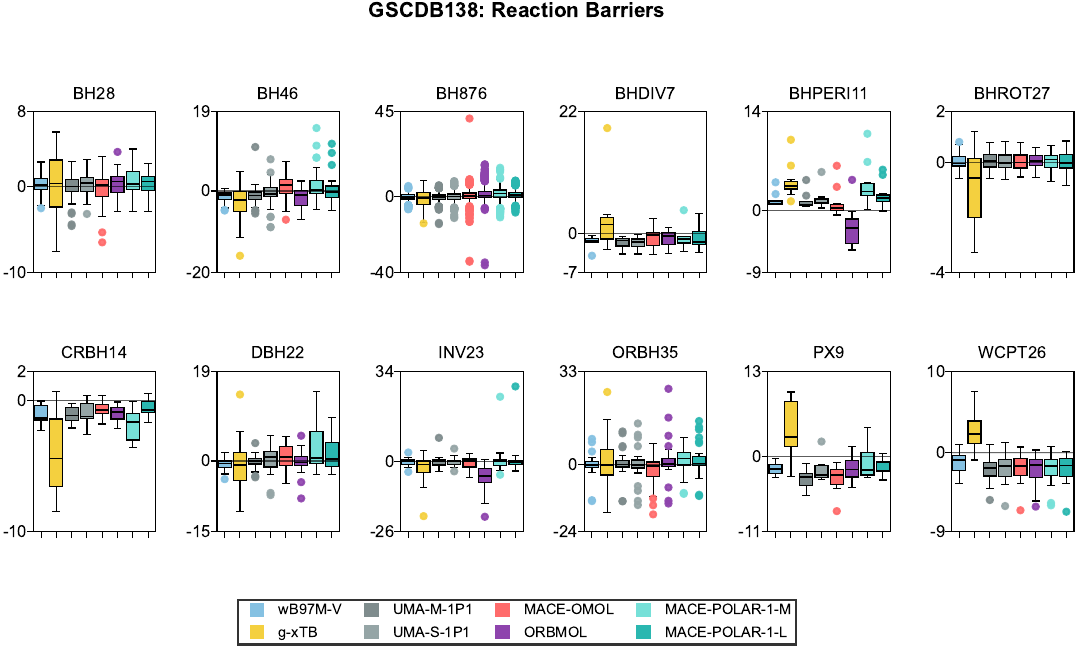}
\caption{Distribution of signed errors (kcal mol$^{-1}$) for the GSCDB138 reaction‑barrier benchmarks. Each panel shows a dataset; boxplots compare six methods ($\omega$B97M‑V, r2SCAN, \UMAM, \UMAS, \MFPL, \MFPM). Boxes
denote the interquartile range with median; whiskers extend to 1.5×IQR; outliers are small, colour‑matched dots. The y‑axis uses a symmetric logarithmic scale when extreme outliers are present; major ticks are placed at the lower bound, 0 (if
within range), and the upper bound.}
\label{fig:gscdb138_barriers}
\end{figure*}

\begin{figure*}[t]
\centering
\includegraphics[width=\linewidth]{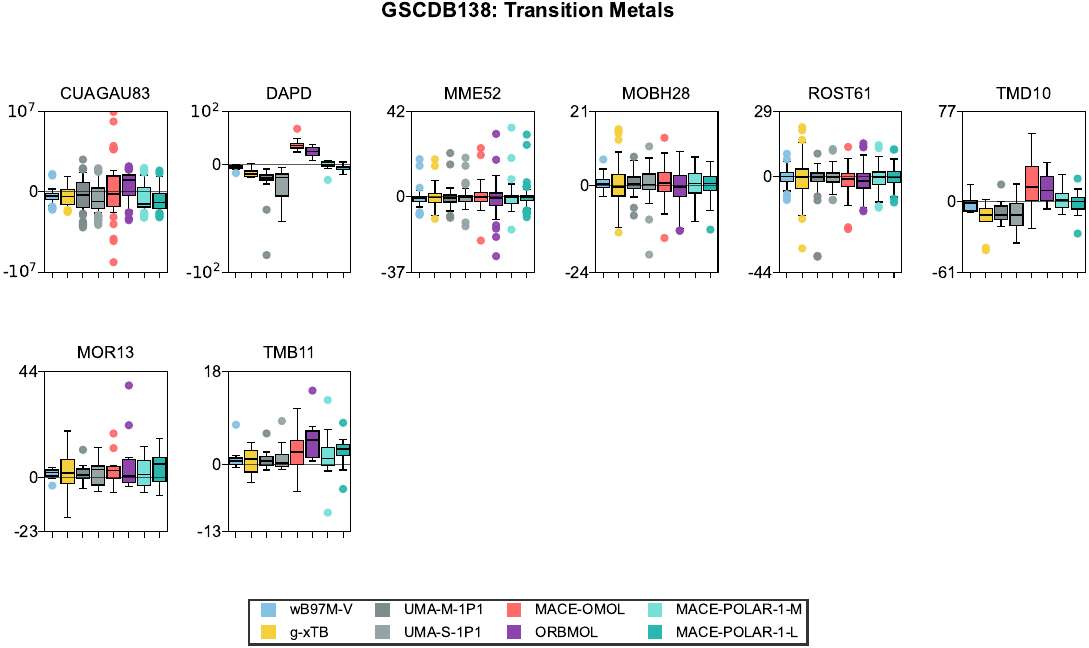}
\caption{Distribution of signed errors (kcal mol$^{-1}$) for transition‑metal datasets in GSCDB138, comparing $\omega$B97M‑V, r2SCAN, \UMAM, \UMAS, \MFPL, and \MFPM.}
\label{fig:gscdb138_tm}
\end{figure*}

\subsection{Fukui functions and conceptual DFT}
\label{sec:fukui-derivation}

This section provides the theoretical basis for the Fukui charge equilibration scheme (Equations~\ref{eq:fukui-eq-up} and \ref{eq:fukui-eq-down} in the main text) by connecting it to the framework of conceptual density functional theory~\cite{parr1983,geerlings2003conceptual}. In conceptual DFT, the Fukui function $f(\mathbf{r})$ describes the local response of the electron density to a change in the total number of electrons $N$ at fixed external potential $v(\mathbf{r})$:
\begin{equation}
f(\mathbf{r}) = \left( \frac{\partial \rho(\mathbf{r})}{\partial N} \right)_{v(\mathbf{r})}
\end{equation}
The Fukui function identifies reactive sites in molecules: regions where $f(\mathbf{r})$ is large are preferential sites for nucleophilic or electrophilic attack, depending on whether electrons are being added or removed.

In our formalism, the charge and spin densities are expanded in atom-centred multipoles (Equations~\ref{eq:rho-up}--\ref{eq:gaussian-basis}), subject to global normalisation constraints:
\begin{align}
    \int s(\mathbf{r})d\mathbf{r} = S, \quad \int \rho(\mathbf{r})d\mathbf{r} = Q \implies S = \sum_i \left(p_{i,00}^{\uparrow} - p_{i,00}^{\downarrow}\right), \quad  Q = \sum_i \left(p_{i,00}^{\uparrow} + p_{i,00}^{\downarrow}\right)
\end{align}
where $Q$ is the total charge, $S$ is the total spin, and $p_{i,00}^{\uparrow\downarrow}$ are the spin-up and spin-down monopole coefficients (atomic charges) for atom $i$. Following the conceptual DFT framework, we define an atom-condensed Fukui function $f_{i,Q}$ that describes how the partial charge on atom $i$ responds to a change in the total charge $Q$. In order to integrate the spin multiplicity, we split the total charge into two quantities, the spin-up total charge $Q^{\uparrow}=\frac{Q + S}{2}$ and the spin-down total charge $Q^{\downarrow}=\frac{Q - S}{2}$, and combine them in the array $Q^{\uparrow\downarrow} = (Q^{\uparrow}, Q^{\downarrow})$ following the convention in the main text. Integrating the Fukui function over some atom-centred partitioning $\Omega_i$ and applying the chain rule through the chemical potential $\mu$:
\begin{align}
    f_{i,Q^{\uparrow}}^{\uparrow} = \int_{\Omega_i} \frac{\partial \rho(\mathbf{r})^{\uparrow}}{\partial Q^{\uparrow}} d\mathbf{r} = \frac{\partial p_{i,00}^{\uparrow}}{\partial Q^{\uparrow}} = \frac{\frac{\partial p_{i,00}^{\uparrow}}{\partial \mu}}{\frac{\partial Q^{\uparrow}}{\partial \mu}} 
    \label{eq:fukui-cdft-up}
    \\
    f_{i,Q^{\downarrow}}^{\downarrow} = \int_{\Omega_i} \frac{\partial \rho(\mathbf{r})^{\downarrow}}{\partial Q^{\downarrow}} d\mathbf{r} = \frac{\partial p_{i,00}^{\downarrow}}{\partial Q^{\downarrow}} = \frac{\frac{\partial p_{i,00}^{\downarrow}}{\partial \mu}}{\frac{\partial Q^{\downarrow}}{\partial \mu}}
    \label{eq:fukui-cdft-down}
\end{align}
The second equality in both \ref{eq:fukui-cdft-up} and \ref{eq:fukui-cdft-down} uses the fact that the monopole coefficient $p_{i,00}^{\uparrow \downarrow}$ represents the integrated spin-charge on atom $i$ for each spin channel, while the third equality applies the chain rule.

We define the unnormalised Fukui coefficient $f_i^{(u), \uparrow \downarrow}$ as the sensitivity of the atomic monopole to the chemical potential:
\begin{align}
    f_i^{(u),\uparrow } = \frac{\partial p_{i,00}^{(u),\uparrow}}{\partial \mu}, \quad   f_i^{(u),\downarrow} = \frac{\partial p_{i,00}^{(u), \downarrow}}{\partial \mu}, \quad f_i^{(u),\uparrow \downarrow} = (f_i^{(u),\uparrow } , f_i^{(u),\downarrow })
\end{align}
In classical electronegativity equalisation models, this quantity is related to the atomic softness~\cite{mortier1986electronegativity,rappe1991charge}. In our model, $f_i^{(u), \uparrow \downarrow}$ is predicted by a neural network and can depend on both local geometry and non-local electrostatic environment for $u>0$. The total charge constraint requires:
\begin{align}
    Q^{\uparrow \downarrow} = \sum_{i} p_{i,00}^{\uparrow \downarrow} \implies \frac{\partial Q^{\uparrow \downarrow}}{\partial \mu} = \sum_i \frac{\partial p_{i,00}^{\uparrow \downarrow}}{\partial \mu} = \sum_i f_i^{(u), \uparrow \downarrow} .
\end{align}
This shows that the global softness (derivative of total charge with respect to chemical potential) is the sum of atomic softnesses. To enforce the charge constraint, we compute how much the current predicted charges deviate from the target spin-resolved total charges $Q^{\uparrow \downarrow}$ and redistribute this error according to the Fukui coefficients. Using a first-order Taylor expansion:
\begin{align}
    \Delta p_{i,00}^{(u), \uparrow \downarrow} \approx \frac{\partial  p_{i,00}^{(u), \uparrow \downarrow}}{\partial Q^{\uparrow \downarrow}} \Delta Q^{\uparrow \downarrow} = \frac{f_i^{(u), \uparrow \downarrow}}{\sum_j f_j^{(u), \uparrow \downarrow}} \Delta Q^{\uparrow \downarrow}
\end{align}
where $\Delta Q^{\uparrow \downarrow} = Q_{\text{target}}^{\uparrow \downarrow} - \sum_j p_{j,00}^{(u), \uparrow \downarrow}$ is the spin-charge deficit or surplus. This yields the equilibration formula used in Equations~\ref{eq:fukui-eq-up} and \ref{eq:fukui-eq-down}, with the normalised spin-resolved Fukui functions $f_i^{(u),\uparrow}/\sum_j f_j^{(u),\uparrow}$ and $f_i^{(u),\downarrow}/\sum_j f_j^{(u),\downarrow}$ acting as learnable weights that distribute charge corrections across atoms according to their chemical softness.

\definecolor{Gray}{gray}{0.85}
\newcolumntype{a}{>{\columncolor{Gray}}c}
\newcolumntype{B}{>{\columncolor{white}}c}

\begin{table*}[b]
\centering
\caption{Calculated densities of organic liquids. Experimental temperatures and densities for each liquid are also given in the table. NPT simulations were performed on systems of around 1000 atoms, with 500 ps equilibration and 500 ps production runs.}
\label{tab:liquid_densities}
\resizebox{0.9\textwidth}{!}{%
\begin{tabular}{@{}laBaBaB@{}}
\toprule
Liquid & Exp. T, K & Exp. density, g cm$^{-3}$ & \MOMOL & \MFPM & \MFPL & \UMAS \\
\midrule

1,2-dichloroethane & 303.0 & 1.245 & 1.354  & 1.370 & 1.333 & 1.369 \\
1,2-dimethoxy-ethane & 303.0 & 0.864 & 0.983  & 0.963 & 0.971 & 0.956 \\
o-xylene & 278.0 & 0.880 & 1.031  & 0.965 & 0.978 & 0.984 \\
1,3-dioxolane & 298.0 & 1.060 & 1.166  & 1.169 & 1.150 & 1.176 \\
1,4-dioxane & 298.0 & 1.034 & 1.168  & 1.134 & 1.147 & 1.160 \\
1-butanol & 298.0 & 0.809 & 0.916  & 0.901 & 0.891 & 0.894 \\
1-octanol & 298.0 & 0.826 & 0.931  & 0.917 & 0.904 & 0.919 \\
    1-propanol & 303.0 & 0.800 & 0.908  & 0.890 & 0.874 & 0.881 \\
2-butanone & 303.0 & 0.800 & 0.936  & 0.895 & 0.849 & 0.895 \\
t-butyl-alcohol & 298.0 & 0.789 & 0.894  & 0.865 & 0.864 & 0.876 \\
Fluoroethylene carbonate & 298.0 & 1.485 & 1.616  & 1.612 & 1.468 & 1.600 \\
N,N-dimethylformamide & 303.0 & 0.945 & 1.044  & 1.028 & 1.026 & 1.031 \\
N-methyl-2-pyrrolidone & 303.0 & 1.023 & 1.130  & 1.115 & 1.108 & 1.111 \\
acetic acid & 303.0 & 1.045 & 1.198  & 1.187 & 1.056 & 1.170 \\
acetone & 303.0 & 0.784 & 0.930  & 0.872 & 0.837 & 0.879 \\
acetonitrile & 298.0 & 0.786 & 0.623  & 0.863 & 0.782 & 0.873 \\
acetophenone & 298.0 & 1.028 & 1.166  & 1.113 & 1.075 & 1.117 \\
benzamide & 408.0 & 1.079 & 1.192  & 1.143 & 1.106 & 1.167 \\
benzene & 298.0 & 0.876 & 0.994  & 0.954 & 0.968 & 0.961 \\
benzenethiol & 298.0 & 1.077 & 1.204  & 1.132 & 1.149 & 1.161 \\
bromobenzene & 298.0 & 1.495 & 1.675  & 1.592 & 1.579 & 1.621 \\
carbon tetrachloride & 298.0 & 1.594 & 1.672  & 1.611 & 1.669 & 1.746 \\
chlorobenzene & 298.0 & 1.106 & 1.242  & 1.197 & 1.187 & 1.221 \\
chloroform & 303.0 & 1.479 & 1.565  & 1.536 & 1.534 & 1.595 \\
cyclohexane & 303.0 & 0.774 & 0.916  & 0.883 & 0.889 & 0.888 \\
dibromomethane & 298.0 & 2.497 & 2.668  & 2.693 & 2.562 & 2.668 \\
dichloromethane & 298.0 & 1.327 & 1.399  & 1.438 & 1.418 & 1.440 \\
diethyl carbonate & 303.0 & 0.969 & 1.113  & 1.082 & 1.060 & 1.079 \\
diethyl ether & 298.0 & 0.714 & 0.841  & 0.801 & 0.815 & 0.809 \\
diethylene glycol & 293.0 & 1.120 & 1.230  & 1.213 & 1.198 & 1.206 \\
diglyme & 298.0 & 0.943 & 1.062  & 1.046 & 1.052 & 1.049 \\
dimethyl carbonate & 303.0 & 1.064 & 1.211  & 1.194 & 1.180 & 1.183 \\
dimethyl sulfide & 298.0 & 0.848 & 0.956  & 0.916 & 0.927 & 0.929 \\
dimethyl sulfoxide & 303.0 & 1.101 & 1.205  & 1.172 & 1.217 & 1.193 \\
ethanol & 298.0 & 0.789 & 0.898  & 0.880 & 0.861 & 0.873 \\
ethyl acetate & 298.0 & 0.900 & 1.071  & 1.036 & 1.002 & 1.038 \\
ethylene carbonate & 317.0 & 1.321 & 1.434  & 1.427 & 1.361 & 1.429 \\
ethylene glycol & 298.0 & 1.113 & 1.212  & 1.201 & 1.174 & 1.191 \\
ethylmethyl carbonate & 298.0 & 1.012 & 1.156  & 1.128 & 1.110 & 1.119 \\
fluorobenzene & 298.0 & 1.022 & 1.185  & 1.091 & 1.087 & 1.129 \\
formaldehyde & 258.0 & 0.815 & 0.958  & 0.997 & 1.058 & 0.983 \\
glycerol & 298.0 & 1.261 & 1.346  & 1.328 & 1.317 & 1.325 \\
hexamethylphosphoramide & 298.0 & 1.030 & 1.152  & 1.114 & 1.150 & 1.116 \\
hexane & 298.0 & 0.661 & 0.773  & 0.742 & 0.741 & 0.749 \\
methanethiol & 298.0 & 0.867 & 0.997  & 0.925 & 0.907 & 0.961 \\
methanol & 298.0 & 0.791 & 0.896  & 0.884 & 0.874 & 0.870 \\
anisole & 298.0 & 0.994 & 1.120  & 1.069 & 1.079 & 1.080 \\
methyl acetate & 298.0 & 0.934 & 1.111  & 1.080 & 1.043 & 1.073 \\
methyl benzoate & 303.0 & 1.084 & 1.230  & 1.175 & 1.139 & 1.184 \\
methyl t-butyl ether & 303.0 & 0.735 & 0.865  & 0.822 & 0.843 & 0.839 \\
morpholine & 298.0 & 1.000 & 1.117  & 1.091 & 1.105 & 1.096 \\
nitromethane & 298.0 & 1.137 & 1.326  & 1.188 & 1.148 & 1.218 \\
nitrobenzene & 298.0 & 1.204 & 1.310  & 1.273 & 1.255 & 1.286 \\
propylene carbonate & 298.0 & 1.205 & 1.324  & 1.295 & 1.240 & 1.312 \\
pyridine & 298.0 & 0.982 & 1.090  & 1.073 & 1.066 & 1.068 \\
quinoline & 293.0 & 1.098 & 1.216  & 1.182 & 1.171 & 1.185 \\
tetrahydrofuran & 303.0 & 0.883 & 1.000  & 0.973 & 0.967 & 0.993 \\
thioanisole & 298.0 & 1.058 & 1.180  & 1.123 & 1.134 & 1.144 \\
thiophene & 298.0 & 1.065 & 1.158  & 1.111 & 1.085 & 1.175 \\
toluene & 298.0 & 0.867 & 0.999  & 0.926 & 0.944 & 0.947 \\
triethylamine & 298.0 & 0.728 & 0.853  & 0.817 & 0.836 & 0.830 \\
vinylene carbonate & 303.0 & 1.350 & 1.522  & 1.412 & 1.386 & 1.495 \\

\bottomrule
\end{tabular}%
}
\hspace{5pt}
\end{table*}

\begin{figure*}[t]
\centering
\includegraphics[width=\linewidth]{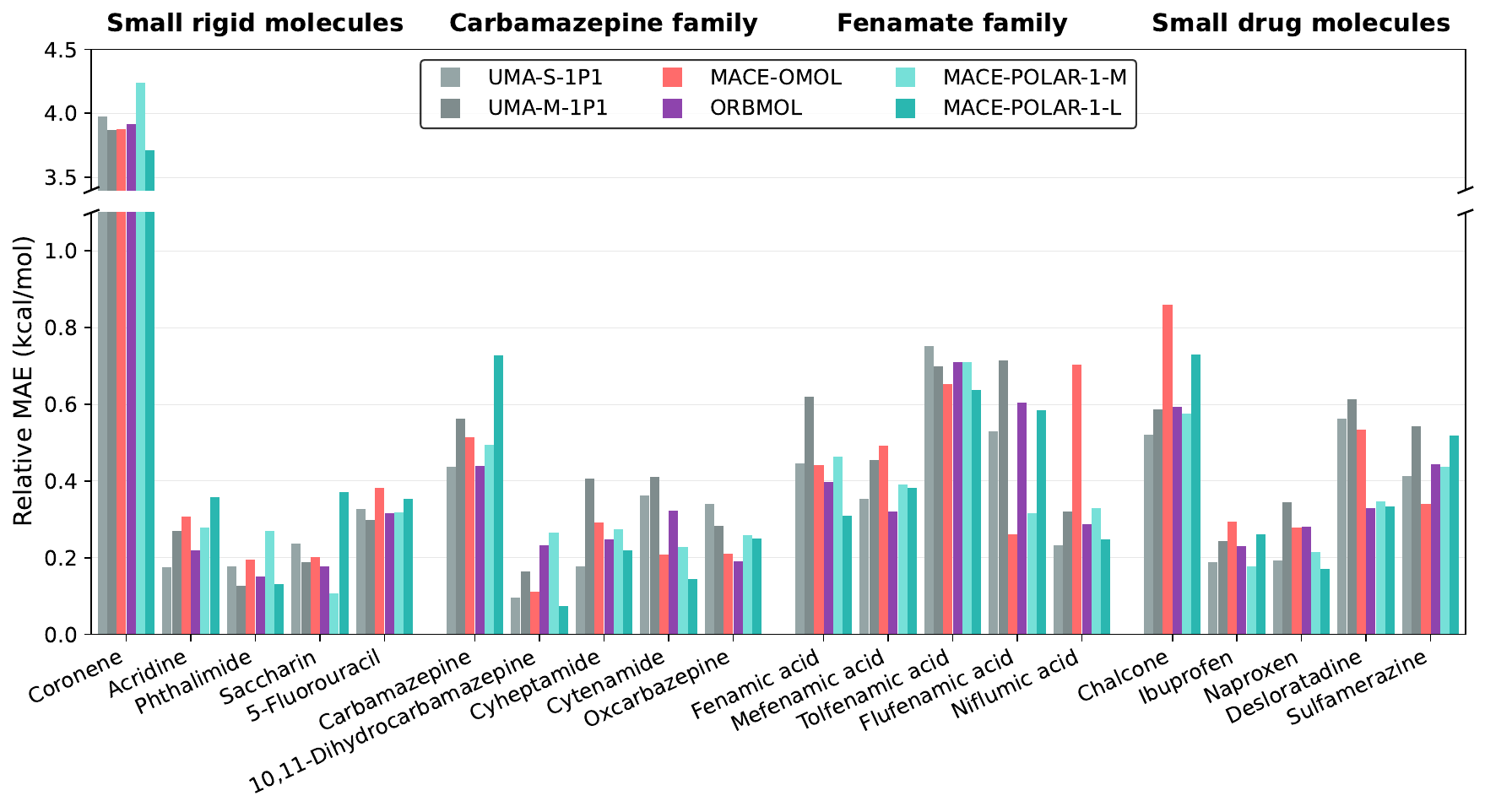}
\caption{\textbf{Relative lattice energy errors for CPOSS209 molecular crystals.}
Bar heights show mean absolute errors in kcal/mol for predicted relative lattice energies (polymorph energy differences), grouped by molecular family. The dataset comprises 209 experimental and predicted polymorphs from 20 small drug molecules. Reference calculations combine $\omega$B97M-D3 crystal-phase energies with 1-body CCSD(T) corrections. Lower values indicate better accuracy.}
\label{fig:cposs-relative-mae}
\end{figure*}

\clearpage

\newcommand{\benchcol}[2]{\parbox[t]{#1}{#2}}
\newcommand{\benchcat}[1]{\benchcol{2.8cm}{#1}}
\newcommand{\benchset}[1]{\benchcol{3.1cm}{#1}}
\newcommand{\benchsize}[1]{\benchcol{2.2cm}{#1}}
\newcommand{\benchdesc}[1]{\benchcol{4.6cm}{#1}}
\newcommand{\benchref}[1]{\benchcol{2.4cm}{#1}}
\newcommand{\benchcite}[1]{\benchcol{1.8cm}{#1}}

\begin{longtable*}{l l l l l l}
\caption{Benchmark datasets used in this work.}
\label{tab:benchmarks_full}\\
\toprule
\benchcat{Category} & \benchset{Dataset} & \benchsize{Size (structures)} & \benchdesc{Description} & \benchref{Reference level} & \benchcite{Citation} \\
\midrule
\endfirsthead
\toprule
\benchcat{Category} & \benchset{Dataset} & \benchsize{Size (structures)} & \benchdesc{Description} & \benchref{Reference level} & \benchcite{Citation} \\
\midrule
\endhead
\midrule
\multicolumn{6}{r}{\emph{Continued on next page}}\\
\midrule
\endfoot
\bottomrule
\endlastfoot

\benchcat{\textbf{Barrier heights (GSCDB)}} & \benchset{BH28 (GSCDB)} & \benchsize{28} & \benchdesc{Barrier heights subset} & \benchref{CCSD(T)/CBS} & \benchcite{\cite{Liang2025,Karton_2019}} \\
\benchcat{} & \benchset{BH46 (GSCDB)} & \benchsize{46} & \benchdesc{Barrier heights} & \benchref{CCSD(T)/CBS} & \benchcite{\cite{Liang2025,Zhao2008BH76,zhao2005multi,zhao2005benchmark}} \\
\benchcat{} & \benchset{BH876 (GSCDB)} & \benchsize{876} & \benchdesc{Barrier heights} & \benchref{CCSD(T)/CBS} & \benchcite{\cite{Liang2025,Prasad2021}} \\
\benchcat{} & \benchset{BHDIV7 (GSCDB)} & \benchsize{7} & \benchdesc{Barrier heights} & \benchref{CCSD(T)/CBS} & \benchcite{\cite{Liang2025,Goerigk2017}} \\
\benchcat{} & \benchset{BHPERI11 (GSCDB)} & \benchsize{11} & \benchdesc{Pericyclic barriers} & \benchref{CCSD(T)/CBS} & \benchcite{\cite{Liang2025,goerigk2010general}} \\
\benchcat{} & \benchset{BHROT27 (GSCDB)} & \benchsize{27} & \benchdesc{Rotational barriers} & \benchref{CCSD(T)/CBS} & \benchcite{\cite{Liang2025,Goerigk2017}} \\
\benchcat{} & \benchset{CRBH14 (GSCDB)} & \benchsize{14} & \benchdesc{Cycloreversion barriers} & \benchref{CCSD(T)/CBS} & \benchcite{\cite{Liang2025,Yu2015CRBH}} \\
\benchcat{} & \benchset{DBH22 (GSCDB)} & \benchsize{22} & \benchdesc{BH76 subset} & \benchref{CCSD(T)/CBS} & \benchcite{\cite{Liang2025,Karton2008DBH,Zheng_2007}} \\
\benchcat{} & \benchset{INV23 (GSCDB)} & \benchsize{23} & \benchdesc{Inversion/racemisation barriers} & \benchref{CCSD(T)/CBS} & \benchcite{\cite{Liang2025,Goerigk_2016}} \\
\benchcat{} & \benchset{ORBH35 (GSCDB)} & \benchsize{35} & \benchdesc{Oxygen reaction barriers} & \benchref{CCSD(T)/CBS} & \benchcite{\cite{Liang2025,Chan2018}} \\
\benchcat{} & \benchset{PX9 (GSCDB)} & \benchsize{9} & \benchdesc{Proton-exchange barriers} & \benchref{CCSD(T)/CBS} & \benchcite{\cite{Liang2025,Karton2012PX}} \\
\benchcat{} & \benchset{WCPT26 (GSCDB)} & \benchsize{26} & \benchdesc{Water-catalyzed proton transfer} & \benchref{CCSD(T)/CBS} & \benchcite{\cite{Liang2025,Karton2012WCPT}} \\

\benchcat{\textbf{Electric field props (GSCDB)}} & \benchset{Dip146 (GSCDB)} & \benchsize{190} & \benchdesc{Dipole moments} & \benchref{CCSD(T)/CBS} & \benchcite{\cite{Liang2025,hait2018dipole}} \\
\benchcat{} & \benchset{HR46 (GSCDB)} & \benchsize{128} & \benchdesc{Polarizabilities} & \benchref{CCSD(T)/CBS} & \benchcite{\cite{Liang2025,hickey2014benchmarking}} \\
\benchcat{\textbf{Isomerisation (GSCDB)}} & \benchset{AlkIsomer11 (GSCDB)} & \benchsize{11} & \benchdesc{Alkane isomerisation} & \benchref{CCSD(T)/CBS} & \benchcite{\cite{Liang2025,Karton_2009}} \\
\benchcat{} & \benchset{C20C246 (GSCDB)} & \benchsize{6} & \benchdesc{C20/C24 isomers} & \benchref{CCSD(T)/CBS} & \benchcite{\cite{Liang2025,Manna_2016}} \\
\benchcat{} & \benchset{C60ISO7 (GSCDB)} & \benchsize{7} & \benchdesc{C60 isomers} & \benchref{CCSD(T)/CBS} & \benchcite{\cite{Liang2025,Sure_2017}} \\
\benchcat{} & \benchset{DIE60 (GSCDB)} & \benchsize{60} & \benchdesc{Conjugated diene migration} & \benchref{CCSD(T)/CBS} & \benchcite{\cite{Liang2025,Yu_2014}} \\
\benchcat{} & \benchset{EIE22 (GSCDB)} & \benchsize{22} & \benchdesc{Enecarbonyl isomers} & \benchref{CCSD(T)/CBS} & \benchcite{\cite{Liang2025,Yu_2015An}} \\
\benchcat{} & \benchset{ISO34 (GSCDB)} & \benchsize{34} & \benchdesc{Small/medium isomers} & \benchref{CCSD(T)/CBS} & \benchcite{\cite{Liang2025,grimme2007compute}} \\
\benchcat{} & \benchset{ISOL23 (GSCDB)} & \benchsize{23} & \benchdesc{Large organic isomers} & \benchref{CCSD(T)/CBS} & \benchcite{\cite{Liang2025,huenerbein2010effects}} \\
\benchcat{} & \benchset{ISOMERIZATION20 (GSCDB)} & \benchsize{20} & \benchdesc{W4-11 isomerisation} & \benchref{CCSD(T)/CBS} & \benchcite{\cite{Liang2025,karton2011w4}} \\
\benchcat{} & \benchset{PArel (GSCDB)} & \benchsize{20} & \benchdesc{Protonated isomers} & \benchref{CCSD(T)/CBS} & \benchcite{\cite{Liang2025,Goerigk2017}} \\
\benchcat{} & \benchset{Styrene42 (GSCDB)} & \benchsize{42} & \benchdesc{C$_8$H$_8$ isomers} & \benchref{CCSD(T)/CBS} & \benchcite{\cite{Liang2025,Karton_2012Explicitly}} \\
\benchcat{} & \benchset{TAUT15 (GSCDB)} & \benchsize{15} & \benchdesc{Tautomer isomerisation} & \benchref{CCSD(T)/CBS} & \benchcite{\cite{Liang2025,Goerigk2017}} \\

\benchcat{\textbf{Intramolecular NCI (GSCDB)}} & \benchset{ACONF (GSCDB)} & \benchsize{15} & \benchdesc{Alkane conformers} & \benchref{CCSD(T)/CBS} & \benchcite{\cite{Liang2025,Gruzman_2009}} \\
\benchcat{} & \benchset{Amino20x4 (GSCDB)} & \benchsize{80} & \benchdesc{Amino-acid conformers} & \benchref{CCSD(T)/CBS} & \benchcite{\cite{Liang2025,Kesharwani_2016}} \\
\benchcat{} & \benchset{BUT14DIOL (GSCDB)} & \benchsize{64} & \benchdesc{Butane-1,4-diol confs} & \benchref{CCSD(T)/CBS} & \benchcite{\cite{Liang2025,kozuch2014conformational}} \\
\benchcat{} & \benchset{ICONF (GSCDB)} & \benchsize{17} & \benchdesc{Inorganic conformers} & \benchref{CCSD(T)/CBS} & \benchcite{\cite{Liang2025,Goerigk2017}} \\
\benchcat{} & \benchset{IDISP (GSCDB)} & \benchsize{6} & \benchdesc{Intramolecular dispersion} & \benchref{CCSD(T)/CBS} & \benchcite{\cite{Liang2025,goerigk2010general}} \\
\benchcat{} & \benchset{MCONF (GSCDB)} & \benchsize{51} & \benchdesc{Melatonin conformers} & \benchref{CCSD(T)/CBS} & \benchcite{\cite{Liang2025,fogueri2013melatonin}} \\
\benchcat{} & \benchset{PCONF21 (GSCDB)} & \benchsize{18} & \benchdesc{Peptide conformers} & \benchref{CCSD(T)/CBS} & \benchcite{\cite{Liang2025,vreha2005structure,goerigk2013accurate}} \\
\benchcat{} & \benchset{Pentane13 (GSCDB)} & \benchsize{13} & \benchdesc{n-Pentane torsions} & \benchref{CCSD(T)/CBS} & \benchcite{\cite{Liang2025,Martin_2013}} \\
\benchcat{} & \benchset{SCONF (GSCDB)} & \benchsize{17} & \benchdesc{Sugar conformers} & \benchref{CCSD(T)/CBS} & \benchcite{\cite{Liang2025,csonka2009evaluation}} \\
\benchcat{} & \benchset{UPU23 (GSCDB)} & \benchsize{23} & \benchdesc{RNA-backbone conformers} & \benchref{CCSD(T)/CBS} & \benchcite{\cite{Liang2025,Kruse_2015}} \\

\benchcat{\textbf{Noncovalent (GSCDB)}} & \benchset{3B-69 (GSCDB)} & \benchsize{69} & \benchdesc{Three-body NCIs} & \benchref{CCSD(T)/CBS} & \benchcite{\cite{Liang2025,Rezac2015Benchmark}} \\
\benchcat{} & \benchset{3BHET (GSCDB)} & \benchsize{20} & \benchdesc{Three-body hetero NCIs} & \benchref{CCSD(T)/CBS} & \benchcite{\cite{Liang2025,Ochieng_2023}} \\
\benchcat{} & \benchset{A19Rel6 (GSCDB)} & \benchsize{114} & \benchdesc{A24 PEC relatives} & \benchref{CCSD(T)/CBS} & \benchcite{\cite{Liang2025,Witte_2015}} \\
\benchcat{} & \benchset{A24 (GSCDB)} & \benchsize{24} & \benchdesc{Small NCI dimers} & \benchref{CCSD(T)/CBS} & \benchcite{\cite{Liang2025,Rezac2013Describing}} \\
\benchcat{} & \benchset{ADIM6 (GSCDB)} & \benchsize{6} & \benchdesc{Alkane dimers} & \benchref{CCSD(T)/CBS} & \benchcite{\cite{Liang2025,grimme2010consistent}} \\
\benchcat{} & \benchset{AHB21 (GSCDB)} & \benchsize{21} & \benchdesc{Anion-neutral dimers} & \benchref{CCSD(T)/CBS} & \benchcite{\cite{Liang2025,Lao_2015}} \\
\benchcat{} & \benchset{Bauza30 (GSCDB)} & \benchsize{30} & \benchdesc{Halogen/chalcogen/pnicogen dimers} & \benchref{CCSD(T)/CBS} & \benchcite{\cite{Liang2025,Bauza_2013}} \\
\benchcat{} & \benchset{BzDC215 (GSCDB)} & \benchsize{215} & \benchdesc{Benzene dimer PECs} & \benchref{CCSD(T)/CBS} & \benchcite{\cite{Liang2025,Crittenden_2009}} \\
\benchcat{} & \benchset{CARBHB8 (GSCDB)} & \benchsize{8} & \benchdesc{Carbene–X hydrogen bonds} & \benchref{CCSD(T)/CBS} & \benchcite{\cite{Liang2025,Goerigk2017}} \\
\benchcat{} & \benchset{CHB6 (GSCDB)} & \benchsize{6} & \benchdesc{Cation-neutral dimers} & \benchref{CCSD(T)/CBS} & \benchcite{\cite{Liang2025,Lao_2015}} \\
\benchcat{} & \benchset{CT20 (GSCDB)} & \benchsize{20} & \benchdesc{Charge transfer} & \benchref{CCSD(T)/CBS} & \benchcite{\cite{Liang2025,Steinmann_2012}} \\
\benchcat{} & \benchset{DS14 (GSCDB)} & \benchsize{14} & \benchdesc{Divalent sulfur NCIs} & \benchref{CCSD(T)/CBS} & \benchcite{\cite{Liang2025,Mintz2012}} \\
\benchcat{} & \benchset{FmH2O10 (GSCDB)} & \benchsize{10} & \benchdesc{F$^-$(H$_2$O)$_{10}$ isomers} & \benchref{CCSD(T)/CBS} & \benchcite{\cite{Liang2025,lao2013improved}} \\
\benchcat{} & \benchset{H2O16Rel4 (GSCDB)} & \benchsize{4} & \benchdesc{(H$_2$O)$_{16}$ conformers} & \benchref{CCSD(T)/CBS} & \benchcite{\cite{Liang2025,yoo2010high}} \\
\benchcat{} & \benchset{H2O20Rel9 (GSCDB)} & \benchsize{9} & \benchdesc{(H$_2$O)$_{20}$ conformers} & \benchref{CCSD(T)/CBS} & \benchcite{\cite{Liang2025,kazimirski2003search}} \\
\benchcat{} & \benchset{HB49 (GSCDB)} & \benchsize{49} & \benchdesc{Hydrogen bonds} & \benchref{CCSD(T)/CBS} & \benchcite{\cite{Liang2025,Boese_2015}} \\
\benchcat{} & \benchset{HB262 (GSCDB)} & \benchsize{262} & \benchdesc{Hydrogen bonds} & \benchref{CCSD(T)/CBS} & \benchcite{\cite{Liang2025,Rezac2020}} \\
\benchcat{} & \benchset{HCP32 (GSCDB)} & \benchsize{32} & \benchdesc{Halogen/chalcogen/pnicogen} & \benchref{CCSD(T)/CBS} & \benchcite{\cite{Liang2025,Oliveira_2017}} \\
\benchcat{} & \benchset{He3 (GSCDB)} & \benchsize{49} & \benchdesc{He trimer 3-body NCIs} & \benchref{CCSD(T)/CBS} & \benchcite{\cite{Liang2025,lang2023three}} \\
\benchcat{} & \benchset{HEAVY28 (GSCDB)} & \benchsize{28} & \benchdesc{Heavy-element NCIs} & \benchref{CCSD(T)/CBS} & \benchcite{\cite{Liang2025,grimme2010consistent}} \\
\benchcat{} & \benchset{HSG (GSCDB)} & \benchsize{21} & \benchdesc{Protein-ligand fragments} & \benchref{CCSD(T)/CBS} & \benchcite{\cite{Liang2025,faver2011formal}} \\
\benchcat{} & \benchset{HW30 (GSCDB)} & \benchsize{30} & \benchdesc{Hydrocarbon–water dimers} & \benchref{CCSD(T)/CBS} & \benchcite{\cite{Liang2025,Copeland_2012}} \\
\benchcat{} & \benchset{HW6Cl5 (GSCDB)} & \benchsize{5} & \benchdesc{Cl$^-$–water clusters} & \benchref{CCSD(T)/CBS} & \benchcite{\cite{Liang2025,lao2013improved}} \\
\benchcat{} & \benchset{HW6F (GSCDB)} & \benchsize{6} & \benchdesc{F$^-$–water clusters} & \benchref{CCSD(T)/CBS} & \benchcite{\cite{Liang2025,lao2013improved}} \\
\benchcat{} & \benchset{IHB100 (GSCDB)} & \benchsize{100} & \benchdesc{Ionic hydrogen bonds} & \benchref{CCSD(T)/CBS} & \benchcite{\cite{Liang2025,Rezac2020}} \\
\benchcat{} & \benchset{IHB100x2 (GSCDB)} & \benchsize{200} & \benchdesc{Ionic H-bond PECs} & \benchref{CCSD(T)/CBS} & \benchcite{\cite{Liang2025,Rezac2020}} \\
\benchcat{} & \benchset{IL16 (GSCDB)} & \benchsize{16} & \benchdesc{Ionic liquids} & \benchref{CCSD(T)/CBS} & \benchcite{\cite{Liang2025,Lao_2015}} \\
\benchcat{} & \benchset{NBC10 (GSCDB)} & \benchsize{184} & \benchdesc{NCI PECs (benzene etc.)} & \benchref{CCSD(T)/CBS} & \benchcite{\cite{Liang2025,takatani2007performance,hohenstein2009effects}} \\
\benchcat{} & \benchset{NC11 (GSCDB)} & \benchsize{11} & \benchdesc{Small NCIs} & \benchref{CCSD(T)/CBS} & \benchcite{\cite{Liang2025,Smith_2014}} \\
\benchcat{} & \benchset{O24 (GSCDB)} & \benchsize{24} & \benchdesc{Open-shell dimers} & \benchref{CCSD(T)/CBS} & \benchcite{\cite{Liang2025,madajczyk2021dataset}} \\
\benchcat{} & \benchset{O24x4 (GSCDB)} & \benchsize{96} & \benchdesc{Open-shell PECs} & \benchref{CCSD(T)/CBS} & \benchcite{\cite{Liang2025,madajczyk2021dataset}} \\
\benchcat{} & \benchset{PNICO23 (GSCDB)} & \benchsize{23} & \benchdesc{Pnicogen NCIs} & \benchref{CCSD(T)/CBS} & \benchcite{\cite{Liang2025,setiawan2015strength}} \\
\benchcat{} & \benchset{RG10N (GSCDB)} & \benchsize{275} & \benchdesc{Rare-gas dimer PECs} & \benchref{CCSD(T)/CBS} & \benchcite{\cite{Liang2025,przybytek2017pair}} \\
\benchcat{} & \benchset{RG18 (GSCDB)} & \benchsize{18} & \benchdesc{Rare-gas complexes} & \benchref{CCSD(T)/CBS} & \benchcite{\cite{Liang2025,Goerigk2017}} \\
\benchcat{} & \benchset{S22 (GSCDB)} & \benchsize{22} & \benchdesc{NCI dimers} & \benchref{CCSD(T)/CBS} & \benchcite{\cite{Liang2025,jurevcka2006benchmark}} \\
\benchcat{} & \benchset{S66 (GSCDB)} & \benchsize{66} & \benchdesc{NCI dimers} & \benchref{CCSD(T)/CBS} & \benchcite{\cite{Liang2025,Rezac2011_1}} \\
\benchcat{} & \benchset{S66Rel7 (GSCDB)} & \benchsize{462} & \benchdesc{PECs for S66} & \benchref{CCSD(T)/CBS} & \benchcite{\cite{Liang2025,Rezac2011_1}} \\
\benchcat{} & \benchset{Shields38 (GSCDB)} & \benchsize{38} & \benchdesc{Water clusters (H$_2$O)$_n$} & \benchref{CCSD(T)/CBS} & \benchcite{\cite{Liang2025,Temelso_2011}} \\
\benchcat{} & \benchset{SW49Bind22 (GSCDB)} & \benchsize{22} & \benchdesc{SO$_4^{2-}$(H$_2$O)$_n$ binding} & \benchref{CCSD(T)/CBS} & \benchcite{\cite{Liang2025,Mardirossian_2013}} \\
\benchcat{} & \benchset{SW49Rel28 (GSCDB)} & \benchsize{28} & \benchdesc{SO$_4^{2-}$(H$_2$O)$_n$ conformers} & \benchref{CCSD(T)/CBS} & \benchcite{\cite{Liang2025,Mardirossian_2013}} \\
\benchcat{} & \benchset{TA13 (GSCDB)} & \benchsize{13} & \benchdesc{Radical dimers} & \benchref{CCSD(T)/CBS} & \benchcite{\cite{Liang2025,Tentscher_2013}} \\
\benchcat{} & \benchset{WATER27 (GSCDB)} & \benchsize{27} & \benchdesc{Water clusters/ions} & \benchref{CCSD(T)/CBS} & \benchcite{\cite{Liang2025,manna2017water27,Bryantsev_2009}} \\
\benchcat{} & \benchset{X40 (GSCDB)} & \benchsize{40} & \benchdesc{Halogenated complexes} & \benchref{CCSD(T)/CBS} & \benchcite{\cite{Liang2025,Rezac2012}} \\
\benchcat{} & \benchset{X40x5 (GSCDB)} & \benchsize{200} & \benchdesc{Halogen PECs} & \benchref{CCSD(T)/CBS} & \benchcite{\cite{Liang2025,Rezac2012}} \\
\benchcat{} & \benchset{XB20 (GSCDB)} & \benchsize{20} & \benchdesc{Halogen-bonded dimers} & \benchref{CCSD(T)/CBS} & \benchcite{\cite{Liang2025,Kozuch_2013}} \\

\benchcat{\textbf{Thermochemistry (GSCDB)}} & \benchset{AE11 (GSCDB)} & \benchsize{11} & \benchdesc{Atomic energies (Ar–Rn)} & \benchref{CCSD(T)/CBS} & \benchcite{\cite{Liang2025,mccarthy2011accurate}} \\
\benchcat{} & \benchset{AE18 (GSCDB)} & \benchsize{18} & \benchdesc{Atomic energies (H–Ar)} & \benchref{CCSD(T)/CBS} & \benchcite{\cite{Liang2025,Chakravorty_1993}} \\
\benchcat{} & \benchset{AL2X6 (GSCDB)} & \benchsize{6} & \benchdesc{AlX$_3$ dimerisation} & \benchref{CCSD(T)/CBS} & \benchcite{\cite{Liang2025,johnson2008delocalization}} \\
\benchcat{} & \benchset{ALK8 (GSCDB)} & \benchsize{8} & \benchdesc{Alkaline reactions} & \benchref{CCSD(T)/CBS} & \benchcite{\cite{Liang2025,Goerigk2017}} \\
\benchcat{} & \benchset{AlkAtom19 (GSCDB)} & \benchsize{19} & \benchdesc{Alkane atomisation} & \benchref{CCSD(T)/CBS} & \benchcite{\cite{Liang2025,Karton_2009}} \\
\benchcat{} & \benchset{ALKBDE10 (GSCDB)} & \benchsize{10} & \benchdesc{Group 1/2 diatomic BDEs} & \benchref{CCSD(T)/CBS} & \benchcite{\cite{Liang2025,Yu_2015Components}} \\
\benchcat{} & \benchset{AlkIsod14 (GSCDB)} & \benchsize{14} & \benchdesc{Alkane isodesmic} & \benchref{CCSD(T)/CBS} & \benchcite{\cite{Liang2025,Karton_2009}} \\
\benchcat{} & \benchset{BDE99MR (GSCDB)} & \benchsize{16} & \benchdesc{W4-11 MR BDEs} & \benchref{CCSD(T)/CBS} & \benchcite{\cite{Liang2025,Chan2017}} \\
\benchcat{} & \benchset{BDE99nonMR (GSCDB)} & \benchsize{83} & \benchdesc{W4-11 SR BDEs} & \benchref{CCSD(T)/CBS} & \benchcite{\cite{Liang2025,Chan2017}} \\
\benchcat{} & \benchset{BH76RC (GSCDB)} & \benchsize{30} & \benchdesc{BH76 reaction energies} & \benchref{CCSD(T)/CBS} & \benchcite{\cite{Liang2025,zhao2005multi,zhao2005benchmark}} \\
\benchcat{} & \benchset{BSR36 (GSCDB)} & \benchsize{36} & \benchdesc{Hydrocarbon bond separations} & \benchref{CCSD(T)/CBS} & \benchcite{\cite{Liang2025,Chan2016BSR,steinmann2009unified}} \\
\benchcat{} & \benchset{CR20 (GSCDB)} & \benchsize{20} & \benchdesc{Cycloreversion energies} & \benchref{CCSD(T)/CBS} & \benchcite{\cite{Liang2025,yu2016can}} \\
\benchcat{} & \benchset{DARC (GSCDB)} & \benchsize{14} & \benchdesc{Diels–Alder energies} & \benchref{CCSD(T)/CBS} & \benchcite{\cite{Liang2025,johnson2008delocalization}} \\
\benchcat{} & \benchset{DC13 (GSCDB)} & \benchsize{13} & \benchdesc{Difficult cases} & \benchref{CCSD(T)/CBS} & \benchcite{\cite{Liang2025,Goerigk2017}} \\
\benchcat{} & \benchset{DIPCS9 (GSCDB)} & \benchsize{9} & \benchdesc{Double ionisation potentials} & \benchref{CCSD(T)/CBS} & \benchcite{\cite{Liang2025,Goerigk2017}} \\
\benchcat{} & \benchset{EA50 (GSCDB)} & \benchsize{50} & \benchdesc{Electron affinities} & \benchref{CCSD(T)/CBS} & \benchcite{\cite{Liang2025,Ermis2021}} \\
\benchcat{} & \benchset{FH51 (GSCDB)} & \benchsize{51} & \benchdesc{Reaction energies} & \benchref{CCSD(T)/CBS} & \benchcite{\cite{Liang2025,Friedrich_2015,friedrich2013incremental}} \\
\benchcat{} & \benchset{G21EA (GSCDB)} & \benchsize{25} & \benchdesc{Electron affinities} & \benchref{CCSD(T)/CBS} & \benchcite{\cite{Liang2025,Parthiban2001}} \\
\benchcat{} & \benchset{G21IP (GSCDB)} & \benchsize{36} & \benchdesc{Ionisation potentials} & \benchref{CCSD(T)/CBS} & \benchcite{\cite{Liang2025,Parthiban2001}} \\
\benchcat{} & \benchset{G2RC24 (GSCDB)} & \benchsize{24} & \benchdesc{G2/97 reaction energies} & \benchref{CCSD(T)/CBS} & \benchcite{\cite{Liang2025,curtiss1991gaussian}} \\
\benchcat{} & \benchset{HAT707MR (GSCDB)} & \benchsize{202} & \benchdesc{Heavy-atom transfer (MR)} & \benchref{CCSD(T)/CBS} & \benchcite{\cite{Liang2025,Karton2017HAT}} \\
\benchcat{} & \benchset{HAT707nonMR (GSCDB)} & \benchsize{505} & \benchdesc{Heavy-atom transfer (SR)} & \benchref{CCSD(T)/CBS} & \benchcite{\cite{Liang2025,Karton2017HAT}} \\
\benchcat{} & \benchset{HEAVYSB11 (GSCDB)} & \benchsize{11} & \benchdesc{Heavy-element dissociation} & \benchref{CCSD(T)/CBS} & \benchcite{\cite{Liang2025,Goerigk2017}} \\
\benchcat{} & \benchset{HNBrBDE18 (GSCDB)} & \benchsize{18} & \benchdesc{N–Br BDEs} & \benchref{CCSD(T)/CBS} & \benchcite{\cite{Liang2025,Chan2018}} \\
\benchcat{} & \benchset{IP23 (GSCDB)} & \benchsize{23} & \benchdesc{Vertical ionisation potentials} & \benchref{CCSD(T)/CBS} & \benchcite{\cite{Liang2025,Cheng2007}} \\
\benchcat{} & \benchset{IP30 (GSCDB)} & \benchsize{30} & \benchdesc{Vertical ionisation potentials} & \benchref{CCSD(T)/CBS} & \benchcite{\cite{Liang2025,Luo2012}} \\
\benchcat{} & \benchset{MB08-165 (GSCDB)} & \benchsize{165} & \benchdesc{Mindless molecules} & \benchref{CCSD(T)/CBS} & \benchcite{\cite{Liang2025,korth2009mindless}} \\
\benchcat{} & \benchset{MB16-43 (GSCDB)} & \benchsize{43} & \benchdesc{Mindless molecules} & \benchref{CCSD(T)/CBS} & \benchcite{\cite{Liang2025,Goerigk2017}} \\
\benchcat{} & \benchset{MX34 (GSCDB)} & \benchsize{34} & \benchdesc{Ionic cluster atomisation} & \benchref{CCSD(T)/CBS} & \benchcite{\cite{Liang2025,Chan_2023}} \\
\benchcat{} & \benchset{NBPRC (GSCDB)} & \benchsize{12} & \benchdesc{Oligomerisation/fragmentation} & \benchref{CCSD(T)/CBS} & \benchcite{\cite{Liang2025,goerigk2010general,goerigk2011efficient}} \\
\benchcat{} & \benchset{P34AE (GSCDB)} & \benchsize{44} & \benchdesc{P-block atomisation} & \benchref{CCSD(T)/CBS} & \benchcite{\cite{Liang2025,Chan_2021}} \\
\benchcat{} & \benchset{P34EA (GSCDB)} & \benchsize{9} & \benchdesc{P-block electron affinities} & \benchref{CCSD(T)/CBS} & \benchcite{\cite{Liang2025,Chan_2021}} \\
\benchcat{} & \benchset{P34IP (GSCDB)} & \benchsize{15} & \benchdesc{P-block ionisation potentials} & \benchref{CCSD(T)/CBS} & \benchcite{\cite{Liang2025,Chan_2021}} \\
\benchcat{} & \benchset{PA26 (GSCDB)} & \benchsize{26} & \benchdesc{Proton affinities} & \benchref{CCSD(T)/CBS} & \benchcite{\cite{Liang2025,Goerigk2017}} \\
\benchcat{} & \benchset{PlatonicRE18 (GSCDB)} & \benchsize{18} & \benchdesc{Platonic reaction energies} & \benchref{CCSD(T)/CBS} & \benchcite{\cite{Liang2025,Karton_2016Heats}} \\
\benchcat{} & \benchset{PlatonicTAE6 (GSCDB)} & \benchsize{6} & \benchdesc{Platonic atomisation energies} & \benchref{CCSD(T)/CBS} & \benchcite{\cite{Liang2025,Karton_2016Heats}} \\
\benchcat{} & \benchset{RC21 (GSCDB)} & \benchsize{21} & \benchdesc{Radical cation reactions} & \benchref{CCSD(T)/CBS} & \benchcite{\cite{Liang2025,grimme2013towards}} \\
\benchcat{} & \benchset{RSE43 (GSCDB)} & \benchsize{43} & \benchdesc{Radical stabilisation} & \benchref{CCSD(T)/CBS} & \benchcite{\cite{Liang2025,Zhao2012RSE,neese2009assessment}} \\

\benchcat{\textbf{Transition metals (GSCDB)}} & \benchset{3dTMV} & \benchsize{reported in Ref.~\cite{Neugebauer2023}} & \benchdesc{3d-metal vertical ionisation energies} & \benchref{ph-AFQMC} & \benchcite{\cite{Liang2025,Neugebauer2023}} \\
\benchcat{} & \benchset{3d4dIPSS (GSCDB)} & \benchsize{32} & \benchdesc{TM ionisation potentials} & \benchref{CCSD(T)} & \benchcite{\cite{Liang2025,balabanov2006basis}} \\
\benchcat{} & \benchset{CUAGAU83 (GSCDB)} & \benchsize{83} & \benchdesc{Coinage complexes} & \benchref{CCSD(T)} & \benchcite{\cite{Liang2025,Chan_2019}} \\
\benchcat{} & \benchset{DAPD (GSCDB)} & \benchsize{12} & \benchdesc{Pd diatomics} & \benchref{CCSD(T)} & \benchcite{\cite{Liang2025,chan2023dapd}} \\
\benchcat{} & \benchset{MME52 (GSCDB)} & \benchsize{52} & \benchdesc{Metalloenzymes} & \benchref{DLPNO-CCSD(T)} & \benchcite{\cite{Liang2025,Wappett2023,Rezac2011_1}} \\
\benchcat{} & \benchset{MOBH28 (GSCDB)} & \benchsize{28} & \benchdesc{Organometallic barriers} & \benchref{CCSD(T)} & \benchcite{\cite{Liang2025,iron2019evaluating}} \\
\benchcat{} & \benchset{ROST61 (GSCDB)} & \benchsize{61} & \benchdesc{Open-shell reactions} & \benchref{CCSD(T)} & \benchcite{\cite{Liang2025,Maurer_2021}} \\
\benchcat{} & \benchset{TMD10 (GSCDB)} & \benchsize{10} & \benchdesc{TM diatomics} & \benchref{CCSD(T)} & \benchcite{\cite{Liang2025,chan2019assessment}} \\
\benchcat{} & \benchset{MOR13 (GSCDB)} & \benchsize{13} & \benchdesc{Closed-shell TM reactions} & \benchref{CCSD(T)} & \benchcite{\cite{Liang2025,chan2019assessment}} \\
\benchcat{} & \benchset{TMB11 (GSCDB)} & \benchsize{11} & \benchdesc{TM barriers} & \benchref{CCSD(T)} & \benchcite{\cite{Liang2025,chan2019assessment}} \\

\benchcat{\textbf{Conformers (external)}} & \benchset{37CONF8} & \benchsize{296} & \benchdesc{Small organic conformers} & \benchref{CCSD(T)/CBS} & \benchcite{\cite{sharapa2019robust}} \\
\benchcat{} & \benchset{ACONFL} & \benchsize{reported in Ref.~\cite{ehlert2022conformational}} & \benchdesc{n-alkane conformers} & \benchref{CCSD(T)/CBS} & \benchcite{\cite{ehlert2022conformational,werner2023accurate}} \\
\benchcat{} & \benchset{DipConfS} & \benchsize{reported in Ref.~\cite{plett2024toward}} & \benchdesc{Amino-acid and dipeptide conformers} & \benchref{CCSD(T)/CBS} & \benchcite{\cite{plett2024toward}} \\
\benchcat{} & \benchset{Maltose222} & \benchsize{222} & \benchdesc{Carbohydrate conformers} & \benchref{CCSD(T)/CBS} & \benchcite{\cite{marianski2016assessing}} \\
\benchcat{} & \benchset{MPCONF196} & \benchsize{196} & \benchdesc{Medicinal-chemistry conformers} & \benchref{CCSD(T)/CBS} & \benchcite{\cite{rezac2018mpconf196,plett2023mpconf196water}} \\
\benchcat{} & \benchset{OpenFF-Tors} & \benchsize{reported in Ref.~\cite{behara2024openff}} & \benchdesc{Drug-like torsional profiles} & \benchref{CCSD(T)/CBS} & \benchcite{\cite{behara2024openff}} \\
\benchcat{} & \benchset{UPU46} & \benchsize{46} & \benchdesc{RNA backbone conformers} & \benchref{CCSD(T)/CBS} & \benchcite{\cite{kruse2015quantum}} \\

\benchcat{\textbf{Non-covalent interactions}} & \benchset{S30L} & \benchsize{30} & \benchdesc{Host–guest supramolecular} & \benchref{DLPNO-CCSD(T)} & \benchcite{\cite{Sure2015}} \\
\benchcat{} & \benchset{IHB100x10} & \benchsize{1000} & \benchdesc{Ionic hydrogen-bond dissociation curves (NCI Atlas)} & \benchref{CCSD(T)/CBS} & \benchcite{\cite{ez2020-1}} \\
\benchcat{} & \benchset{PLA15} & \benchsize{15} & \benchdesc{Protein active sites} & \benchref{MP2-F12 + DLPNO-CCSD(T)} & \benchcite{\cite{kriz2020benchmarking}} \\
\benchcat{} & \benchset{PLF547} & \benchsize{547} & \benchdesc{Protein–fragment interactions} & \benchref{MP2-F12/cc-pVDZ-F12 + DLPNO-CCSD(T)} & \benchcite{\cite{kriz2020benchmarking}} \\
\benchcat{} & \benchset{QUID} & \benchsize{170} & \benchdesc{Protein-pocket dimers} & \benchref{LNO-CCSD(T)} & \benchcite{\cite{puleva2025quid}} \\
\benchcat{} & \benchset{Alkali-halide PECs (LiCl/NaCl/KBr)} & \benchsize{3 PECs} & \benchdesc{Dissociation curves} & \benchref{DFT ($\omega$B97M-V)} & \benchcite{This work} \\
\benchcat{\textbf{Molecular Crystals}} & \benchset{X23-DMC} & \benchsize{23} & \benchdesc{Molecular crystal lattice energies} & \benchref{DMC} & \benchcite{\cite{DellaPia2024}} \\
\benchcat{} & \benchset{CPOSS209} & \benchsize{209} & \benchdesc{Molecular crystal lattice energies} & \benchref{\wb + 1-body CCSD(T) correction} & \benchcite{\cite{cposs209}, This work} \\
\benchcat{\textbf{Liquids \& Water}} & \benchset{Water density/RDF} & \benchsize{333} & \benchdesc{Liquid water (NPT MD)} & \benchref{Experiment targets} & \benchcite{\cite{soper_radial_2013}} \\
\benchcat{} & \benchset{Organic liquids densities} & \benchsize{62} & \benchdesc{Experimental densities (NPT MD)} & \benchref{Experiment targets} & \benchcite{\cite{Weber2025MPNICE}} \\
\benchcat{\textbf{Solvation / Redox}} & \benchset{Fe/Cl aqueous redox clusters} & \benchsize{traj. (ps)} & \benchdesc{Solvated Fe/Cl chlorides} & \benchref{DFT} & \benchcite{\cite{kocer2024machinelearningpotentialsredox}} \\
\benchcat{} & \benchset{Solvated Cl$_2$/Cl$^-$ dissociation} & \benchsize{traj. (ps)} & \benchdesc{Charge-localisation test in water clusters} & \benchref{DFT ($\omega$B97M-V)} & \benchcite{This work} \\
\benchcat{} & \benchset{Water-cluster dissociation (charge transfer)} & \benchsize{scan of separations} & \benchdesc{Fragment-charge localisation during dissociation} & \benchref{DFT ($\omega$B97M-V)} & \benchcite{This work} \\
\benchcat{} & \benchset{Hydrated TM ionisation (M-W6/M-W18)} & \benchsize{14} & \benchdesc{TM ionisation in water} & \benchref{DLPNO-CCSD(T)} & \benchcite{\cite{bhattacharjee2022dlpno}} \\
\benchcat{} & \benchset{Redox Potentials in solution} & \benchsize{traj. (ps)} & \benchdesc{Solvated Fe/Co chlorides} & \benchref{DFT} & \benchcite{\cite{kocer2024machinelearningpotentialsredox}} \\
\benchcat{\textbf{Lanthanides}} & \benchset{Lanthanide isomers} & \benchsize{18} & \benchdesc{Lanthanide complex isomers} & \benchref{r2SCAN-3c} & \benchcite{\cite{Rose2024}} \\
\end{longtable*}

\end{document}